\documentclass[notitlepage,onecolumn,aps,showpacs,pra,superscriptaddress,nofootinbib]{revtex4-1}

\usepackage{graphicx}
\usepackage{amsfonts}
\usepackage{amssymb}
\usepackage{amsmath}
\usepackage{color}
\usepackage{subfigure}
\newcommand{\ii}{\mathrm{i}}
\newcommand{\rr}{\vec{r}}

\begin{document}

\title{Existence, stability and nonlinear dynamics of vortices and vortex clusters \\in anisotropic Bose-Einstein condensates}
\author{J. Stockhofe}
\email{jstockho@physnet.uni-hamburg.de}
\affiliation{Zentrum f\"ur Optische Quantentechnologien, Universit\"at Hamburg, Luruper Chaussee 149, 22761 Hamburg, Germany}
\author{P. G.\ Kevrekidis}
\affiliation{Department of Mathematics and Statistics, University of Massachusetts,
Amherst MA 01003-4515, USA}
\author{P.\ Schmelcher}
\affiliation{Zentrum f\"ur Optische Quantentechnologien, Universit\"at Hamburg, Luruper Chaussee 149, 22761 Hamburg, Germany}

\begin{abstract}

This chapter is devoted to the study of vortex excitations in one-component Bose-Einstein condensates, with a special emphasis on the impact of 
anisotropic confinement on the existence, stability and dynamical properties of 
vortices and particularly few-vortex clusters. Symmetry breaking features
are pervasive within this system even in its isotropic installment,
where cascades of symmetry breaking bifurcations give rise to the
multi-vortex clusters, but also within the anisotropic realm which
naturally breaks the rotational symmetry of the multi-vortex states.
Our first main tool for analyzing the system consists of a weakly
nonlinear (bifurcation) approach which starts from the linear
states of the problem and examines their continuation and bifurcation
into novel symmetry-broken configurations in the nonlinear case. This is
first done in the isotropic limit and the modifications introduced
by the anisotropy are subsequently presented. The second main
tool concerns the highly nonlinear regime where the vortices can
be considered as individual topologically charged 
``particles'' which precess within the parabolic trap and interact 
with each other, similarly to fluid vortices.
The conclusions stemming from both the bifurcation and the interacting
particle picture are corroborated by numerical computations which
are also used to bridge the gap between these two opposite-end regimes.
\end{abstract}

\maketitle

\section{Introduction}

The study of topologically charged vortex states is a theme of wide
appeal in numerous fields such as superfluid helium \cite{Donnelly2005}, type-II superconductors \cite{Blatter1994}, nonlinear optics~\cite{Kivshar1998,desya} and atomic Bose-Einstein condensates 
(BECs)~\cite{9780198507192,pethick,book_kevrekidis},
among many others~\cite{book_pismen}. Admittedly, BECs constitute one of the 
most pristine settings where structural and
dynamical properties of single- and multi-vortex (both of the same
and of opposite charge) and multi-charged-vortex
states can be investigated not only theoretically and computationally but also 
by means of a wide array of experimental techniques. 

The main focus of study within this theme of vortex dynamics over the
past decade has been the examination of single- and multi-charge
vortices, as well as of highly structured vortex 
lattices~\cite{book_kevrekidis}. On the other hand, far less attention
has been paid to the dynamics of small clusters of (few) vortices.
In this context, it is particularly relevant to understand 
both the potential existence of stationary or periodic orbits in
such systems, and their dynamical stability, as well as the motion
near such ``equilibria'', but also the generic motion of the vortices
in these clusters far from their equilibrium configurations. Questions
concerning also the integrable or non-integrable structure of the 
vortex trajectories and the existence of chaotic dynamics therein
constitute a fascinating topic for further investigation.
Much of the relevant literature has been reserved to the fundamental
(beyond the single vortex) building block of the 
vortex dipole~\cite{PhysRevE.66.036612,PhysRevA.68.063609,PhysRevA.65.043612,PhysRevA.71.033626,PhysRevA.74.023603,PhysRevA.83.011603,PhysRevA.77.053610,PhysRevA.82.013646,Torres20113044}. Recently, this state has
also received considerable experimental 
attention~\cite{PhysRevLett.104.160401,D.V.Freilich09032010,PhysRevA.84.011605}
but other configurations have been 
considered theoretically as well~\cite{PhysRevA.71.033626,PhysRevA.74.023603,PhysRevA.82.013646,Middelkamp2010b} and
are becoming amenable to experiments~\cite{Seman2009}.
Let us remark that from the theoretical side small vortex clusters have also been studied in the presence of periodic lattice potentials \cite{Mayteevarunyoo20091439}, and it was demonstrated how introducing different lattice parameters along the different axes can significantly affect their stability properties.
In the presence of a periodic lattice potential, also more complex entities such as
super-vortices have been constructed~\cite{sakaguchi2005}.
Recently, there has also been an increase in interest in so-called azimuthon excitations, i.e. vortices accompanied by azimuthal density modulations around their cores \cite{PhysRevLett.95.203904,PhysRevA.77.025602}. Clusters of these have been studied in \cite{PhysRevA.85.013620}.

Our aim here, in considerable extension of the recent discussion
of~\cite{aniso}, is to examine the context where vortex clusters
become subject to anisotropy in the harmonic trapping potential. Although some effort has been invested
in such investigations in the context of co-rotating vortex systems
by the 
works of~\cite{PhysRevA.79.053616,PhysRevA.82.013628,PhysRevA.83.053612},
very little attention has been paid to this subject in the case of
counter-rotating vortices. In that light, we offer a perspective 
encompassing two complementary approaches: close to the linear
limit, we develop a weakly
nonlinear (bifurcation) approach
which examines the continuation of various states and their bifurcation
into novel symmetry-broken states in the nonlinear regime;
far from the linear regime (in fact, in the highly nonlinear regime),
the vortices can be
considered as individual topologically charged 
``particles'' for which (ordinary differential) equations of motion
describing their precessions and interactions are devised
and analyzed. Our numerical analysis of the system corroborates these limits
and bridges them by means of detailed computations~\footnote{The connection 
provided through the numerical results is often
essential as some configurations may e.g. be stable in the above
two limits but possess instabilities in finite intermediate ranges
of parameter values that would not be observable by restricting our
view to the analytically tractable limits. A notable example of
this type is offered by the vortex quadrupole configuration 
(see e.g. Fig.~8 of~\cite{PhysRevA.82.013646} and equivalently
the isotropic limit of both Fig.~\ref{fig:quadA} and~\ref{fig:quadB} below).
Such a state is found to be linearly stable in both of the above
quasi-analytical limits and its intermediate range of instability
parameter values is only detected by the bridging numerical continuation.}.

The presentation of this chapter will be structured as follows.
In section 2, we present the model and its theoretical setup. 
Upon a discussion of the highly nonlinear particle-type description of vortex dynamics in anisotropic traps in section 3, we
focus on aligned vortex clusters in sections 4 and 5 from the
particle and bifurcation perspectives, respectively. In section 6,
we address non-aligned vortex clusters, which are examined in the
presence of anisotropy in section 7. Finally, in section 8, we
summarize our findings and present our conclusions.

\section{Model and Theoretical Setup}
Our prototypical model for the pancake-shaped Bose-Einstein condensates
under consideration is the 2D Gross-Pitaevskii equation (GPE) for the condensate wavefunction $\psi(x,y,t)$
which can be cast into a convenient dimensionless form by measuring length, time, energy and density $|\psi|^2$ in units of $a_z$ (harmonic
oscillator length in the $z$-direction), $\omega_z^{-1}$ (inverse
trap strength in the $z$-direction), 
$\hbar \omega_z$ and $(2\sqrt{2\pi}|a|a_z)^{-1}$, respectively.
Here, $a$ denotes the s-wave scattering length, encoding the low-energy limit of the interaction between the bosons \cite{pethick}.
In the rescaled variables the model of interest is given by:
\begin{equation}
 \ii \partial_t \psi(x,y,t) = \left[ -\frac{1}{2} {\nabla_\perp}^2 + V(x,y) + \sigma | \psi(x,y,t) |^2 \right] \psi(x,y,t),
\label{th:eq: GPE2d2}
\end{equation}
where $\sigma$ denotes the sign of the s-wave scattering length $a$. In the following we will exclusively discuss the case of repulsive interaction, $\sigma = +1$, which ensures stability of the condensate against collapse.

In Eq.~(\ref{th:eq: GPE2d2}), the harmonic (parabolic trap) potential 
that will be considered in this work is 
given by $V(x,y)=(\omega_x^2 x^2 + \omega_y^2 y^2)/2$,
where the trapping frequencies in the plane
have already been rescaled by the
trap strength in the $z$-direction. 
All equations will be
presented in dimensionless units for simplicity.

Below, we will analyze the existence and
linear stability of the stationary modes of
Eq.~(\ref{th:eq: GPE2d2}). These are obtained in the form $\psi(x,y,t) = \phi(x,y)\exp(-\ii\mu t)$, where $\mu$ denotes the chemical potential.
Substituting into Eq.~(\ref{th:eq: GPE2d2}) yields the stationary 2D Gross-Pitaevskii equation for $\phi(x,y)$.
Numerically the relevant nonlinear stationary states will be identified as a
function of the chemical potential $\mu$ and often as a function
of the anisotropy by means of a Newton-Krylov
scheme over a rectangular two-dimensional domain with suitably
small spacing.
Linear stability will be explored
by means of the Bogoliubov-de Gennes (BdG) analysis.
This involves the derivation of the BdG equations, which stem from a 
linearization
of the GPE Eq.~(\ref{th:eq: GPE2d2}) around the
stationary solution $\phi(x,y)$ by using the ansatz \cite{book_kevrekidis}
\begin{eqnarray}
\psi(x,y,t) =\left[ \phi(x,y) + a(x,y) e^{-\ii \omega t} + b^{\ast}(x,y) e^{+\ii \omega^{\ast} t} \right] e^{-\ii \mu t},
\end{eqnarray}
where $\ast$ indicates complex conjugation, and expanding to first order in $a, b$. The
solution of the ensuing BdG eigenvalue
problem yields the eigenvectors $\left( a(x,y), b(x,y)\right)$ and
eigenfrequencies $\omega$. As concerns the latter, we note that due
to the Hamiltonian nature of the system, if $\omega$ is an eigenfrequency
of the Bogoliubov-de Gennes spectrum, so are $-\omega$, $\omega^{\ast}$ and
$-\omega^{\ast}$. Notice that a linearly stable configuration is
tantamount to ${\rm Im}(\omega) =0$, i.e., all eigenfrequencies being real.

An important quantity resulting from the BdG analysis is the energy
carried by the normal mode with eigenfrequency $\omega$, namely,
\begin{equation}
E=\omega\int{\text{d}x\text{d}y(|a|^2-|b|^2)}.
\label{energy}
\end{equation}
The sign of this quantity, known as {\it Krein sign} \cite{Kapitula2004263},
is a topological property of each eigenmode.
Let us remark that for a real mode $\omega$, both $(\omega,a,b)$ and $(-\omega,b^\ast,a^\ast)$ solve the BdG equations, such that both modes of this pair have the same $E$ and the same Krein signature according to Eq.~(\ref{energy}).
For eigenfrequencies with a nonvanishing imaginary part, one can show that $E=0$ \cite{book_kevrekidis}.

A BdG mode for which the sign of $E$ is negative is called {\it anomalous mode} \cite{9780198507192},
or {\it negative energy mode}
\cite{PhysRevA.63.013602}, or mode with {\it negative Krein signature}
\cite{Kapitula2004263}.
If in the course of tuning one of the system's parameters such a mode becomes resonant with a mode with positive Krein signature
then, typically,
complex frequencies appear in the excitation
spectrum, i.e., a dynamical instability arises \cite{MacKay}.

\section{A vortex particle picture in anisotropic traps}
Let us in the following consider a harmonic trapping potential $V(x,y) = (\omega_x^2 x^2 + \omega_y^2 y^2)/2$, where the two dimensionless trapping frequencies $\omega_x$, $\omega_y$ can be different.
The parameter $\alpha=\omega_y/\omega_x$ will be used to quantify the anisotropy of the trap.
For simplicity, we restrict the discussion to singly-charged vortices with charge $s \in \{ \pm 1 \}$, although generalizations to multiply-quantized vortices are available \cite{Fetter2001}. 
In \cite{Svidzinsky2000,Fetter2001} a set of vortex precession ODEs is derived within the full three-dimensional Gross-Pitaevskii framework which also holds in the anisotropic regime of $\omega_x \neq \omega_y$.
Matching this result to the expressions obtained in \cite{0953-4075-43-15-155303} for the two-dimensional GPE in the isotropic limit, we find the following set of ordinary differential equations governing the precessional dynamics of a single vortex of charge~$s$:
\begin{equation}
 \dot{x} = - s \omega_y^2 Q y , \ \ \ \ \, \, \, \dot{y} = s \omega_x^2 Q x.
\label{eq:vordynaniso}
\end{equation}
where $Q(\mu, \omega_x, \omega_y) = \ln \left( A \mu/\omega_\text{eff} \right)/({2\mu})$, $\omega_\text{eff} = \sqrt{(\omega_x^2 + \omega_y^2)/2}$, and the numerical constant $A \approx 8.88$.

It is straightforward to check that $(x^2/\omega_y^2 + y^2/\omega_x^2)$ is a constant of motion for this dynamical system.
Thus, these equations describe elliptical vortex orbits in the anisotropic trap, where the precession frequency at which both the $x$ and $y$ coordinates perform harmonic oscillations is given by $\omega_\text{pr} = \omega_x \omega_y Q$.
We note that in isotropic traps the precession frequency is known to increase as a function of displacement from the trap center \cite{Fetter2001}, but this correction is small for vortices close to the center and we will neglect it 
here for simplicity. Additionally, we point out that the above mentioned
elliptical orbits of a single vortex inside the parabolic trap 
naturally degenerate into circular ones in the isotropic limit 
of $\omega_x=\omega_y$.

In the presence of more than one vortex, there is an additional interaction contribution to the equations of motion: Each vortex moves with the local velocity field created by all the other vortices \cite{Kevrekidis2004c,Newton2009}.
Neglecting modifications of the velocity field profiles due to the inhomogeneous condensate background in the presence of the trap, we can employ the interaction term that has also been used in \cite{PhysRevA.82.013646,PhysRevA.84.011605}, yielding
\begin{align}
\dot{x}_k = - s_k \omega_y^2 Q y_k + B \sum_{j \neq k} s_j \frac{y_j-y_k}{2 \rho_{jk}^2}, \quad
 \dot{y}_k = s_k \omega_x^2 Q x_k - B \sum_{j \neq k} s_j \frac{x_j-x_k}{2 \rho_{jk}^2}, 
\label{eq:partpict}
\end{align}
where $\rho_{jk}^2 = (x_j - x_k)^2 + (y_j - y_k)^2$. $B$ is a numerical constant, which in earlier works for isotropically trapped quasi two-dimensional condensates of aspect ratio $\omega_r/\omega_z = 0.2$ (which coincides with the isotropic limit in our simulations) has been found to be $B \approx 1.35$ from fits of the dipole's equilibrium position \cite{PhysRevA.82.013646,PhysRevA.84.011605}. We will use this value of $B$ whenever numerically evaluating results from the above ODE system.

The ``particle picture'' of Eqs.~\ref{eq:partpict} will be used throughout this work to study equilibrium positions, linearization frequencies and dynamics of few-vortex arrangements in anisotropic traps.

Before concluding this section, let us remark that the above equations of motion for the vortices can also be obtained from a suitable Hamiltonian with a logarithmic interaction potential \cite{Newton2009}.
A similar type of interaction term has also been used in the study of mesoscopic systems, in particular to model electrostatically interacting charged balls of millimetre size free to move on a plane conductor \cite{0295-5075-55-1-045,PhysRevE.72.046122}.

\section{Aligned vortex states in the particle picture}
\label{aniso:aligned}
One particular class of stationary vortex clusters in Bose-Einstein condensates has received considerable attention in the past years, namely configurations where a number of singly-charged vortices is located along one of the symmetry axes of the trap, and the sign of the vortex charges is alternating between adjacent vortices, see e.g. \cite{PhysRevA.71.033626,PhysRevA.74.023603,PhysRevA.82.013646}.
We will refer to these solutions of the Gross-Pitaevskii equation as ``aligned vortex states''.
In the following, we will apply the particle picture ODEs to determine equilibrium positions and linearization frequencies of the aligned vortex cluster states. In contrast to numerous previous works on the theme of counter-rotating 
vortices 
(see \cite{PhysRevA.82.013646} for a recent discussion of the relevant
literature), we will not restrict the analysis to isotropic traps, but 
allow for different trapping frequencies in the $x$- and $y$-direction.
This will turn out to have important consequences for the stability of the vortex clusters.
It should be noted again, on the other hand, that the subject of co-rotating
vortices in the presence of anisotropy has been considered in some
detail in the series of 
works~\cite{PhysRevA.79.053616,PhysRevA.82.013628,PhysRevA.83.053612}.

\subsection{Single vortex}
\begin{figure}[ht]
\centering
\subfigure{\includegraphics[height=0.15\textheight]{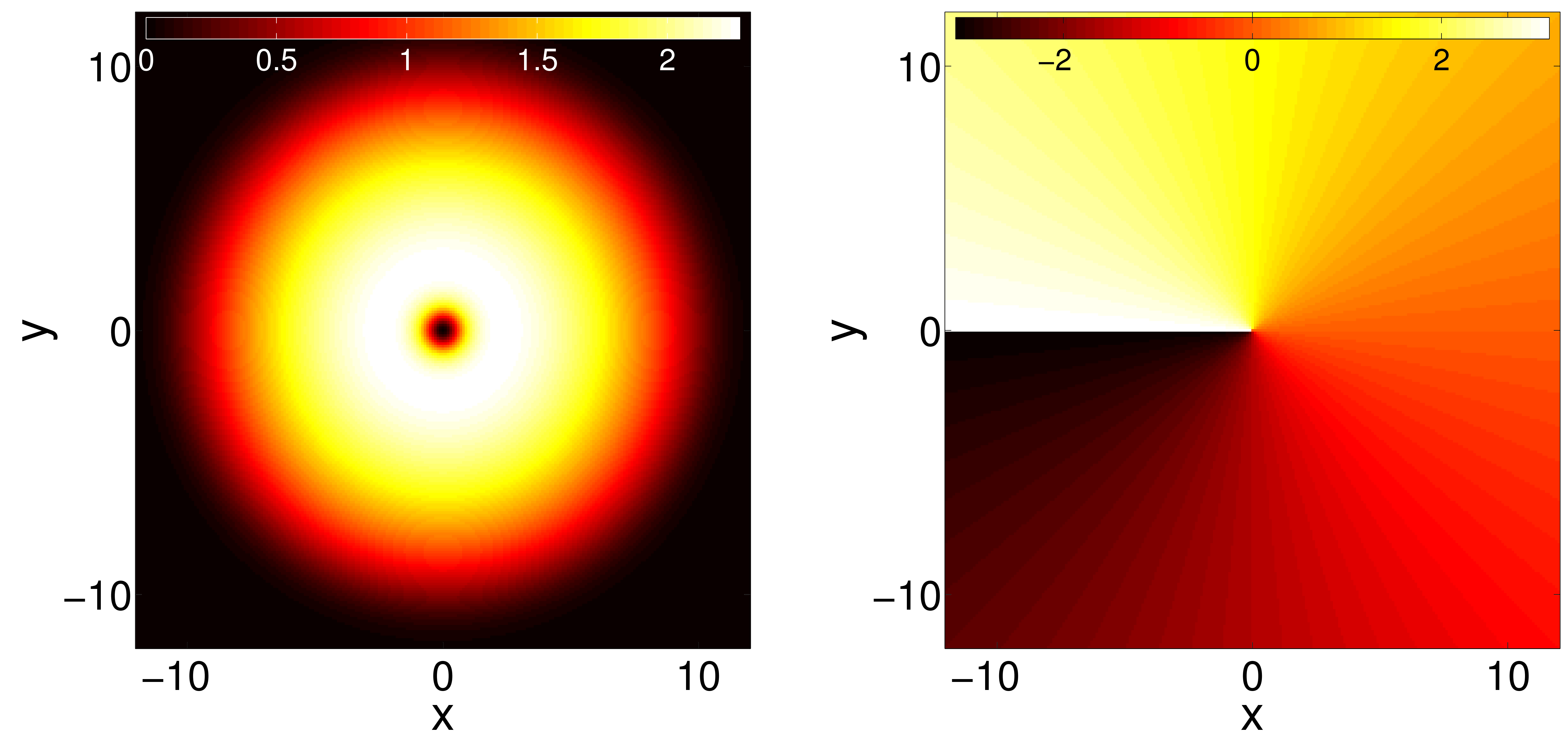}}
\subfigure{\includegraphics[height=0.15\textheight]{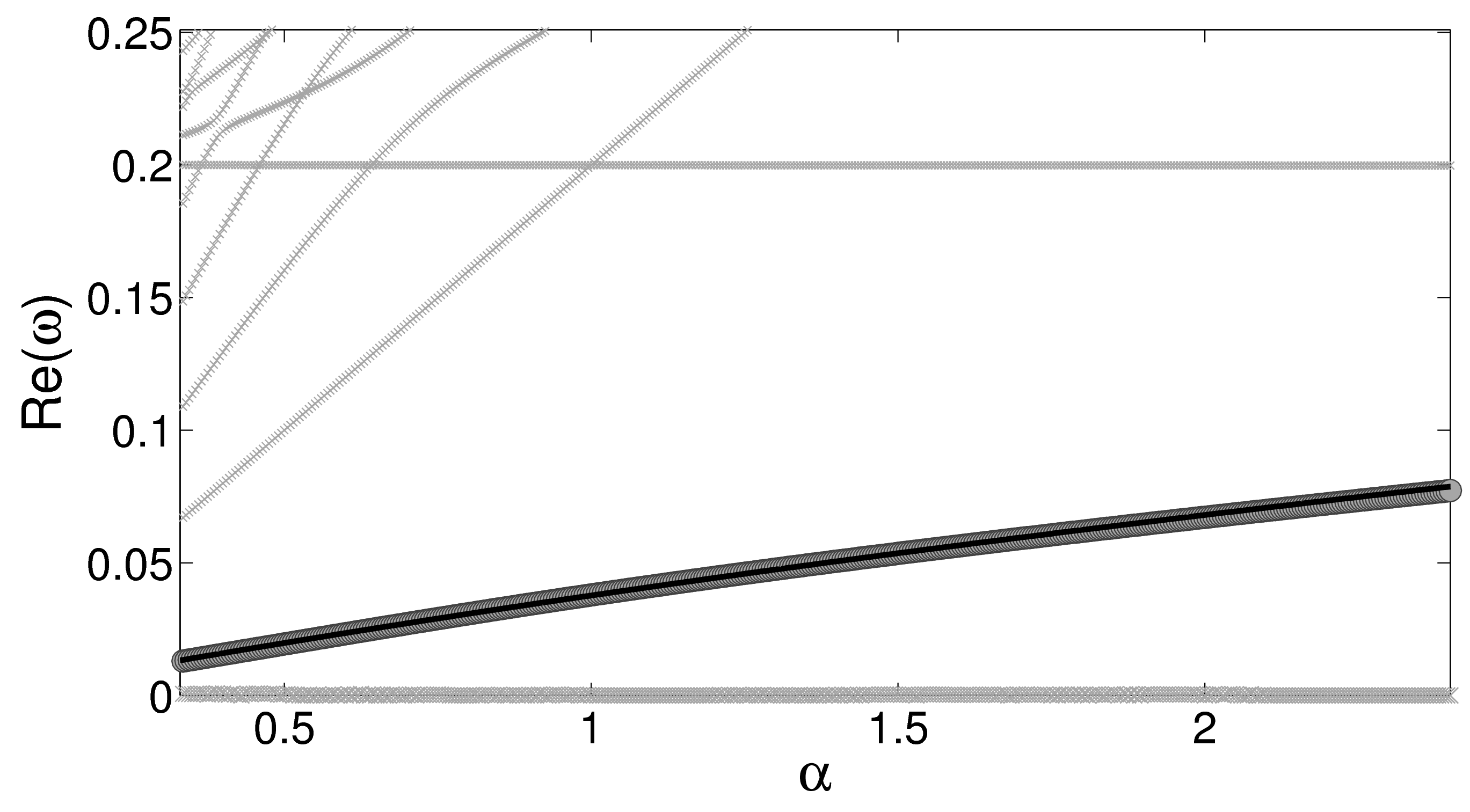}}
\caption{\label{fig:bdg_1vor} Profiles (density in the left panel
and phase in the middle panel) and BdG spectrum (right panel) of the single vortex state, precession frequency from the particle picture in black, coinciding with the anomalous BdG mode (dark gray line made up of circular markers). Chemical potential fixed at $\mu = 2.5$.}
\end{figure}

To start out, we numerically determine the stationary single vortex solution to the full Gross-Pitaevskii equation for different values of $\alpha = \omega_y/\omega_x$ by representing the Laplacian in terms of finite differences on a spatial grid and employing a Newton-Krylov method as described in \cite{kelley}.
Technically, to scan $\alpha$, we fix $\omega_x = 0.2$ throughout this work and vary $\omega_y$. 
The chemical potential is held fixed at $\mu = 2.5$, high enough for the 
Thomas-Fermi (TF) large-density approximation to be applicable.
As expected, and predicted by the particle picture, the stationary single vortex is located at the trap center $x=y=0$ for any value of $\alpha$.
A typical profile of the state's density and phase structure is shown in the left and middle panel of Fig. \ref{fig:bdg_1vor}. Let us remark that these profiles do not show the full size of the grid used in our numerical simulations.

Having numerically identified the vortex solution in various anisotropic settings, we calculate its BdG spectrum as a function of $\alpha$ by diagonalizing the ensuing BdG matrix.
The resulting spectrum is shown in Fig.~\ref{fig:bdg_1vor}, together with the linearization frequency calculated from the particle picture (black line).
The spectrum contains one anomalous (negative Krein) mode indicated by the gray circles.
This anomalous mode in the BdG spectrum is connected to the precessional motion of the vortex \cite{PhysRevLett.86.564}.
Exciting it slightly shifts the vortex from its equilibrium position and makes it precess around the trap center.
Obviously, this mode's numerically found dependence  on $\alpha$ is very well described by the precession frequency of the particle picture, both in the isotropic limit and in anisotropic settings.
Finally, we stress that the BdG spectrum contains no eigenfrequencies with non-zero imaginary part, indicating that the single vortex state is dynamically stable in arbitrary anisotropic traps.
 
Apart from the anomalous mode responsible for vortex precession, the full BdG spectrum exhibits a large number of ``background modes'' which are not captured by the vortex particle picture. 
The $\omega = 0$ mode present for any value of $\alpha$ can be identified as the Goldstone mode related to the $U(1)$ invariance of the Gross-Pitaevskii equation. In addition, there is always a so-called Kohn or dipolar mode with 
frequency $\omega_x$ (which assumes the value $0.2$ independent of $\alpha$
in Fig.~\ref{fig:bdg_1vor}). Similarly there is a dipolar mode (linear
in $\alpha$ in Fig.~\ref{fig:bdg_1vor}) with frequency $\omega_y$.
These modes involve a collective oscillation of the entire cloud
around the center of the trap in each of these two directions.

\subsection{Vortex dipole}
Let us perform the same analysis for the so-called vortex ``dipole'', i.e. two vortices with opposite charges, say $s_1 = +1$, $s_2 = -1$. 
The existence of stationary configurations of such a vortex-antivortex pair (as it is sometimes called) was first demonstrated in \cite{PhysRevA.68.063609}, followed by more detailed discussions \cite{PhysRevA.71.033626, PhysRevA.74.023603, PhysRevA.77.053610}.
Stability properties of the dipole have been studied in a number of works, but the results were partially incoherent, in particular when anisotropic trapping was taken into account \cite{PhysRevA.74.023603,PhysRevA.82.013646}. 
Recently, interest in the vortex dipole has been renewed by experimental progress in the field, allowing to controllably create and observe such structures with unprecedented precision \cite{PhysRevLett.104.160401,D.V.Freilich09032010,PhysRevA.83.011603,PhysRevA.84.011605}.

The particle picture predicts two equilibrium positions 
of the vortices along each of the trap's main axes. Fixing the dipole along the $y$-axis, these read $x_1 = 0$, $y_1 = \pm \sqrt{B/(4 \omega_y^2 Q)}$, $x_2=0$, $y_2 = \mp \sqrt{B/(4 \omega_y^2 Q)}$.
Naturally, the solutions associated with the two different signs 
can be transformed into each other by simply interchanging the 
positive- and the negative-charge vortex.

\begin{figure}
\centering
 \subfigure{
\includegraphics[height=0.25\textheight]{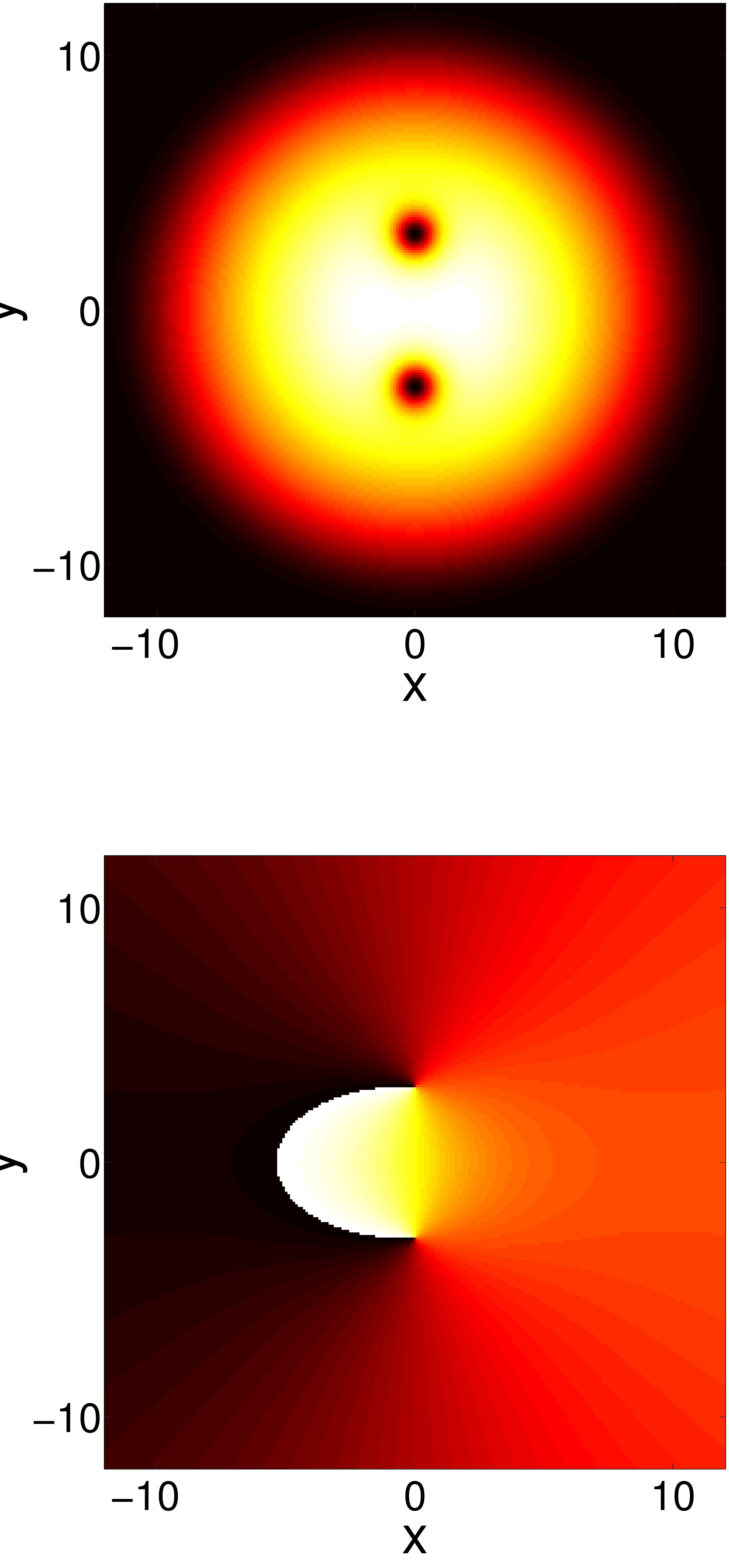}
}
\subfigure{
 \includegraphics[height=0.25\textheight]{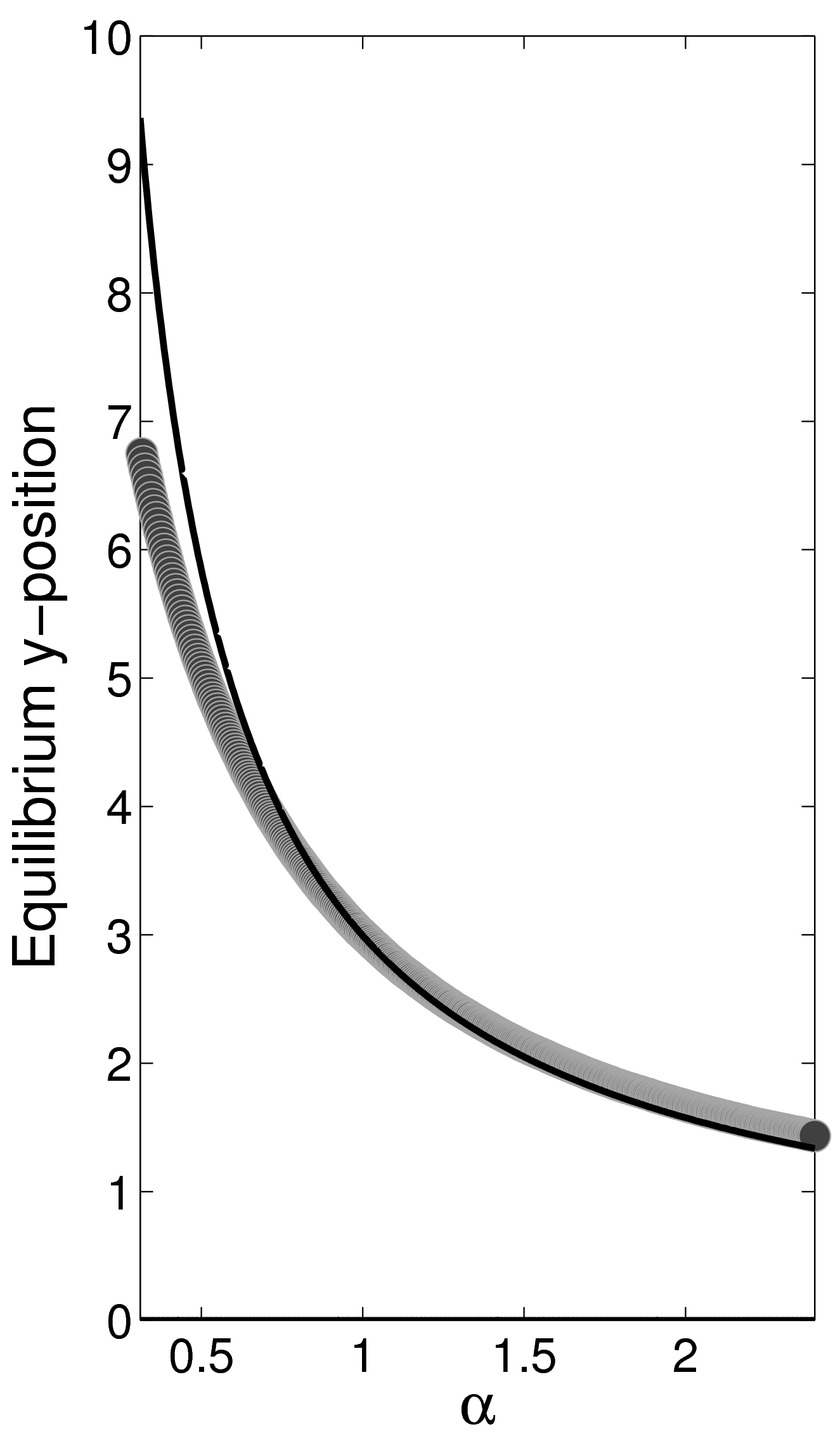}}
 \subfigure{
\includegraphics[height=0.25\textheight]{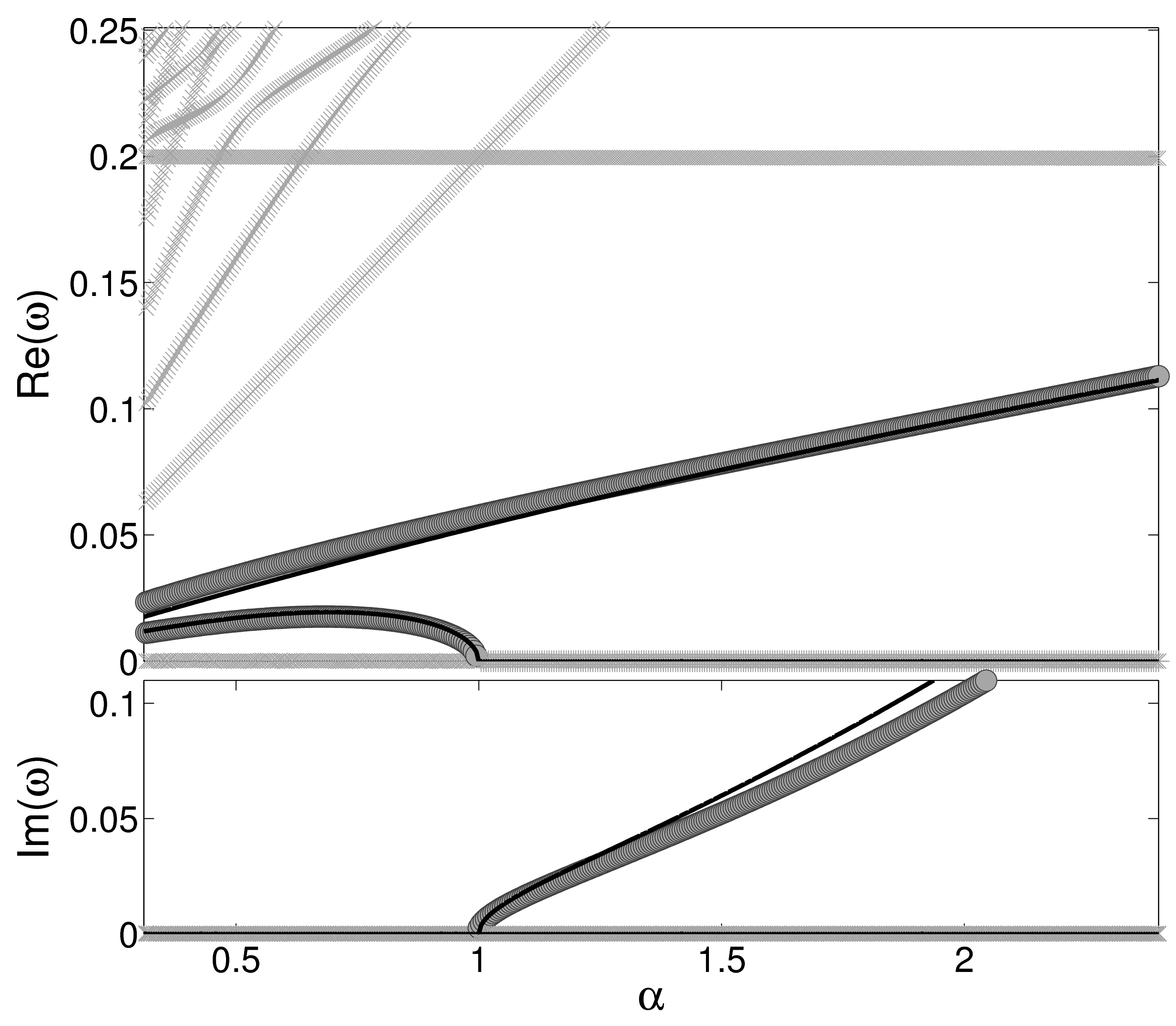}}
\caption{The left panels show a typical example of the density and
phase of a vortex dipole for an equilibrium position of the vortices within 
the PDE. The middle panel shows the equilibrium position of the 
dipole configuration as a function of the 2D aspect ratio: Numerical data (line of gray circles) and ODE prediction (black line). 
The right panel shows the results of the linearization around such a dipole
as a function of $\alpha$. For the anomalous (internal) modes of the
two vortices the prediction from the particle equation theory is given
again by the solid black lines, while the numerical findings are
given by lines of dark gray circles. Purely imaginary modes are also denoted by gray circles. Notice the instability emerging for
$\alpha>1$. The chemical potential is fixed at $\mu = 2.5$ 
throughout. \label{fig:vd} }
\end{figure}

The middle panel of Fig.~\ref{fig:vd} compares the numerically found equilibrium positions of the vortices forming the stationary dipole to this prediction.
Technically, from the numerically calculated wavefunctions the vortex locations are extracted by evaluating the $z$-component of the superfluid vorticity.
The vortices then show up as sharp, well localized extrema.
It can be observed that the agreement between the numerical data and the ODE prediction is better in the $\alpha > 1$ regime, and for small $\alpha$ the errors become larger.
Partially, this inaccuracy of the particle picture in the low $\alpha$ regime can presumably be attributed to the radial dependence of the vortex precession frequency which we do not take into account. Another factor that should be
taken into consideration here and accounts for the observed discrepancy is
the modification of the vortex-vortex interaction 
due to the non-homogeneous, modulated TF density, 
especially near the condensate boundaries. The latter effect has been
implicitly included in the equations through a shift of the $B$ factor away from its background value. 
However, the deviations in the middle panel of Fig.~\ref{fig:vd} suggest that including effects due to the presence of the trap 
(and the ensuing non-homogeneous condensate background) on the vortex-vortex interaction by using an effective $B$ is not sufficiently accurate for full quantitative agreement.
A refined description of vortex interaction in the trapped condensate, taking into account not only the vortices' positions but also the non-trivial shape of the background density distribution and the ensuing deformation of the velocity fields around the vortices, would be 
an interesting direction for further studies and could presumably, when included into our particle picture equations, significantly improve their quantitative predictions.

The dipole's linearization frequencies are found to be
 $\omega_{1,2} = \pm \sqrt{2} \omega_\text{pr}$, $ \omega_{3,4} = \pm \omega_\text{pr} \sqrt{1 - \alpha^2}$.
A key consequence of this prediction is that for $\alpha > 1$, the spectrum 
of the vortex dipole exhibits a purely imaginary mode, indicating 
instability in this regime.
On the other hand, for $\alpha \leq 1$ the dipole is stable.
In this stable interval, the spectrum contains two anomalous modes (indicated by the gray line made up of circular markers in the right panel of Fig. \ref{fig:vd}) whose functional dependence is well described by the linearization frequencies predicted in the particle picture.
At $\alpha = 1$, one of these two anomalous modes vanishes.
The existence of such a zero mode in the isotropic limit is a general feature that will be found for all subsequent aligned vortex states:
One can think of the presence of the aligned vortex configuration breaking the rotational symmetry of the $\alpha = 1$ system, which leads to the emergence of a corresponding Goldstone mode.
For $\alpha > 1$, this former Goldstone mode in the dipole's spectrum becomes 
a purely imaginary eigenfrequency, 
which again is predicted correctly by the particle picture.
Intuitively, in the isotropic setting of $\alpha=1$ the dipole as a whole can be arbitrarily rotated. This neutrality is represented by the zero mode in the BdG spectrum.
When $\alpha>1$, in our current setup, the vortex pair is compressed along the axis of the vortex dipole, which favours ``buckling'' of the axial structure and leads to instability, while $\alpha < 1$ has the opposite effect.
A typical example of the dipole's decay at $\alpha>1$ is shown below, in Fig. \ref{dyn_anis}.

\subsection{Vortex tripole}
Let us now turn to the next aligned vortex state, consisting of three vortices of alternating charge, i.e. $s_1 = \pm 1$, $s_2=\mp 1$, $s_3 = \pm 1$.
As for the vortex dipole, we take the stationary vortices to be aligned along the $y$-axis, i.e. ${x}_{1,2,3}=0$, and for their equilibrium $y$-coordinates we consider the symmetric ansatz ${y}_2=0$, ${y}_1=-{y}_3={y}$.
Inserting this into the particle picture ODEs yields a fixed point for ${y}=\sqrt{B/(4 \omega_y^2 Q)}$.
Thus, the particle picture predicts a stationary ``vortex tripole'' state, where two vortices of the same charge are placed along one of the trap's main axes, while the oppositely charged third vortex rests between them, at the trap center.
Previous theoretical discussions of this tripole configuration can be found in \cite{PhysRevA.71.033626,PhysRevA.74.023603,Middelkamp2010b}, 
while a recent experimental observation has been reported in~\cite{Seman2009}.

\begin{figure}
\centering
 \subfigure{
\includegraphics[height=0.25\textheight]{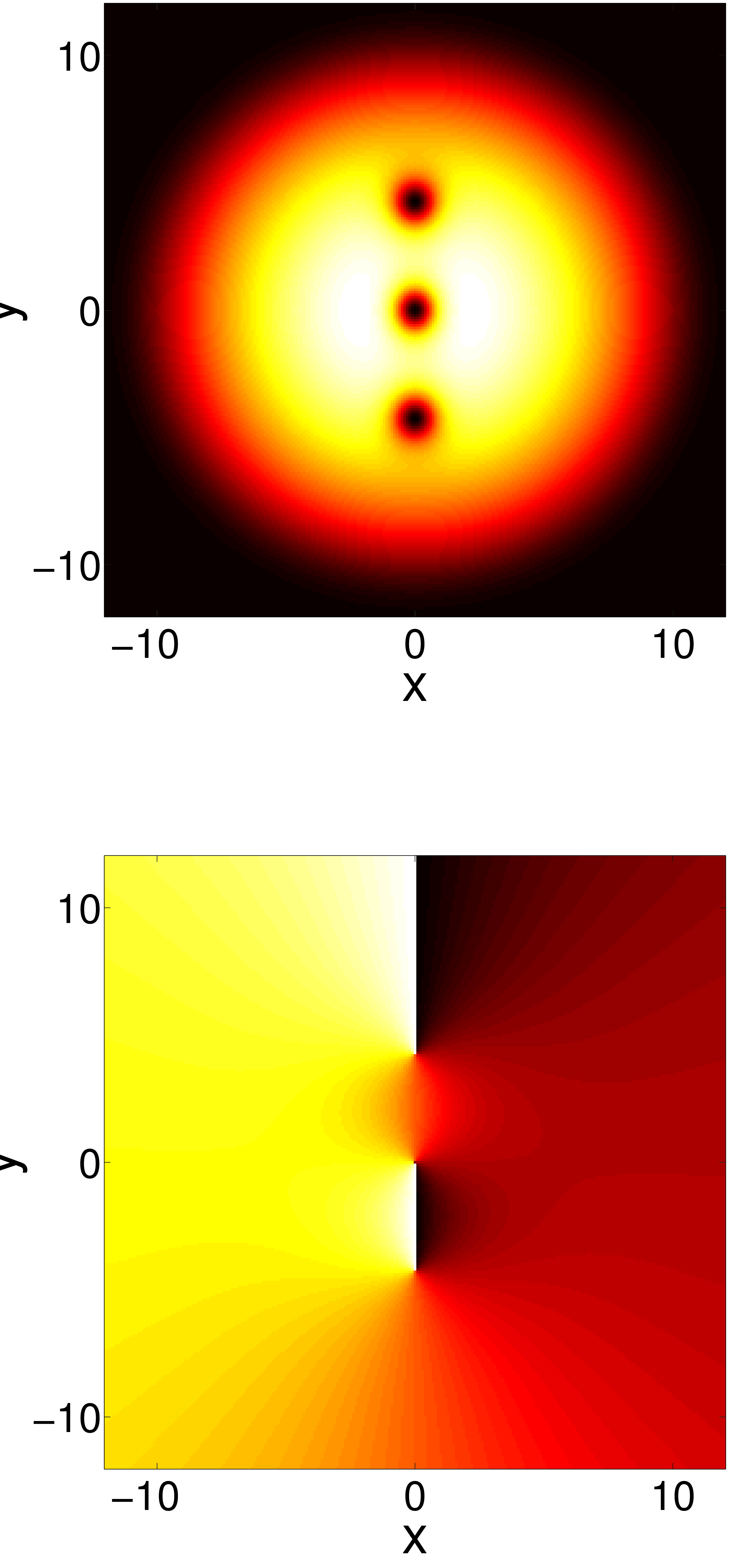}
}
\subfigure{
 \includegraphics[height=0.25\textheight]{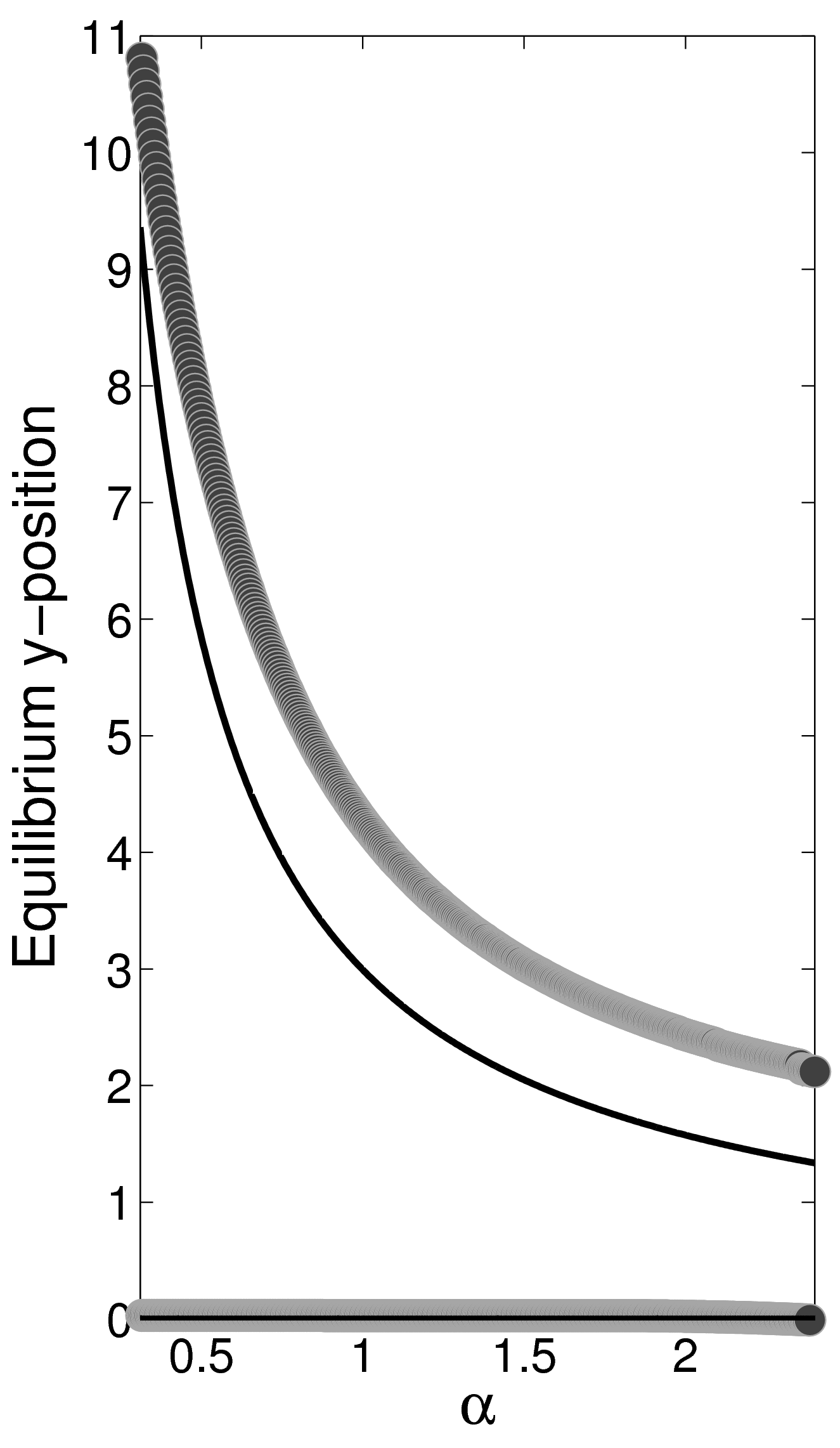}}
 \subfigure{
\includegraphics[height=0.25\textheight]{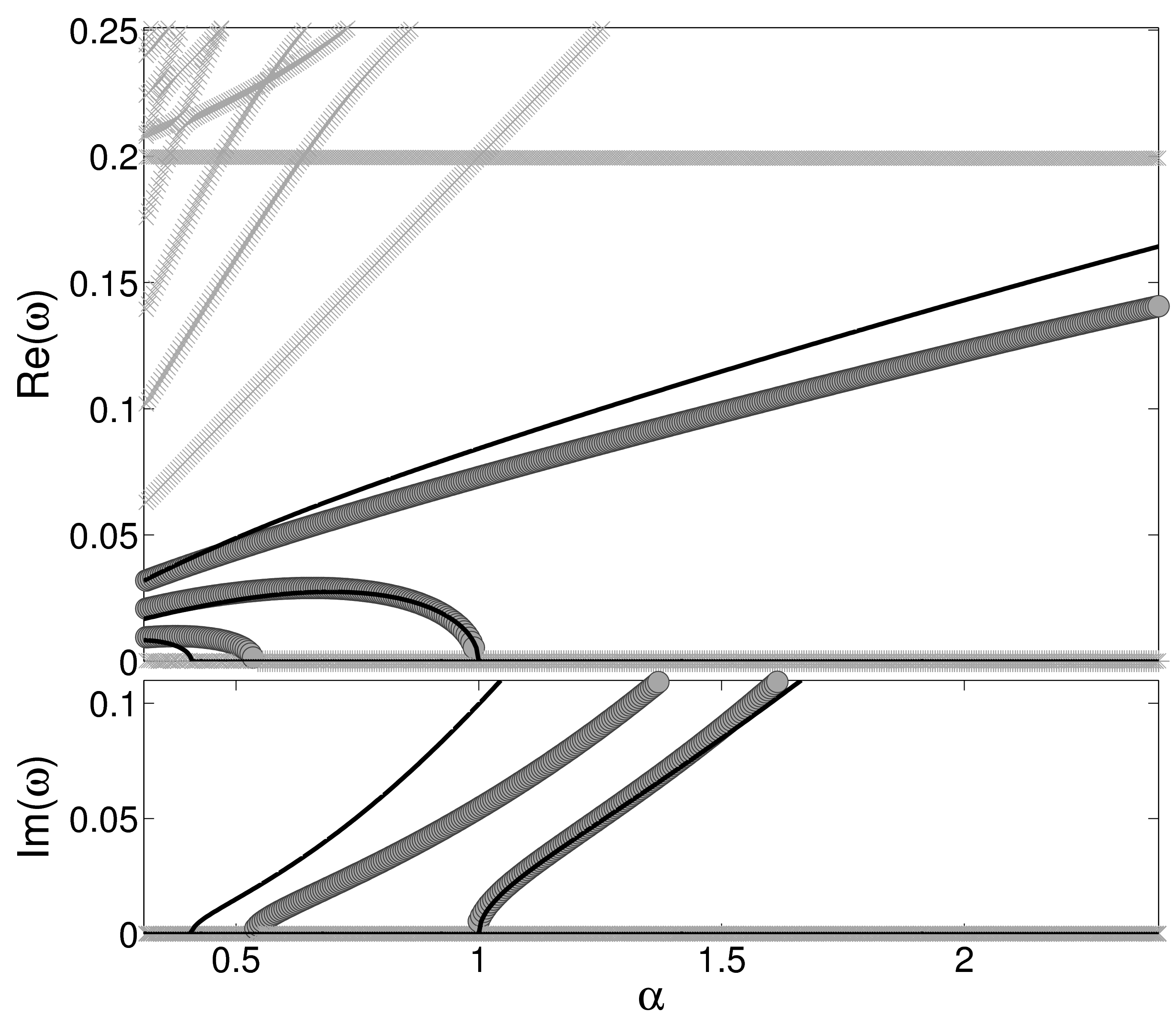}}
\caption{The left panels show a typical example of the density and
phase of an equilibrium vortex tripole. The position of the vortices in the 
tripole configuration as a function of the 2D aspect ratio is shown in the
middle panel with numerical data shown as line of gray circles and the ODE prediction shown in black solid line. The right panel
shows the eigenmodes of linearization around the tripole: ODE theoretical predictions represented by solid lines and anomalous modes and purely imaginary modes by lines of gray circles.
Chemical potential $\mu = 2.5$.\label{fig:vt} }
\end{figure}

Again, we can compare the fixed point coordinates calculated from the vortex particle picture to the vortex positions obtained from numerically identifying the vortex tripole solution of the stationary GPE.
The results are shown in Fig. \ref{fig:vt}.
 
Interestingly, we observe that while choosing $B=1.35$ led to good agreement for the dipole configuration, this is no longer fully the case for the tripole. Here, better quantitative agreement is achieved if the interaction constant $B$ is taken at its background value $B=1.95$, valid for vortex interaction in a homogeneous condensate \cite{PhysRevA.82.013646} (data not shown).
This discrepancy may again be regarded as a warning sign that accounting for condensate inhomogeneities due to the trap by rescaling $B$ is insufficient
in the general case and it is certainly desirable to take
into account the effects of the nonhomogeneous background discussed above. 

Next, we calculate the linearization frequencies around the tripole equilibrium. 
From the particle picture ODEs we obtain $\omega_{1,2} = \pm \sqrt{2}  \omega_{\text{pr}} \sqrt{1-\alpha^2}$, $\omega_{3,4,5,6} = \pm \omega_{\text{pr}} \sqrt{4-5 \alpha^2 \pm \sqrt{9+2\alpha^2 +25 \alpha^4}}$.
In the isotropic limit of $\alpha=1$, these equations reproduce the result of \cite{PhysRevA.82.013646}.
Concerning stability, the most important conclusion to be drawn is that below the critical anisotropy $\alpha_\text{cr}=1/\sqrt{6} \approx 0.408$, all linearization frequencies become real, i.e., the tripole can be completely stabilized by 
means of strong enough transversal confinement.
Comparing with the numerically found BdG spectrum of the tripole essentially confirms this prediction.
The particle picture still captures the overall behaviour of the relevant vortex modes.
In detail, however, the predictions are less exact than for the vortex dipole.
In particular, while stabilization in general is correctly predicted by the particle picture, the critical value of $\alpha$ for which it occurs is found to be $\alpha_\text{cr} \approx 0.53$. 
It should also be stressed that similarly to the dipole case, the neutral
mode present for $\alpha=1$ (due to isotropic rotation of the tripole) can be tipped towards stability (for $\alpha<1$) or instability (for
$a>1$) depending on the direction of anisotropic compression of the
condensate with respect to the axis of the multi-vortex state
(perpendicular, or parallel, respectively).
A typical example of the dynamics following the tripole's decay in an isotropic trap, triggered by the imaginary linearization modes, will be shown in Fig. \ref{dyn_anis} below.

\begin{figure}
 \centering
\includegraphics[width=0.25\textwidth]{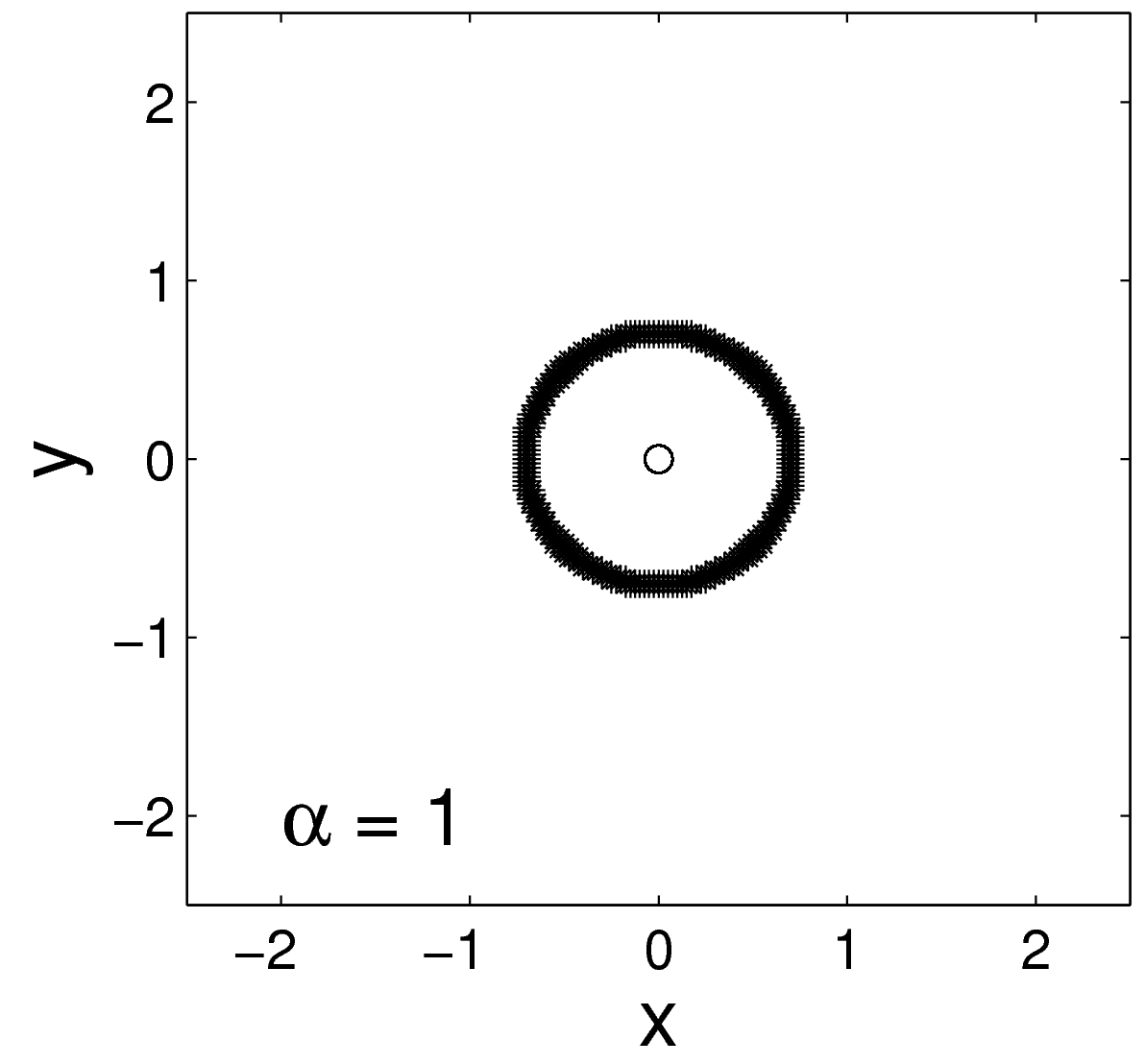}
\hspace{5mm}
\includegraphics[width=0.25\textwidth]{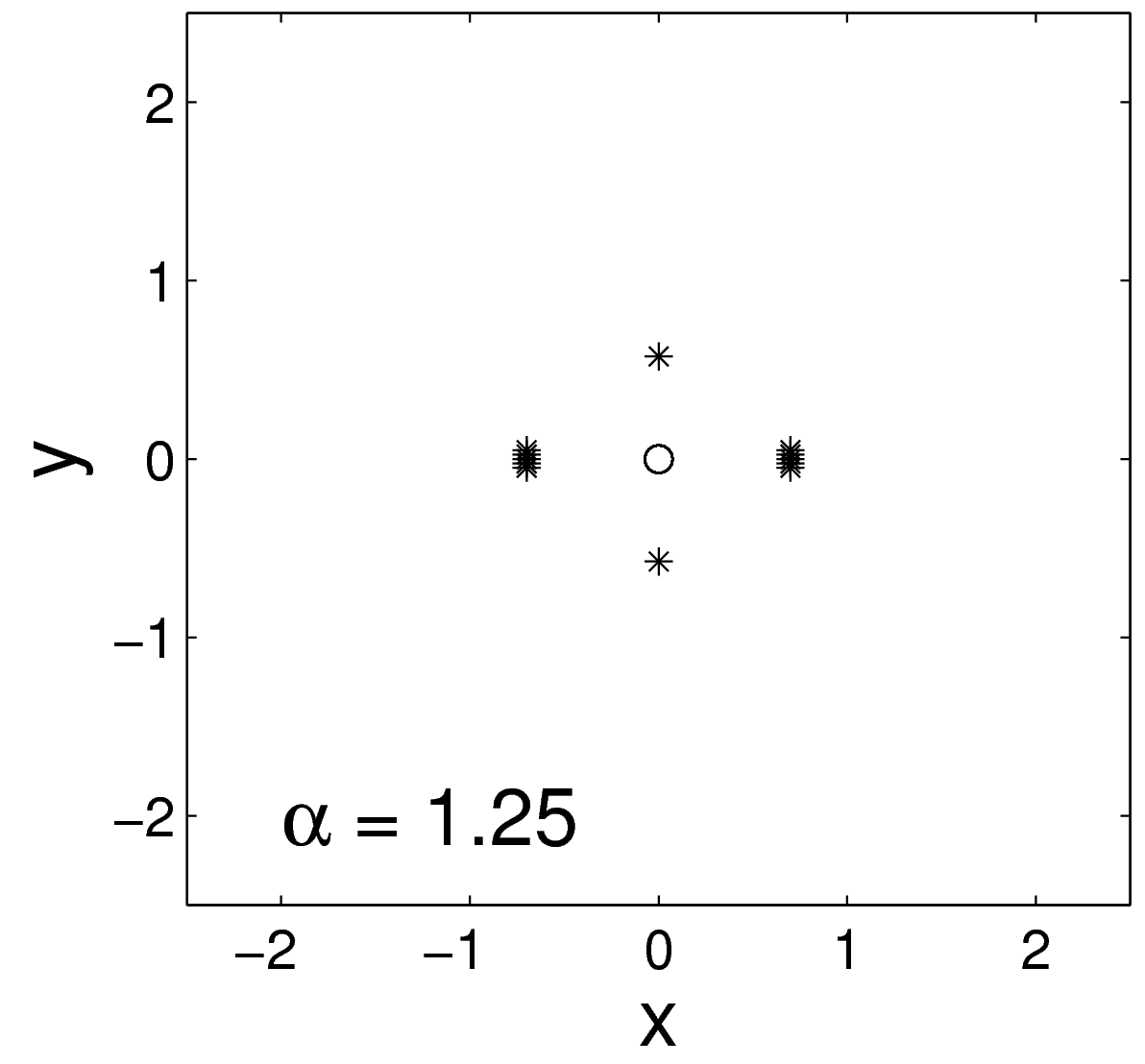}
\caption{Numerically calculated fixed points of the three vortex ($s_1 = s_3 = -s_2$) ODE system. Coordinates $x_1$, $y_1$, $x_2$, $y_2$ were each scanned in the interval $[-2.5, 2.5]$ (divided into 200 steps). The plots show positions of the three vortices for configurations at which $\max\{|\dot x_j|,|\dot y_j|\} < 0.03$ for each $j \in \{1, 2, 3\}$. Vortex 1 and 3 denoted by $\times$ and $+$ symbol, respectively, vortex 2 by an open circle. Note that, as anticipated, an exchange of vortices 1 and 3 does not lead to a different configuration, so symbols $\times$ and $+$ appear on top of each other. Position variables are scaled by a factor of $\sqrt{2 \omega_x^2 Q/B}$\label{fig:tripoleODEfix}.}
\end{figure}

Concerning the three vortex case, one more remark is in order here.
For more than two vortices, we found it impossible to determine all fixed points of the vortex equations of motion analytically.
This is why we had to make an ansatz motivated by the expected symmetry properties to find the tripole fixed point of the ODE system.
One result that can still be proved in full generality is that for any fixed point $({x}_1, {y}_1, \dots, {x}_K, {y}_K)$ of a $K$ vortex system, the $x$- and $y$-coordinates have to sum up to zero independently.
Of course, this by itself does not rule out other equilibrium positions than the aligned tripole for the three vortex system. 
Thus, we checked numerically that no other stationary vortex solutions are predicted by the particle picture.
To do so, we evaluated the functions $\dot x_j$, $\dot y_j$, $j \in \{1, 2, 3\}$, within a large region of configuration space, i.e. for a large number of test coordinates $(x_1, y_1, x_2, y_2, x_3=-x_1-x_2, y_3=-y_1-y_2)$ and identified their roots.
The results are shown in Fig. \ref{fig:tripoleODEfix}, confirming that the tripole along one of the trap's main axes really is the only equilibrium solution predicted by the particle picture ODE system. It should be mentioned in 
passing that this result is only true if the precession frequency is
identical for all vortices; if it depends on the distance of the vortex
from the center of the trap, then the above result no longer holds.

Having discussed both the vortex dipole and the vortex tripole case
should render evident the fact that anisotropy presents a remarkable
handle for controlling the stability and dynamics of multi-vortex
clusters at will away from the isotropic limit. In particular, it
is evident that configurations such as the vortex dipole which
are structurally robust in the isotropic limit can be immediately
rendered unstable when departing from that limit for values of
$\alpha>1$. On the flip side, any configuration which is more
highly excited and unstable in the isotropic limit,
can instead be stabilized when operating in a sufficiently
anisotropic regime for $\alpha < 1$. Examples of all four
of these scenaria: perturbed but stable isotropic dipole,
perturbed unstable anisotropic dipole for $\alpha>1$, perturbed
unstable isotropic tripole and finally, perturbed but stable
sufficiently anisotropic tripole for $\alpha \ll 1$ are shown 
systematically in Fig.~\ref{dyn_anis}.\\
We would like to point out that a systematic study of vortex dynamics far from equilibrium in the presence of the anisotropic trap, similar to the work of \cite{Torres20113044}, promises to be a very interesting direction for further investigations.
At first sight, the dynamics triggered by the decay of the vortex dipole at $\alpha > 1$ seem to be periodic, i.e., continuing the propagation we observe a regular sequence of revivals and decays of the dipole (Fig.~\ref{dyn_anis} only shows part of the first half period). 
For the decaying tripole, on the other hand, no such periodicity seems to be present. A general investigation of these dynamics far away from the fixed points, and the potentially different types of dynamics triggered by the different imaginary BdG modes, is beyond the scope of this work where our main aim
is to  identify and understand the parameter regimes of linear (in)stability for the equilibrium vortex clusters. 
Nevertheless, we should add that the 
dynamical evolution results of Fig.~\ref{dyn_anis} afford the more general
expectation that the instability of the vortex clusters will evolve
towards smaller, more stable ``building blocks'' of the configuration.
In this sense, the anisotropic dipole can only break up towards a stable single vortex (and one in the periphery of the cloud).
Following the decay of larger vortex clusters we particularly often observe the formation of transient vortex dipoles,
i.e. pairs of vortices of opposite charge moving together over comparably long timescales and only unbinding to pair with other vortices.
This resembles observations made in studies of large-scale superfluid turbulence in Bose gases, see e.g. \cite{PhysRevA.85.043627}.
The occurence of such transient dipoles can be observed e.g. in the isotropic tripole's decay in Fig.~\ref{dyn_anis}(c) and has also 
been checked numerically for more complex states such as the aligned vortex quadrupole discussed below in Fig.~\ref{fig:vq}.
The same feature is also present in the vortex dynamics triggered by the decay of non-aligned clusters, see Fig.~\ref{fig:prop_vqa_2-5} in section 7.

\begin{figure}[ht]
\centering
\fbox{
\subfigure{
\includegraphics[width=8cm]{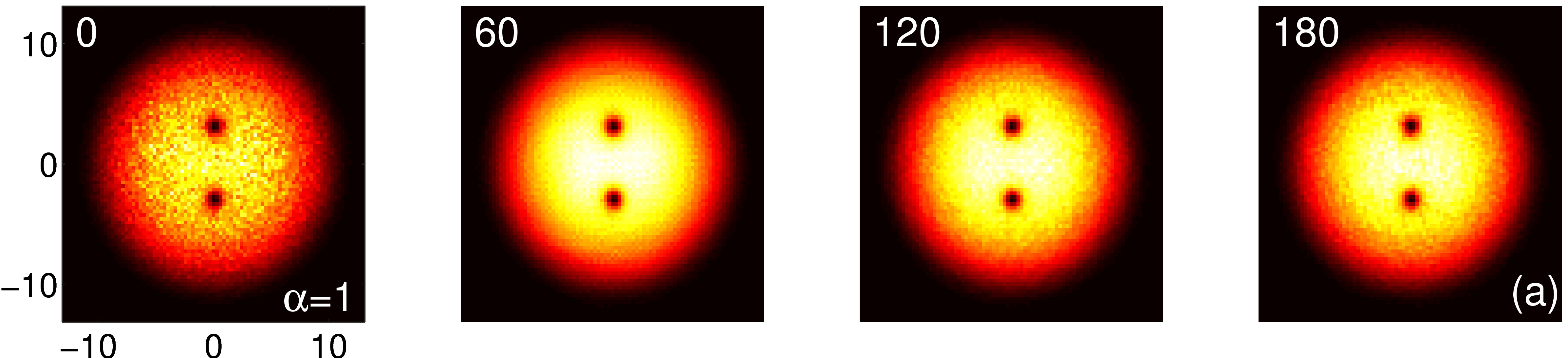}
}
}
\fbox{
\subfigure{
\includegraphics[width=8cm]{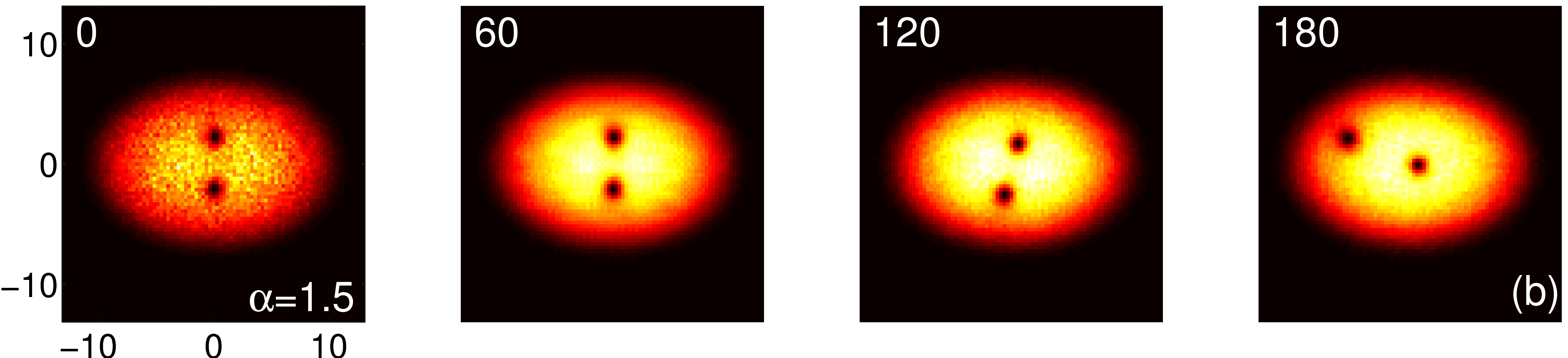}
}
}
\fbox{
\subfigure{
\includegraphics[width=8cm]{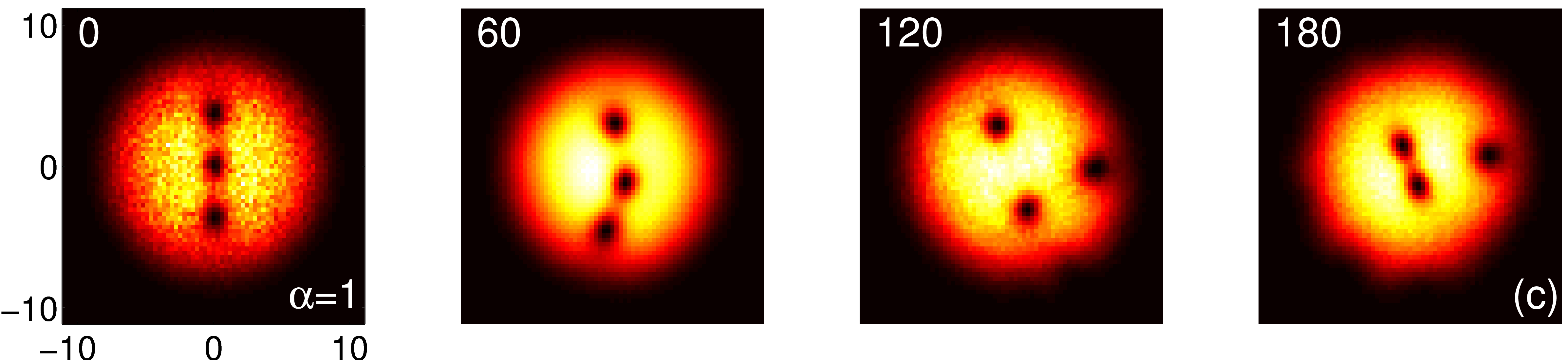}
}
}
\fbox{
\subfigure{
\includegraphics[width=8cm]{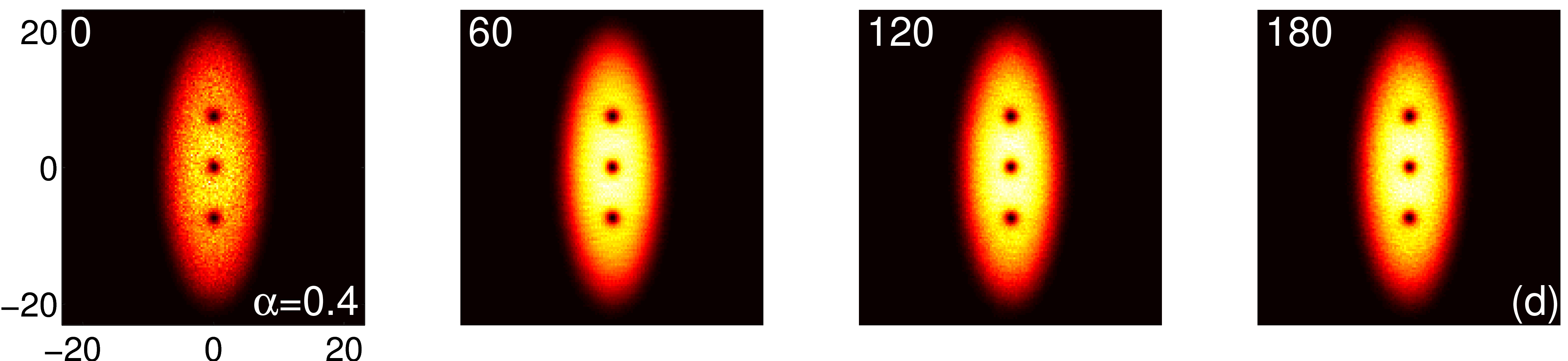}
}
}
\caption[Optional caption for list of figures]{\label{dyn_anis} 
Dynamical evolution of aligned vortex states 
perturbed by white noise for different values of 
$\alpha$: (a) $\alpha=1$ dipole, (b) $\alpha=1.5$ dipole,
(c) $\alpha=1$ tripole, (d) $\alpha=0.4$ tripole. All plots show 
$|\psi|^2$, and the elapsed dimensionless time is given in the upper left corners. Reprinted with permission from \cite{aniso}.}
\end{figure}

\subsection{Aligned vortex quadrupole}
\begin{figure}
\centering
 \subfigure{
\includegraphics[height=0.25\textheight]{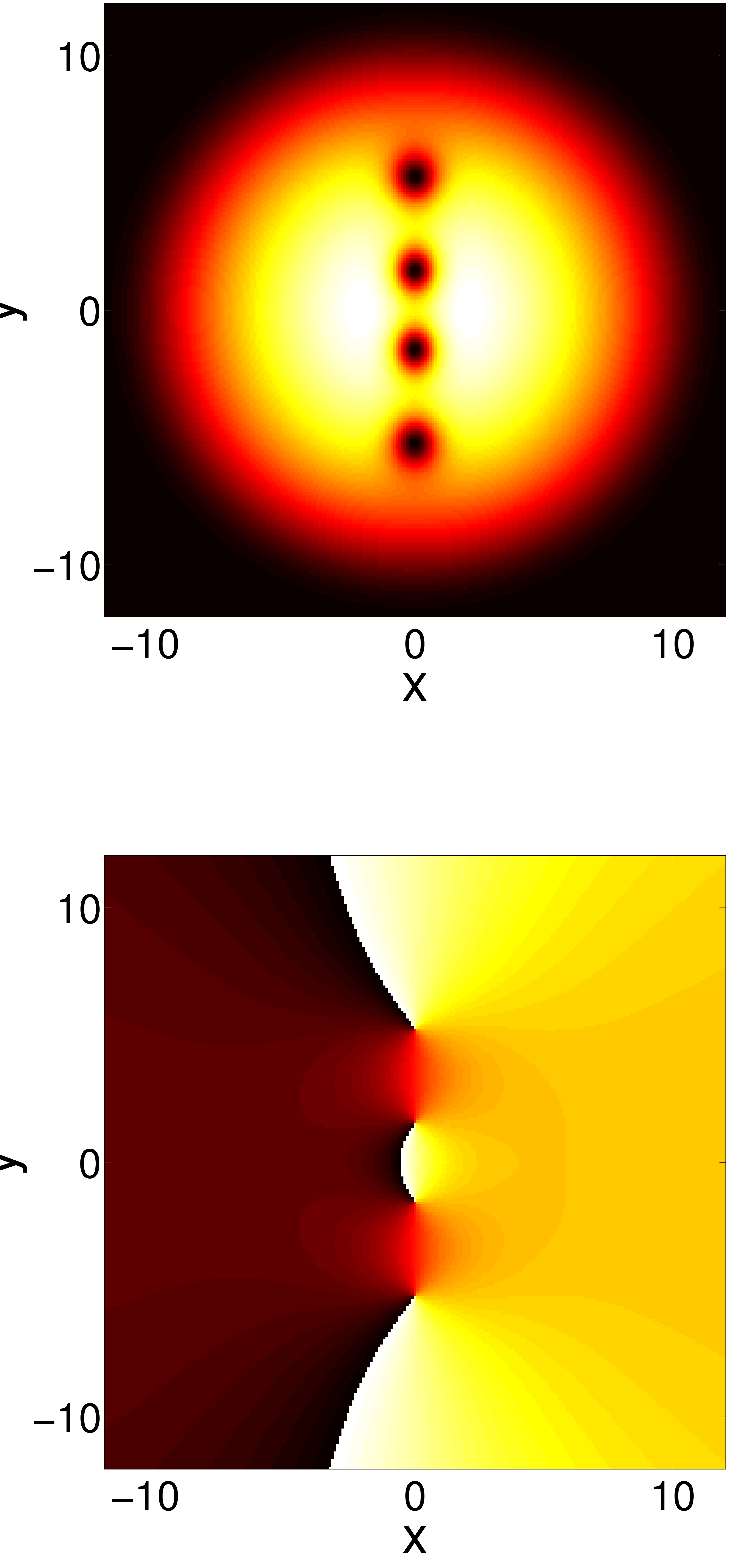}}
\subfigure{
 \includegraphics[height=0.25\textheight]{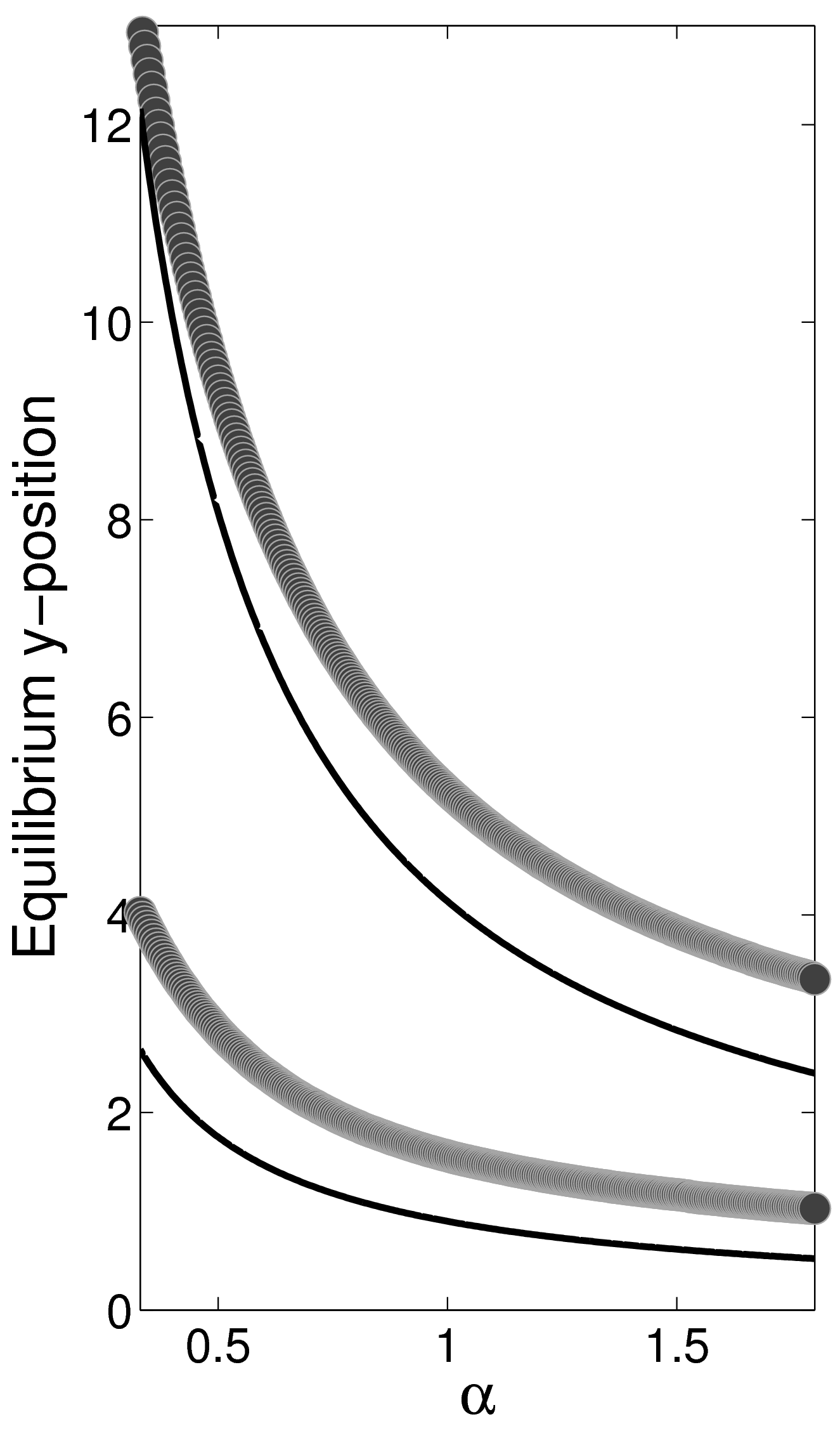}}
 \subfigure{
\includegraphics[height=0.25\textheight]{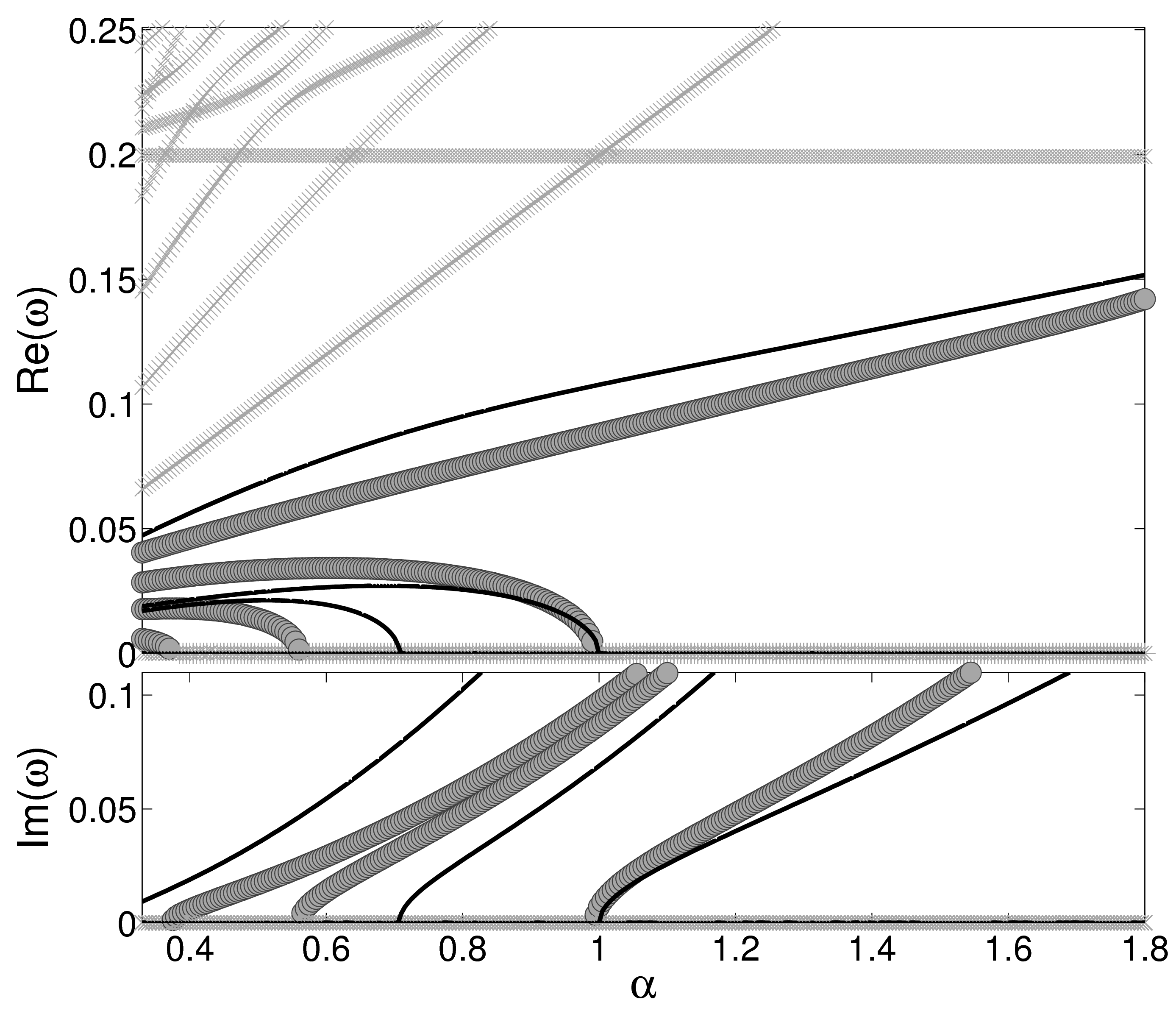}}
\caption{Same as Figs.~\ref{fig:vd} and \ref{fig:vt} but now for the aligned
vortex quadrupole.
\label{fig:vq} }
\end{figure}

As a final example, let us consider the case of four vortices with $s_1 = s_3 = \pm 1$, $s_2 = s_4 = \mp 1$.
An ansatz where these four vortices are aligned along the $y$-axis, symmetrically with respect to the origin, allows us to determine the following equilibrium positions of the aligned vortex quadrupole along the $y$-axis:
\begin{align*}
{y}_1 = -{y}_4 = \sqrt{\frac{B}{4 \omega_y^2 Q}} \cdot \sqrt{1+\sqrt{2\sqrt{2}-2}}, \quad
{y}_2 = -{y}_3 = \sqrt{\frac{B}{4 \omega_y^2 Q}} \cdot \sqrt{1-\sqrt{2\sqrt{2}-2}}.
\end{align*}
The central panel of Fig. \ref{fig:vq} compares these predictions to the numerically calculated equilibrium positions. Again, qualitative (and to some extent also quantitative) agreement is very good.

The linearization frequencies from the particle equations of motion could not be obtained analytically in this case.
The results calculated by numerical diagonalization of the ODE system's Jacobian at the 
aligned quadrupole fixed point, together with the full BdG spectrum, are shown in Fig. \ref{fig:vq}.

Once again, we observe the onset of stabilization for small enough values of $\alpha$.
This feature is contained in the particle picture predictions as well, but the critical value $\alpha_\text{cr}$ theoretically identified is below the corresponding numerical one.. 
From the BdG spectrum we obtain $\alpha_\text{cr} \approx 0.37$.
In summary, quantitative agreement between our particle picture predictions and the BdG modes calculated from the full Gross-Pitaevskii theory somewhat
deteriorates the more vortices are considered.
We assume that mainly two effects are responsible for this:
On the one hand, the more vortices are present, the further the vortex 
cluster stretches out into the condensate and the off-center correction 
term to the precession frequency which we neglect becomes increasingly 
important.
On the other hand, our simple, semiclassically justified modeling of vortex 
interaction is not fully accurate in that it does not adequately account for 
density inhomogeneities due to the trap i.e., the background-induced effect mentioned
above. The higher the number of vortices, the further outward the vortex
cluster extends and hence the more the density variation at the
rims of the cloud affects the result. Nevertheless, for all the
cases considered the qualitative agreement between the conclusions
of the particle picture and those of the full PDE has been excellent.

As an aside, it should be mentioned at this point that for four vortices of 
alternating charge, the aligned quadrupole does not form the only possible 
equilibrium position. Other quadrupole configurations, where the vortices are 
located at the vertices of a non-degenerate parallelogram, will be discussed 
in section \ref{sec:aniso2}.

\section{Aligned vortex states: Bifurcations}
In this section, we will show that valuable insight into aligned vortex states, and in particular into their stability, can be obtained from the point of view of bifurcation theory.
In particular, solutions to the stationary GPE depend non-trivially on their norm, or on the total particle number $N = \int \text{d}x\text{d}y |\psi|^2$, physically speaking.
In the following, we study branches of vortex cluster (and solitonic) solutions to the stationary GPE, varying the chemical potential $\mu$. 
This amounts to examining a state's parametric dependence on the particle number $N$, as $\mu$ is a strictly increasing function of $N$ and vice versa.
Relevant bifurcations are identified and related to the stability properties of the different branches.
This will prove to be a useful complementary tool to better understand the changes in stability induced by anisotropy that we observed for the vortex states of section \ref{aniso:aligned}.

In a sense, the above presented perspective of ``particle theory'' is the
one of the highly nonlinear limit where the individual coherent structures
(the vortices) can be clearly identified as distinct, highly
localized objects which form an effective interacting particle system.
On the other hand, the discussion of the present section will focus
on the opposite limit, namely that of the weakly nonlinear regime.
In the latter, the states bifurcate from the eigenstates of underlying linear
operators which constitute the canonical starting point for the relevant
bifurcation analysis that will be presented below. Lastly, the aim
of the overall program is to connect this weakly nonlinear analysis
with the strongly nonlinear particle regime by means of numerical
computations that bridge the two limits.

\subsection{Bifurcation approach in the isotropic case}

In this section, we review the bifurcation analysis put forward in \cite{PhysRevA.82.013646} to study aligned vortex states within an isotropic trap for which $\omega_x = \omega_y \equiv \omega_r$.
In that work, it was argued that aligned vortex clusters are intimately related to the dark soliton stripe solution of the two-dimensional Gross-Pitaevskii equation.
For reference, density and phase profiles of a dark soliton state (in an anisotropic setting) are presented below, in Fig. \ref{fig:sv}.
In an isotropic trap, following the soliton stripe branch of solutions as the chemical potential $\mu$ is increased, one finds that subsequently new branches of fixed points bifurcate from it.
These emerging branches are identified as the aligned vortex solutions, with the dipole branch bifurcating first, then the tripole branch and so on. 
Similar observations, concerning the dipole branch only, had previously been made in \cite{PhysRevA.68.063609,PhysRevA.77.053610}.

It has been argued in \cite{PhysRevA.82.013646} that the bifurcations leading from the soliton branch to the aligned vortex branches are of the supercritical pitchfork type.
Such pitchfork bifurcations generically occur in systems with some internal symmetry.
The transfer of stability from a symmetric (parental) branch of fixed points to two non-symmetric branches can then be thought of as a symmetry-breaking process:
beyond the bifurcation point, the stable equilibra do not exhibit the system's full symmetry anymore.

Let us apply these statements to the dipole's bifurcation from the dark soliton stripe branch.
For small values of $\mu$, the soliton is linearly stable. One can check that its BdG spectrum exhibits no imaginary mode.
Increasing the chemical potential, for our choice of $\omega_r = 0.2$ at a critical value of $\mu \approx 0.68$ the vortex dipole branch bifurcates from the soliton branch.
More precisely, there are two different dipole branches coming into existence at this critical $\mu$: These two can be transformed into each other by interchanging the roles of the vortex and antivortex, i.e. by globally flipping the vorticity.
This should be thought of as a time-reversal transformation: As known from (linear) quantum mechanics, applying the antiunitary time-reversal operator is essentially tantamount to complex conjugation.
Thus, under a time-reversal transformation, the wavefunction phase changes its sign, which in turn means that the velocity field $\mathbf{v}(\rr)$, proportional to the gradient of the phase, changes its sign, too. The same goes for the vorticity field $\nabla \times \mathbf{v}$. Physically speaking, the superfluid flow changes its direction.
Note now that the soliton stripe state is purely real, i.e. it is invariant with respect to time-reversal. The vortex dipoles, on the other hand, are described by complex wavefunctions. Thus, they are not invariant under the action of time-reversal, instead they are transformed into each other. This is the characteristic symmetry-breaking feature expected in a supercritical pitchfork bifurcation.
Furthermore, one can observe that the two dipole branches ``inherit'' the soliton's stability, while in the soliton's BdG spectrum an imaginary mode occurs.
The corresponding decay mechanism is well-known as the transversal (or ``snaking'') instability of the soliton stripe in two dimensions, see e.g. the 
recent review \cite{Frantzeskakis2010} and references therein.

Having outlined the close connection between bifurcation theory and stability analysis, a natural question to ask is whether one can understand why certain bifurcations occur, and at which particular critical value of $\mu$ or $N$.
This problem has been addressed in the slightly different context of a condensate trapped in a 1D double-well potential in \cite{PhysRevE.74.056608}. In \cite{PhysRevA.82.013646}, the same technique has been demonstrated to be of use for the study of vortex states in isotropic traps as well.

Lying at the heart of this approach to bifurcations in the Gross-Pitaevskii equation is the observation that in the limit of $N \rightarrow 0$ the nonlinear interaction term can be neglected, and the stationary GPE reduces to the familiar (linear) Schr\"odinger equation (with the chemical potential $\mu$ playing the role of energy). In the presence of a harmonic trap, the  solutions of this equation are the well-known 2D harmonic oscillator eigenfunctions $\gamma_{mn}(x,y)=\gamma_m(x) \gamma_n(y)$, where the quantum numbers $m, n$ are non-negative integers and the energy eigenvalue of state $\gamma_{mn}$ is given by $E_{mn} = (m + n + 1) \omega_r$.

Thus, in the limit of $N \rightarrow 0$, the stationary solutions of the GPE have to reduce to harmonic oscillator eigenstates asymptotically.
For the soliton stripe branch and the single vortex branch, which both exist in the linear limit of infinitesimally small particle numbers, these linear counterparts are readily identified.
Taking the direction of its density minimum along the $y$-axis, the soliton stripe can be traced back to the eigenstate $\gamma_{10}$, which is purely real and has a nodal line at $x=0$. The vortex, on the other hand, approaches the complex linear combination $\gamma_{10} \pm \ii \gamma_{01}$, which in polar coordinates $(r,\theta)$ leads to the characteristic $\exp(\pm \ii \theta)$ phase profile. 
Due to the isotropic trap, $\gamma_{01}$ and $\gamma_{10}$ are degenerate, and 
thus their superposition is also a solution of 
the linear Schr\"odinger equation.
Let us remark that a continuation of this linear solution into the vortex state for the case of an {\it attractive} cubic nonlinearity has been performed in \cite{PhysRevA.73.043615}, where also the doubly charged vortex we will encounter in the next section was studied.

We now turn to branches of states which do not exist in the limit of vanishing $N$, but bifurcate close to it, such as the vortex dipole.
While these states cannot be expected to reduce to a single eigenfunction of the Schr\"odinger equation, it has been demonstrated that they can be approximated as linear combinations of non-degenerate harmonic oscillator functions \cite{PhysRevA.68.063609,PhysRevE.66.036612,PhysRevA.82.013646}.
One can think of this as a Galerkin-type method, where the nonlinear GPE problem is approximately discretized by projecting onto suitable eigenspaces of the Schr\"odinger Hamiltonian. 
From the point of view of such a few-mode expansion, the dipole is described as a superposition of $\gamma_{10}$ and $\gamma_{02}$, with a constant relative phase of $\pm \pi/2$. 
In other words, as $\mu$ and $N$ are increased away from the linear limit, the soliton stripe branch is still approximated by $\gamma_{10}$, and the dipole's bifurcation is then attributed to an admixture of $\pm \ii \gamma_{02}$ that sets in at a critical particle number (or chemical potential, equivalently).

In such a setting, considering a linear combination of the two relevant linear modes $\varphi_0$, $\varphi_1$ and requiring that it be stationary can be shown to lead to a prediction for the critical values where the bifurcation from the branch starting as $\varphi_0$ due to an admixture of $\varphi_1$ at a relative phase of $\pm \pi/2$ occurs \cite{PhysRevE.74.056608,PhysRevA.82.013646}
\begin{align}
 N_\text{cr} = \frac{E_1 - E_0 }{A_{0000}-A_{0011}}, \label{eq:anisoncr} \\
\mu_\text{cr} = E_0 + A_{0000} N_\text{cr} \label{eq:anisomucr}.
\end{align}
Here, $A_{0000} = \int \text{d}x \text{d}y \varphi_0^4$, $A_{0011} = \int \text{d}x \text{d}y \varphi_0^2 \varphi_1^2$ and $A_{1111} = \int \text{d}x \text{d}y \varphi_1^4$ denote the two modes' nonlinear overlap integrals, $E_0$, $E_1$ their energies.
Several generic assumptions are made in the derivation of Eqs. (\ref{eq:anisoncr}, \ref{eq:anisomucr}). The linear modes are taken to be real, ``mixed'' overlap integrals $A_{0001}=\int \text{d}x \text{d}y \varphi_0^3 \varphi_1$, $A_{0111}=\int \text{d}x \text{d}y \varphi_0 \varphi_1^3$ are assumed to vanish, and use of the inequalities $A_{0000}>A_{0011}$, $A_{1111}>A_{0011}$ is made.  

Let us now return to the aligned vortex states bifurcating from the soliton branch.
Generally, the bifurcation of the aligned vortex state with $n$ vortices ($n \geq 2$) can be attributed to an admixture of $\pm \ii \gamma_{0n}$ to the soliton's $\gamma_{10}$ mode.
With this two-mode picture in mind, one can apply the Galerkin approach to predict critical particle numbers and chemical potentials for the bifurcations.
In \cite{PhysRevA.82.013646}, it has been demonstrated that excellent agreement with the numerical data is obtained for the lowest-lying bifurcations, leading to the dipole and tripole. For higher numbers of vortices, the Galerkin predictions tend to be less exact. This is understandable, as the corresponding bifurcations happen at comparably large values of $N$, far away from the linear limit, which impairs the applicability of the near-linear few-mode expansion.

\subsection{Modified bifurcation approach in anisotropic settings}
Extending this analysis to the anisotropic regime is essentially straightforward.
Let us in the following explain how insight into qualitative changes in the bifurcation diagram due to $\alpha \neq 1$ (and the ensuing implications for stability) can be gained using the Galerkin approach. Linear eigenfunctions in the anisotropic trap still factorize according to $\gamma_{mn}(x,y)=\gamma_m(x) \gamma_n(y)$, where the one-dimensional modes now read
\begin{align*}
 \gamma_m(x) &\propto H_m(\sqrt{\omega_x} x) \exp(-\omega_x x^2/2), \\
 \gamma_n(y) &\propto H_n(\sqrt{\omega_y} y) \exp(-\omega_y y^2/2),
\end{align*}
where normalization constants have been omitted and $H_n$ denotes the $n$-th Hermite polynomial. The energy eigenvalue of state $\gamma_{mn}$ is given by 
\begin{equation*}
E_{mn} = \left( m + 1/2 + \alpha(n + 1/2) \right) \omega_x.
\end{equation*}

Thus, for the two linear modes used to predict the bifurcation of
aligned vortex clusters, $\gamma_{10}$ and $\gamma_{0n}$, the difference between the eigenenergies is found to be $E_{0n} - E_{10} = ( n \alpha - 1) \omega_x$.
This energy difference crucially enters the expression for the critical particle number of the bifurcation predicted by the Galerkin approach, Eq.~(\ref{eq:anisoncr}), and in turn the position of the critical point is controlled by the anisotropy parameter $\alpha$.

\begin{figure}[ht]
\centering
\subfigure{
\includegraphics[height=0.26\textheight]{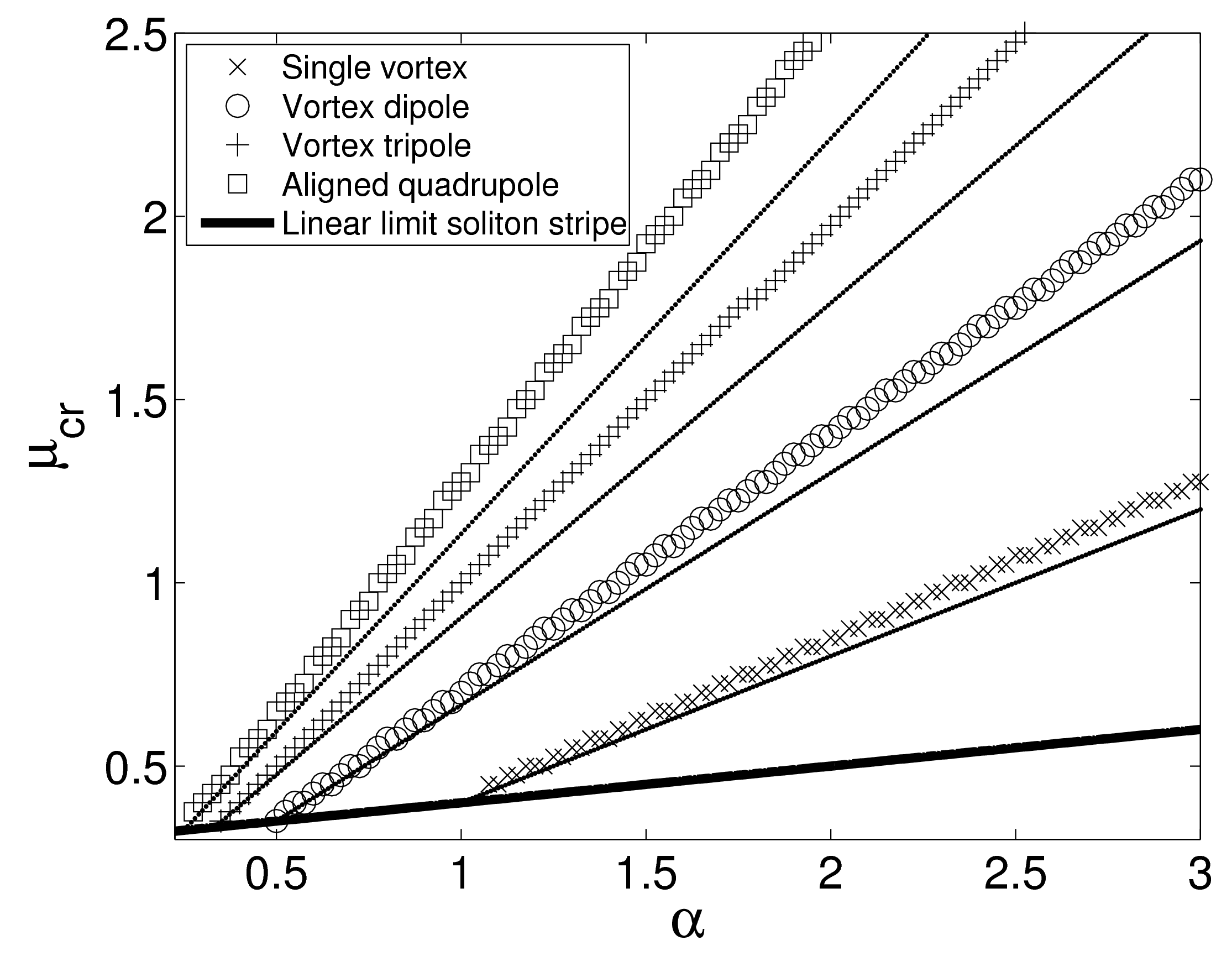}}
\subfigure{
\includegraphics[height=0.26\textheight]{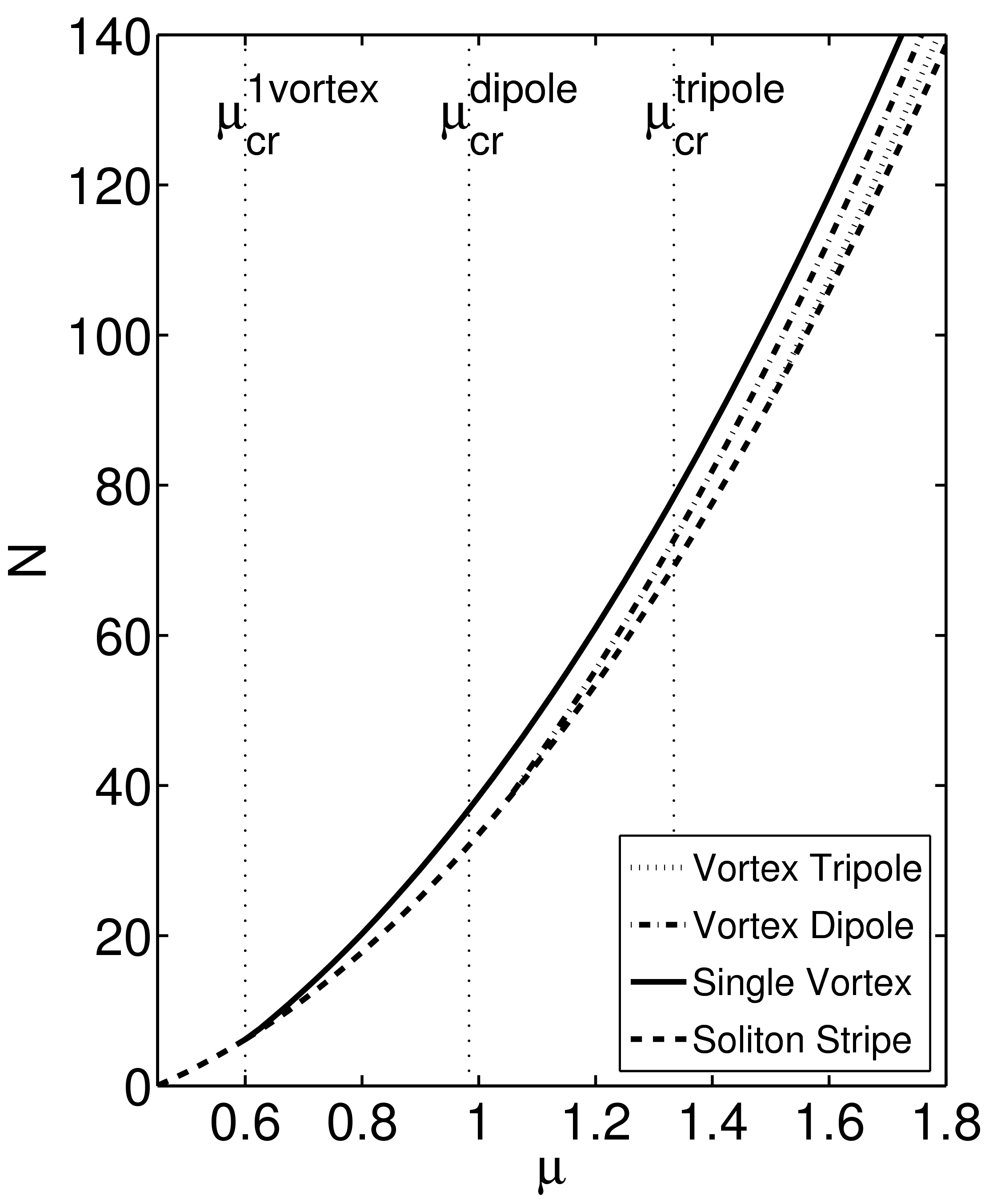}}
\caption[Optional caption for list of figures]{\label{fig:bif_1sol_alpha} (Left) Critical values of $\mu$ at which bifurcations leading to aligned vortex states occur, as a function of the anisotropy parameter $\alpha$: predictions of the Galerkin approach (thin dotted lines) and numerically found values (symbols as shown in legend). 
(Right) Example of an $N(\mu)$ bifurcation diagram, $\alpha = 1.5$, predicted critical values indicated by dotted vertical lines. Reprinted with permission from \cite{aniso}.}
\end{figure}

The theoretical predictions for $\mu_{\text{cr}}$ are shown in Fig. \ref{fig:bif_1sol_alpha}, together with the bifurcation points obtained from our numerical simulations.
We find that for any of the bifurcations considered the Galerkin approach gives correct results as long as the bifurcation happens sufficiently close to the linear limit. 
As $\alpha$ is increased, the bifurcation points are shifted to higher values of the particle number $N$, and the Galerkin approximation is less accurate.

Intuitively, for $\alpha > 1$, that is $\omega_y > \omega_x$, the energy of the $\gamma_{0n}$ ($n \geq 1$) states is higher than in the isotropic limit, while the energy of the $\gamma_{10}$ state is only weakly affected. 
In particular, the degeneracy of $\gamma_{10}$ and $\gamma_{01}$ (that constitute the one-vortex state) is lifted: 
In contrast to the isotropic case, the admixture of a $\gamma_{01}$ component to the $\gamma_{10}$ soliton state is suppressed, and thus the bifurcation of the single vortex is shifted away from the linear limit to a nonzero value of $N$.
By the same reasoning, all the other aligned vortex states bifurcate further away from the linear limit, too. We note that this has implications 
also for the stability of the higher aligned multi-vortex states, since the
only stable bifurcating state will be the single charge vortex inheriting the soliton's initial stability and
the remaining aligned states will, by necessity, be more unstable (by
one eigenmode) than before, a feature corroborated in the previous
section by the particle picture for $\alpha>1$.

On the other hand, for $\alpha < 1$ the energy of the $\gamma_{0n}$ states is lower than in the isotropic case, that is their admixture to the $\gamma_{10}$ soliton state is favored.
For values of $\alpha$ just below 1, the consequences are most drastic for the single vortex state.
As the energy of the $\gamma_{01}$ state is now lower than that of the $\gamma_{10}$ soliton state, it is no longer the case that the vortex emerges by an admixture of $\pm \ii \gamma_{01}$ to $\gamma_{10}$.
Rather, this picture is reversed, with the vortex emerging by an admixture of $\pm \ii \gamma_{10}$ to the (energetically favourable) $\gamma_{01}$ state.
In other words, the vortex branch now bifurcates from the soliton stripe oriented along the x-axis.
Similar arguments apply to the other aligned vortex states if the value of $\alpha$ is further decreased. 
For $\alpha=1/2$, the energies of the $\gamma_{10}$ soliton state and the $\gamma_{02}$ state (whose admixture leads to the vortex dipole) are the same, and thus the vortex dipole emerges from the linear limit.
For even smaller values of $\alpha$, the vortex dipole no longer bifurcates from the single soliton, but rather from the two soliton state $\gamma_{02}$ parallel to the x-axis (through an admixture of $\pm \ii \gamma_{10}$).
Obviously, this goes for any aligned vortex state: For $\alpha=1/n$, the states $\gamma_{10}$ and $\gamma_{0n}$ are degenerate and the vortex state emerges from the linear limit.
For smaller values than this, the bifurcation picture is reversed and the vortex state no longer bifurcates from the single soliton stripe along the $y$-axis,
but rather $\gamma_{0n}$ becomes the lower energy state to which
$\pm \ii \gamma_{10}$ gets admixed.

Now that we have understood the structural dependence of the bifurcation diagram on $\alpha$, let us summarize the conclusions on the stability properties of aligned vortex states in the presence of anisotropy.

We consider the regime of $\alpha > 1$ first.
As in the isotropic case, the soliton stripe is stable when it emerges from the linear limit. 
The first bifurcation (which now leads to the single vortex state) renders it unstable, with the vortex inheriting the soliton's stability.
The vortex dipole then bifurcates from this already unstable soliton stripe and is thus unstable as well (in contrast to the isotropic case). The higher aligned vortex states (tripole, quadrupole...) are all more unstable by the same reasoning.
Remember that, as stated above, (further) destabilization of the dipole, tripole and aligned quadrupole for $\alpha > 1$ is also predicted by the particle picture and confirmed by our numerical computations.

We can also draw some conclusions on the stability of the soliton stripe itself. 
For $\alpha > 1$, we find that the length of the interval between the emergence of the soliton from the linear limit and the first bifurcation point increases as a function of $\alpha$, see Fig. \ref{fig:bif_1sol_alpha}.
The Galerkin approach predicts a linear increase.
This corresponds to a growing range of values of the chemical potential for which the soliton stripe is stable.
This, in turn, reflects the fact that for $\alpha \gg 1$ we progressively
approach the 1D regime where the soliton stripe -- ultimately, the 1D dark soliton -- 
is stable for all values of $\mu$ for which it exists (which is
consonant with the prediction of the 1D Gross-Pitaevskii theory).

\begin{figure}[ht]
\centering
\fbox{\includegraphics[width=0.65\textwidth]{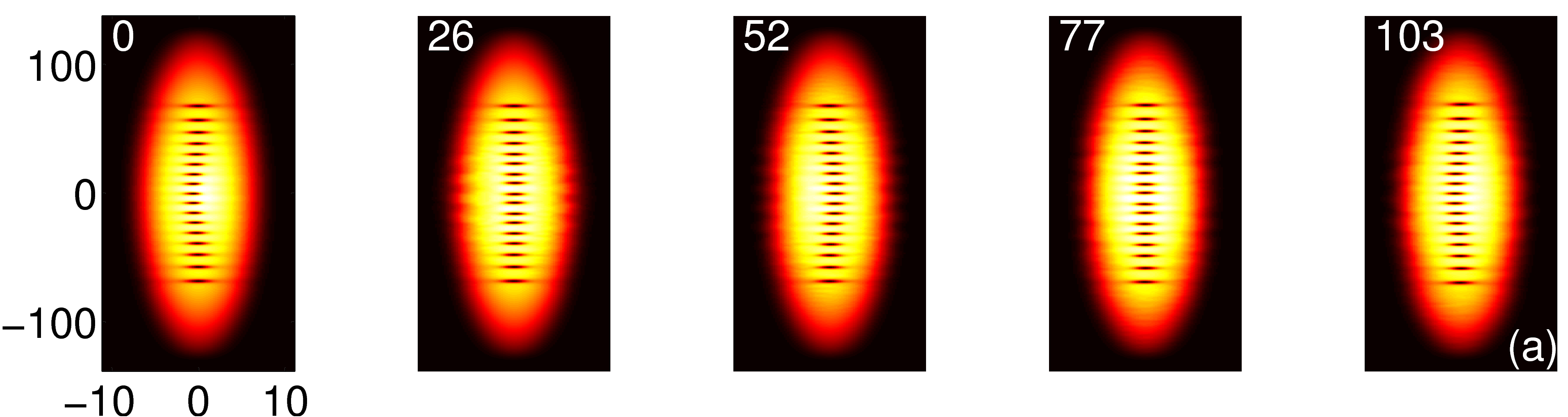}}
\fbox{\includegraphics[width=0.65\textwidth]{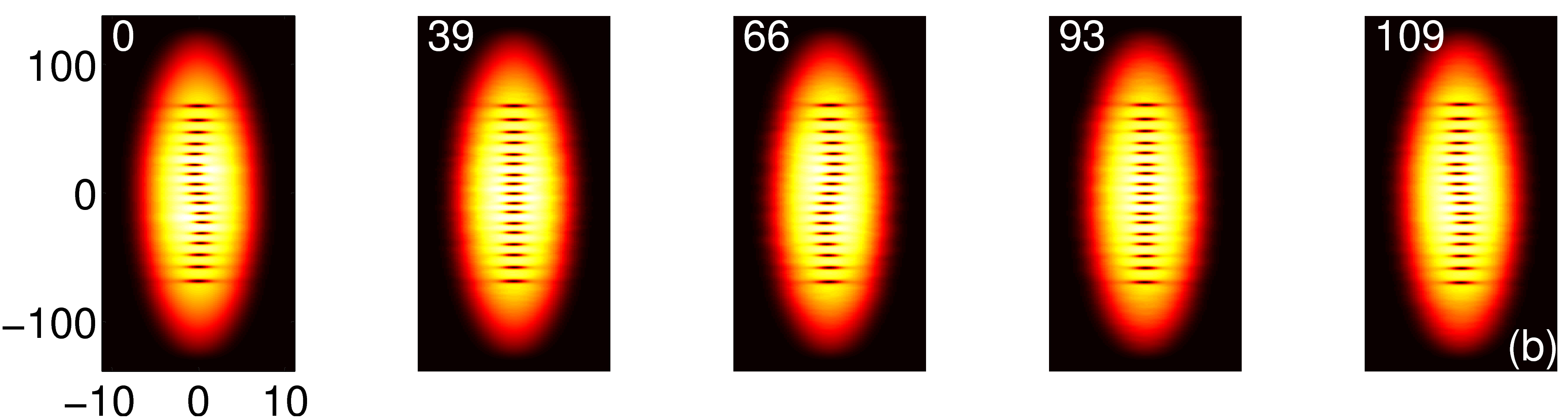}}
\fbox{\includegraphics[width=0.65\textwidth]{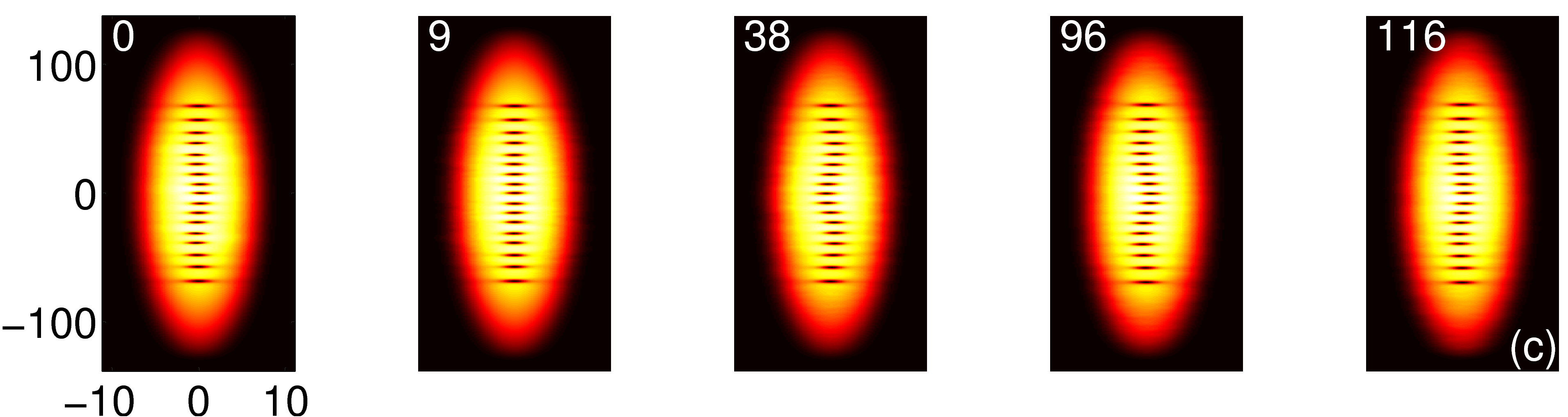}}
\caption[Optional caption for list of figures]{\label{fig:17vortex}Anomalous modes of a stabilized 17-vortex cluster, resembling the fundamental and first harmonic modes of a classical string. Note the different scaling in the $x$- and $y$-axes. Reprinted with permission from \cite{aniso}.}
\end{figure}

On the other hand, for $\alpha < 1$ the aligned vortex clusters 
with the vortices located along the $y$-axis tend to get stabilized.
In the interval $1/2 \leq \alpha \leq 1$ the vortex dipole is the first state bifurcating from the soliton and is thus stable, while the soliton stripe gets 
destabilized.
For $1/3 \leq \alpha \leq 1/2$ the vortex dipole no longer bifurcates from the soliton and the tripole takes its place (and its stability properties) as the
first emerging state.
So for $\alpha \leq 1/2$ the vortex tripole (which is unstable in isotropic traps) is expected to be stabilized.
By the same reasoning, the aligned quadrupole gets stabilized for $\alpha \leq 1/3$, the vortex quintupole for $\alpha \leq 1/4$, and so on.
It can further be argued that the aligned vortex states are still stable when $\alpha$ is so small that they no longer bifurcate from the soliton stripe along the $y$-axis.
In this case (as discussed above) the state with $n$ vortices aligned bifurcates from the $\gamma_{0n}$ branch through an admixture of $\pm \ii \gamma_{10}$.
Similarly to the $\gamma_{01}$ soliton stripe branch, the general $\gamma_{0n}$ solitonic branch can be expected to be stable when it emerges from the linear limit, and the admixture of $\pm \ii \gamma_{10}$ will induce the first bifurcation from it, thus leading to a stable vortex cluster state (while the $n$ soliton state gets destabilized).
Thus, in total, the states with $n$ vortices aligned along the $y$-axis are expected to be stable for any $\alpha \leq 1/(n-1)$, which is in very good agreement with our numerical results for the single vortex, vortex dipole, tripole and quadrupole shown above.
It has been demonstrated numerically in \cite{aniso} that even large aligned vortex clusters of up to $n=17$ vortices can be stabilized by strong enough transversal confinement, and that the anomalous BdG eigenmodes of such a linear cluster are reminiscent of standing waves on a classical string, see Fig. \ref{fig:17vortex}. Interestingly, in such a vortex string, there will be
$n$ (i.e., in the above example $17$) internal (anomalous) modes of 
vibration in its spectrum, which will, in turn, correspond to the 
$n$ normal modes of such a vortex lattice.

At this point, a remark is in order. We have demonstrated in this section how on the one hand insight into the stability of vortex clusters can be gained by identifying the relevant bifurcations from solitonic branch of states. Tuning the anisotropy parameter $\alpha$ can lead to qualitative changes in the bifurcation diagram, which carry over to changes in the stability properties of vortex clusters close to their emergence from the soliton. From this, one can learn about the stability of vortex clusters in the limit of small values of $N$ or $\mu$. On the other hand, in the limit of large chemical potentials we have the results from the particle picture ODEs and the corresponding stability analysis. In the preceding section it has been shown that predictions concerning the stability obtained from these two opposite end regimes agree very well. However, it is not to be taken for granted that in the course of tuning the chemical potential $\mu$ from small values to larger ones the stability properties of each state have to be preserved. In fact, the BdG spectra, of course, do not only depend on $\alpha$ but in general also on $\mu$. We have performed extensive numerical scans to cover the whole parameter space and make sure that no essential dependence of the stability on $\mu$ is missed.
Indeed, we find that in almost all cases tuning the chemical potential at fixed $\alpha$ only weakly affects the BdG spectrum and does not lead to the appearance or vanishing of purely imaginary modes (which are the ones that the bifurcation approach can tell us about). However, in numerous cases we observe collisions between modes of positive and negative Krein signature, resulting in the emergence of complex mode quartets that persist for a limited range of values of $\mu$ and then split again into two real, stable modes (see also the relevant
footnote at the end of the Introduction). For the vortex dipole in isotropic traps, e.g., the presence of these complex ``bubbles'' at intermediate $\mu$ is well-known \cite{PhysRevA.82.013646}, and it is no surprise that such intervals of weak oscillatory instability can also be found in anisotropic settings. Let us note that in general neither the near-linear bifurcation approach nor the highly nonlinear particle picture can provide information about these complex quartets at intermediate chemical potentials, and these are only captured by the 
detailed numerical continuations discussed (wherever relevant) herein.

Finally, let us point out that our findings presented in this section are consistent with previous results on the stability of the one soliton state in anisotropic settings \cite{PhysRevA.65.043612}.
In this work it was found (employing box boundary conditions and keeping the particle density fixed) that in the regime corresponding to our $\alpha \gg 1$ relaxing the confinement in the $y$-direction (i.e. approaching $\alpha = 1$) opens up an increasing number of decay channels for the soliton, with the first one leading to a single vortex, the second one leading to a vortex dipole and so on: In our analysis, these ``decay channels'' correspond to the imaginary modes in the soliton's BdG spectrum, induced by supercritical pitchfork bifurcations from the soliton branch having happened at lower values of $N$ (or $\mu$) than the one under consideration.

\begin{figure}[ht]
\centering
\includegraphics[width=0.4\textwidth]{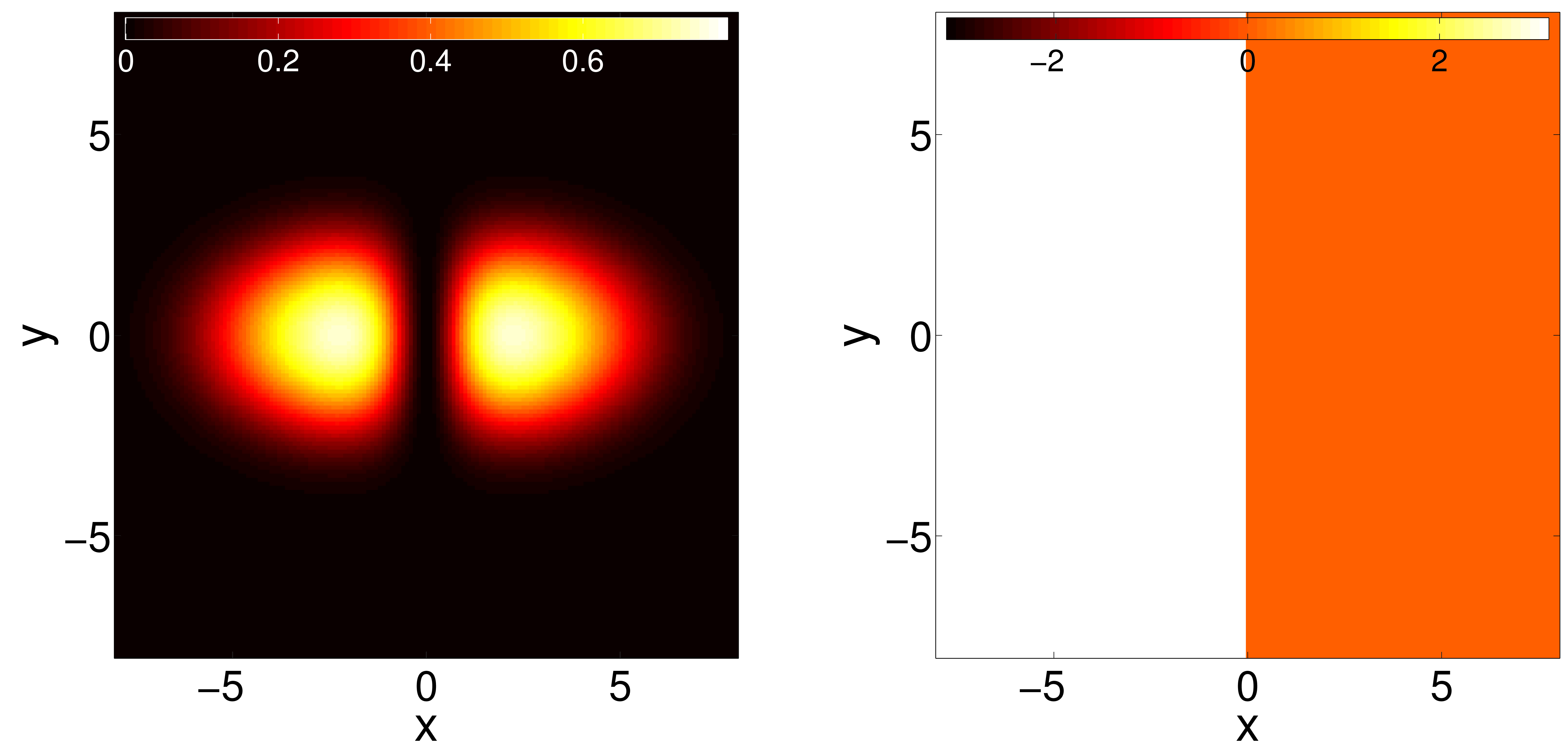}\hspace{0.04\textwidth}
\includegraphics[width=0.4\textwidth]{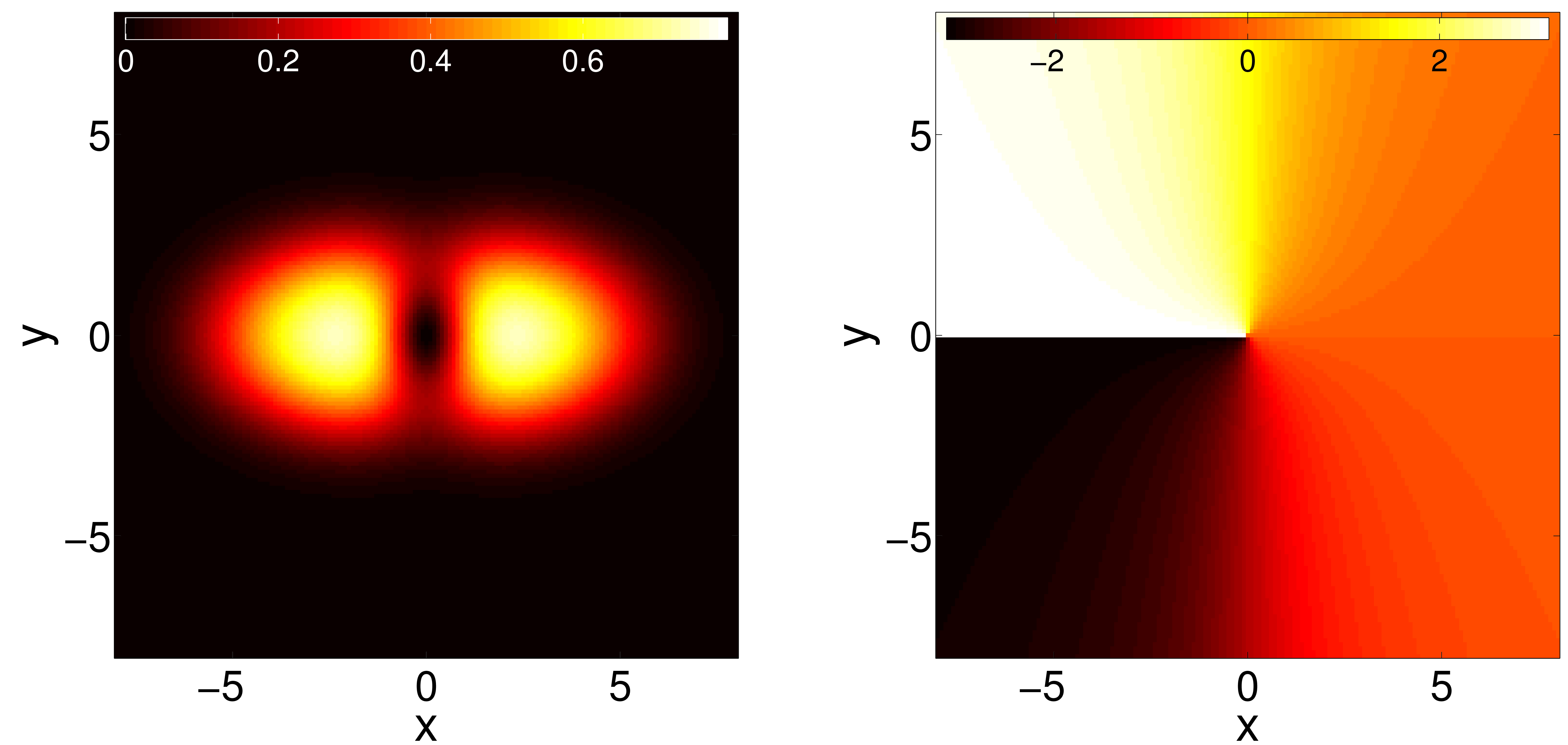}
\caption[Optional caption for list of figures]{\label{fig:sv} Soliton stripe (left panels) and single vortex state (right panels) profiles at $\mu = 1$, $\alpha = 2.1$ (``solitonic vortex'').}
\end{figure}

Furthermore, the authors of \cite{PhysRevA.65.043612} report the numerical observation of a ``solitonic vortex'' solution to the GPE in anisotropic settings, i.e. a stationary state with its density and phase properties in between those of a soliton stripe and a single vortex. This is reported to bifurcate from the dark soliton stripe as the anisotropy of the confinement is varied. Even though different boundary conditions are employed, this agrees well with our findings:
If $\alpha \neq 1$, the single vortex state bifurcates from the soliton stripe at a finite particle number (due to the non-degeneracy of $\gamma_{01}$ and $\gamma_{10}$).
This, in turn, necessarily means that close to the bifurcation point the vortex state will still show some similarity to the soliton solution from which it just bifurcated.
Fig. \ref{fig:sv} shows an example of the single vortex state close to its bifurcation from the soliton stripe.
Clearly, the density shows remnants of the soliton stripe.
The phase runs from $-\pi$ to $+\pi$ continuously, a characteristic of the singly charged vortex, however the azimuthal phase gradient is not constant (as it is for the known vortex solution in isotropic settings).
Rather, the phase changes very sharply in the region around the $y$-axis, again owing to the state's former solitonic properties.

\section{Non-aligned vortex clusters: The isotropic limit}
\label{sec:aniso2}
Let us now apply the methods developed so far to stationary clusters of vortices that are not necessarily aligned along one axis.
Having extensively discussed aligned vortex states bifurcating from the one soliton stripe branch (which reduces to $\gamma_{01}$ or $\gamma_{10}$ in the linear limit), the natural next step is to turn to the next higher excited harmonic oscillator states, namely $\gamma_{20}$, $\gamma_{02}$ and $\gamma_{11}$, and study solitonic and vortex-type branches of states that reduce to linear combinations of these modes in the linear limit.
Again, the strategy is to gain insight into the stability properties of vortex clusters under the influence of anisotropy by locating their bifurcations from solitonic branches.
Subsequently, calculations within the particle picture ODE system and full dynamical simulations will be employed in section 7 to access the problem from different directions and to provide a unifying picture.

\begin{figure}[ht]
\centering
\subfigure[Two parallel soliton stripes]{
\includegraphics[width=0.4\textwidth]{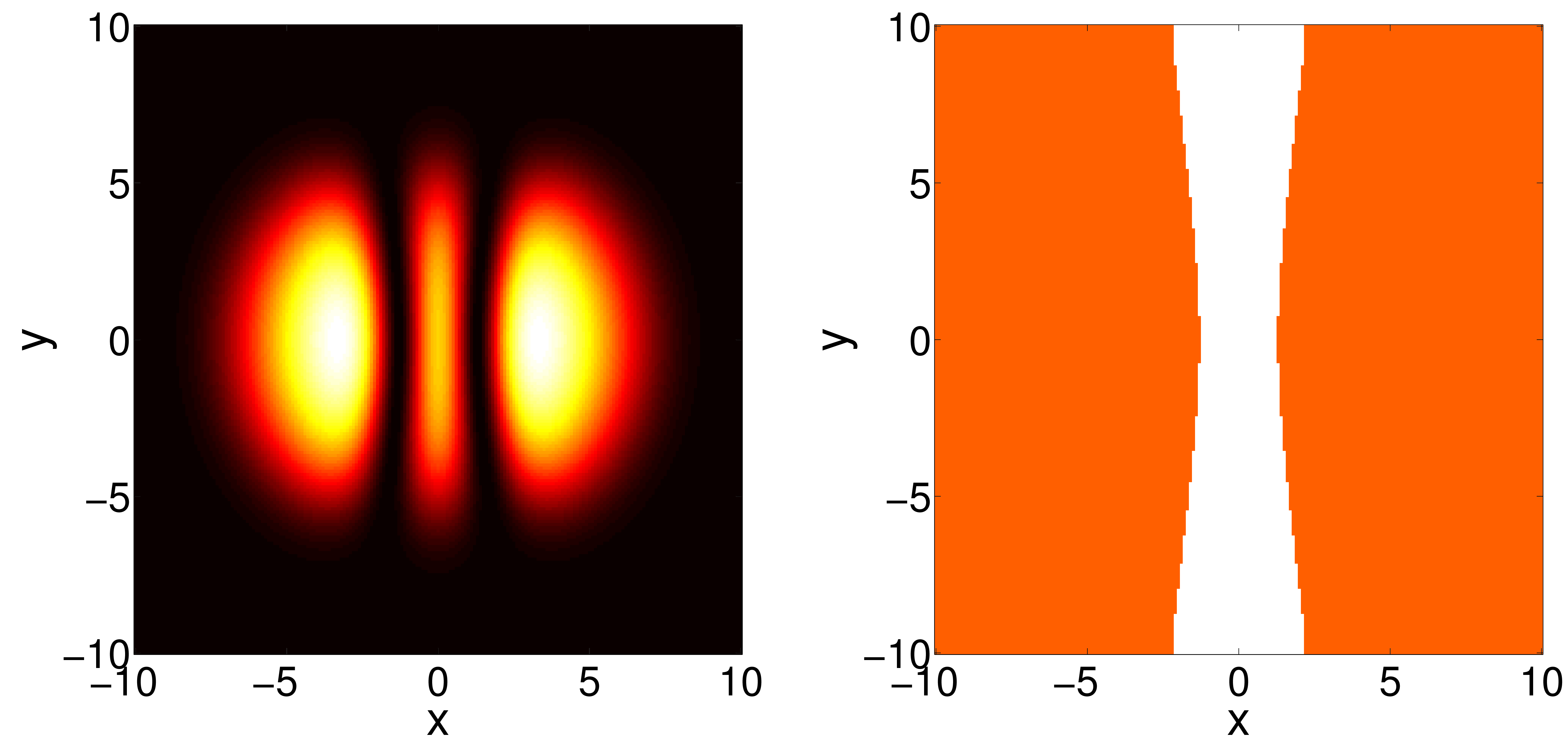}
\label{fig:2sol}
}
\subfigure[Soliton cross]{
\includegraphics[width=0.4\textwidth]{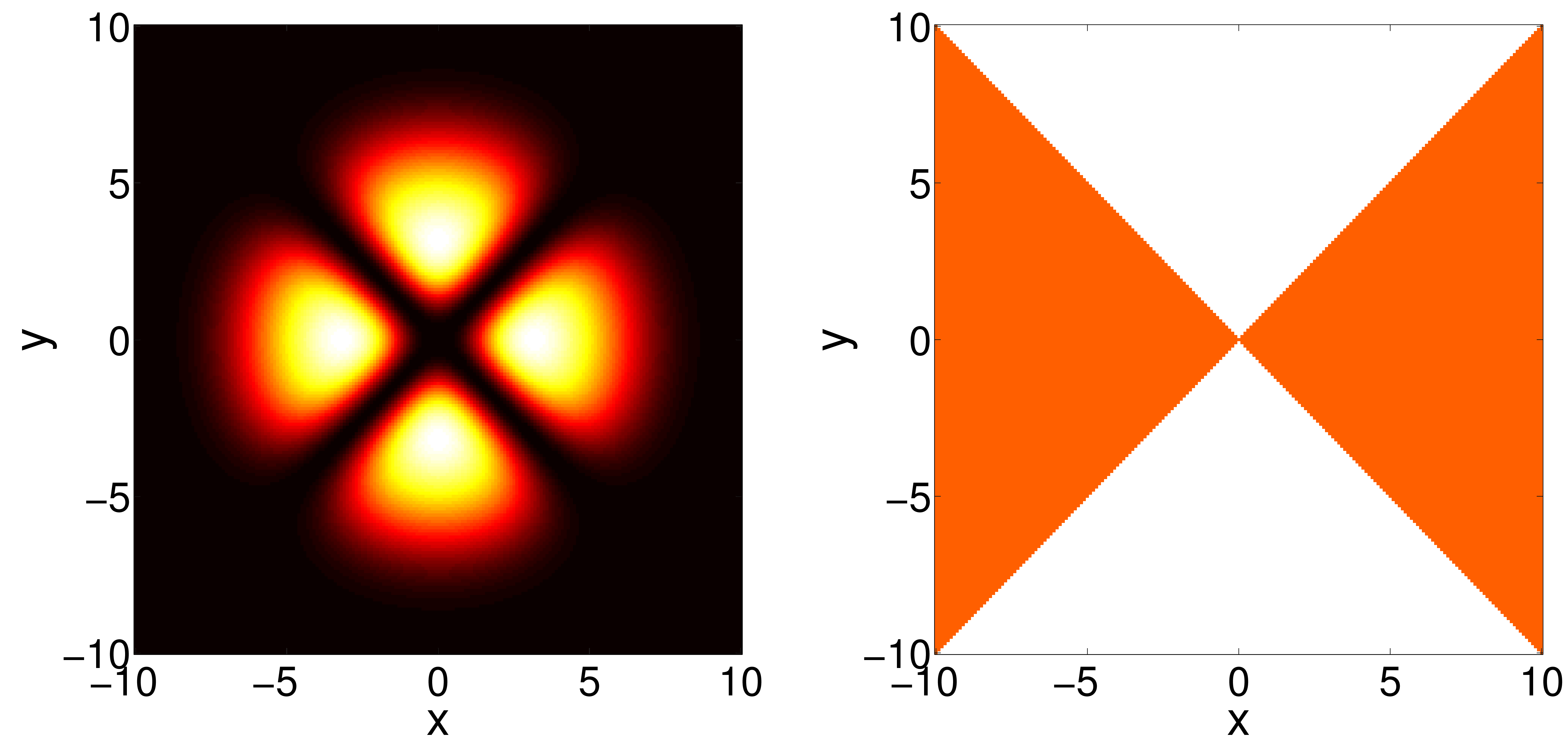}
\label{fig:cross}
}
\subfigure[Ring soliton]{
\includegraphics[width=0.4\textwidth]{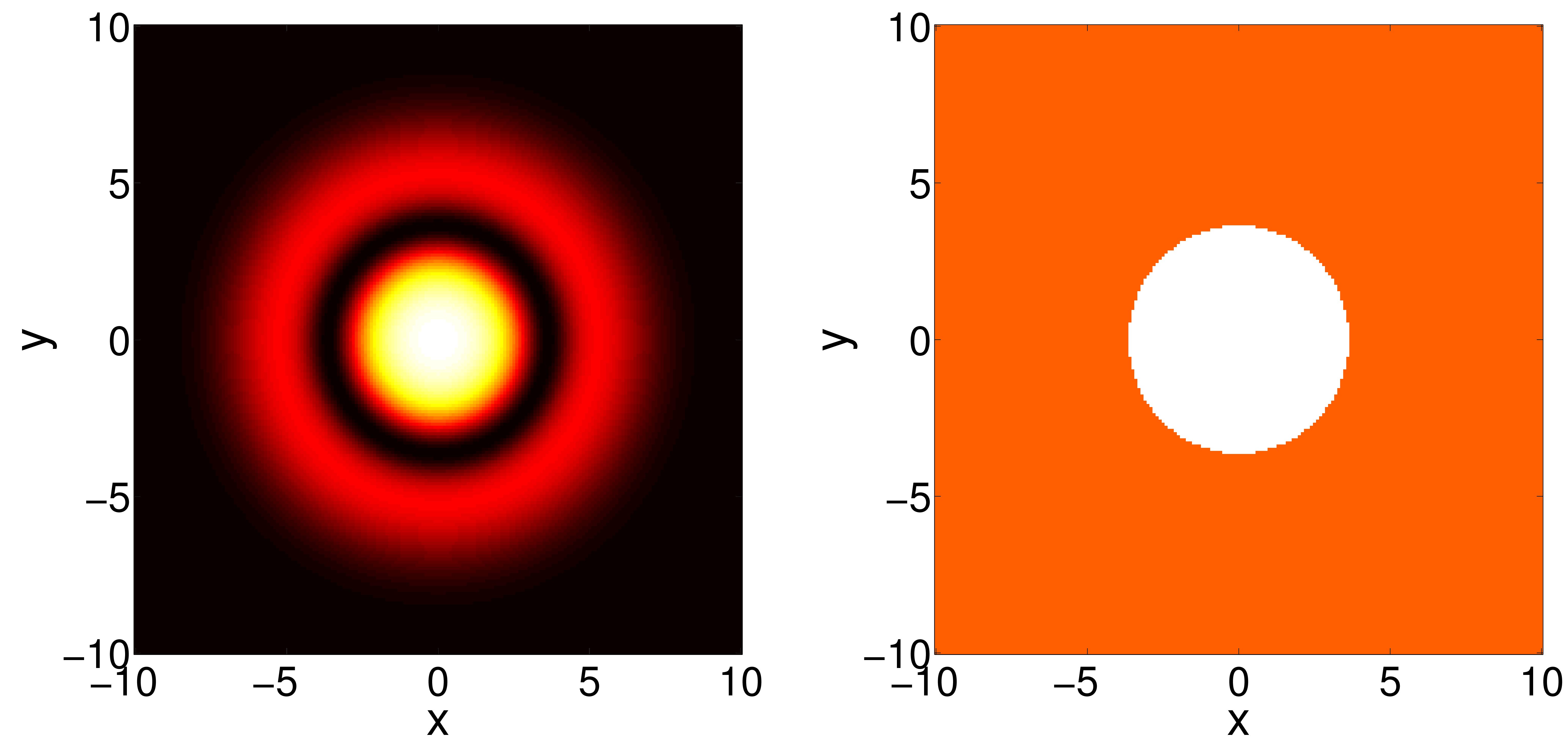}
\label{fig:ring}
}
\subfigure[Vortex quadrupole, orientation A]{
\includegraphics[width=0.4\textwidth]{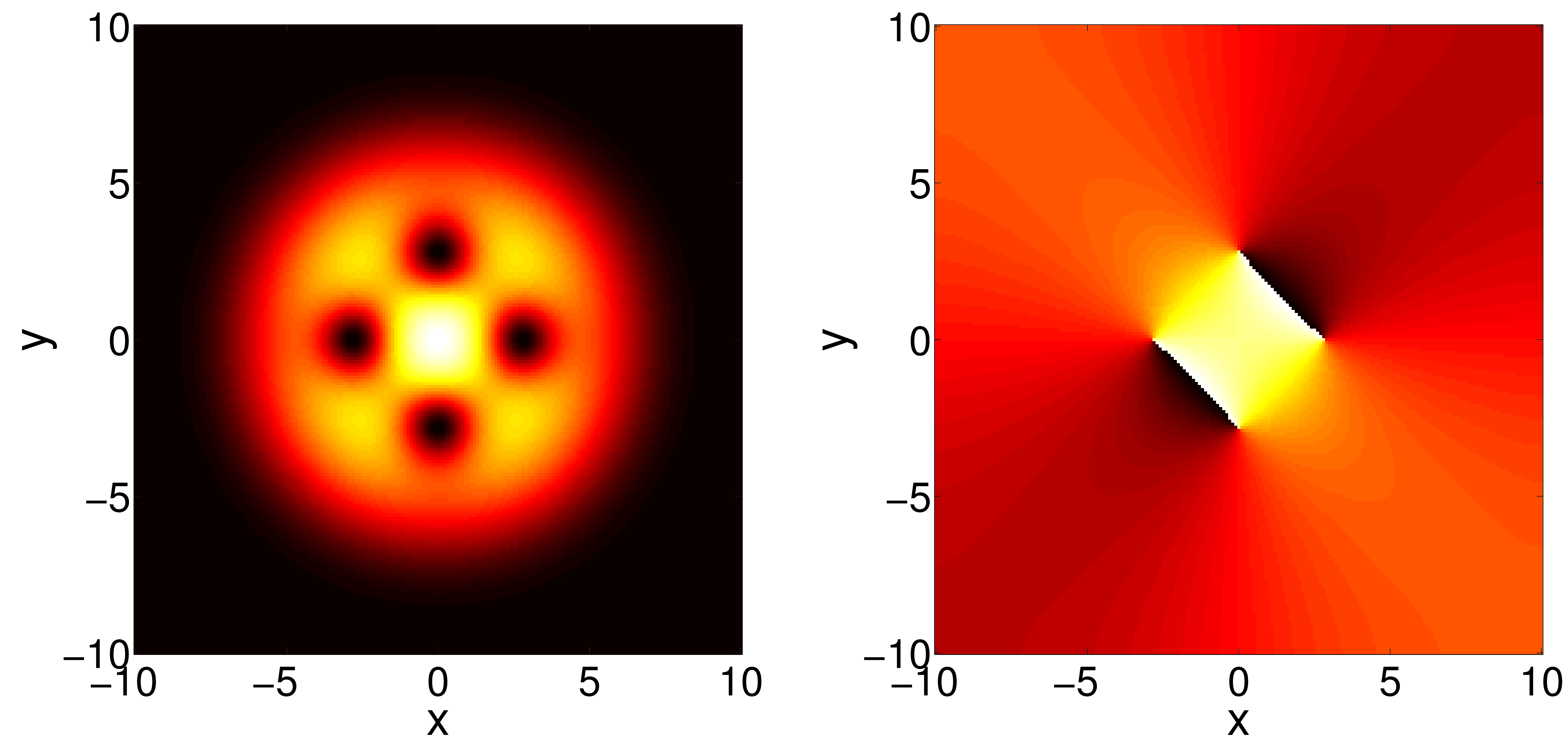}
\label{fig:quadA}
}
 \subfigure[Vortex quadrupole, orientation B]{
\includegraphics[width=0.4\textwidth]{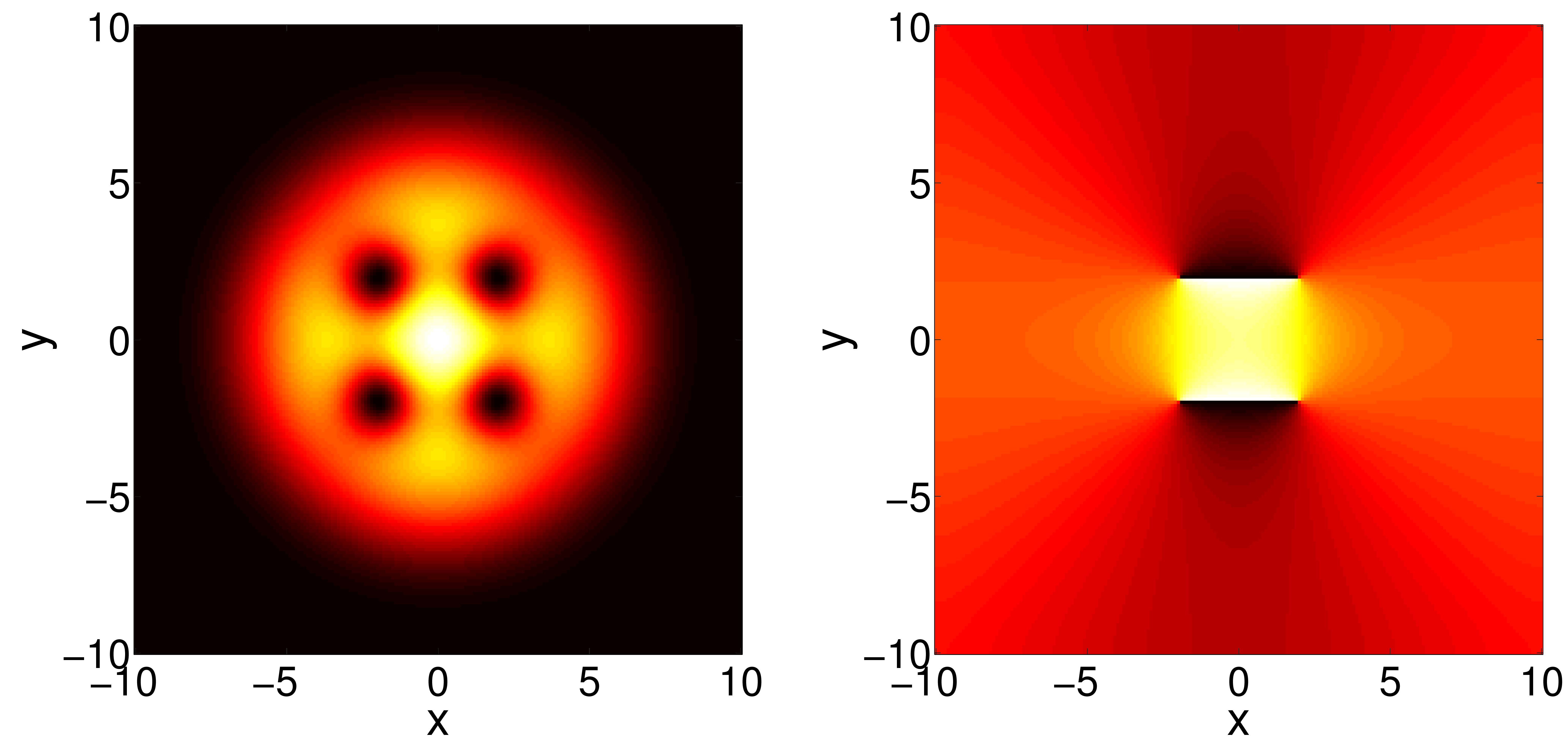}
\label{fig:quadB}
}
\subfigure[Doubly charged vortex]{
\includegraphics[width=0.4\textwidth]{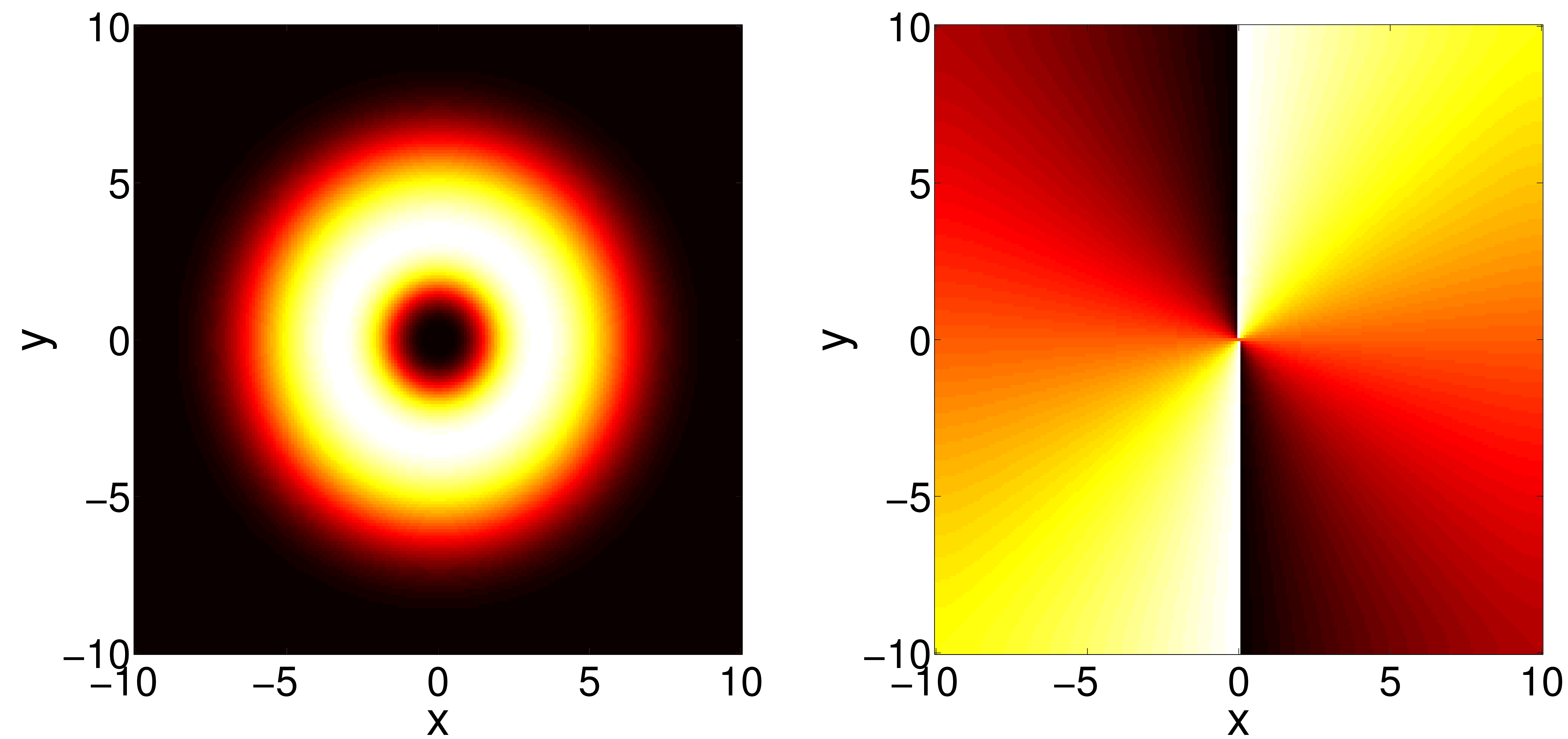}
\label{fig:ch2vorpic}
}
\subfigure[Tripole with doubly charged center]{
\includegraphics[width=0.4\textwidth]{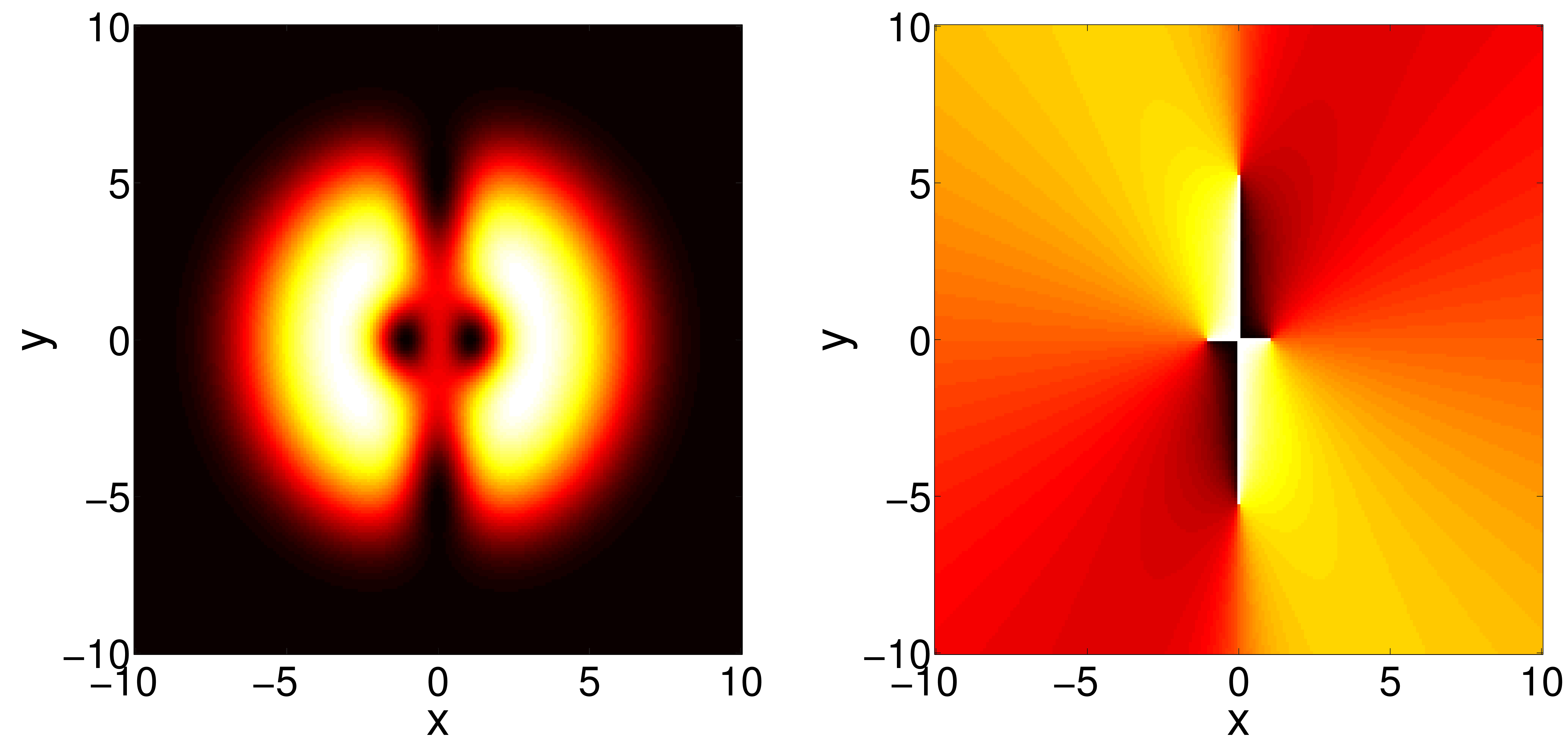}
\label{fig:20i11}
}
\subfigure[Tripole with doubly charged center, $\mu = 3$]{
\includegraphics[width=0.4\textwidth]{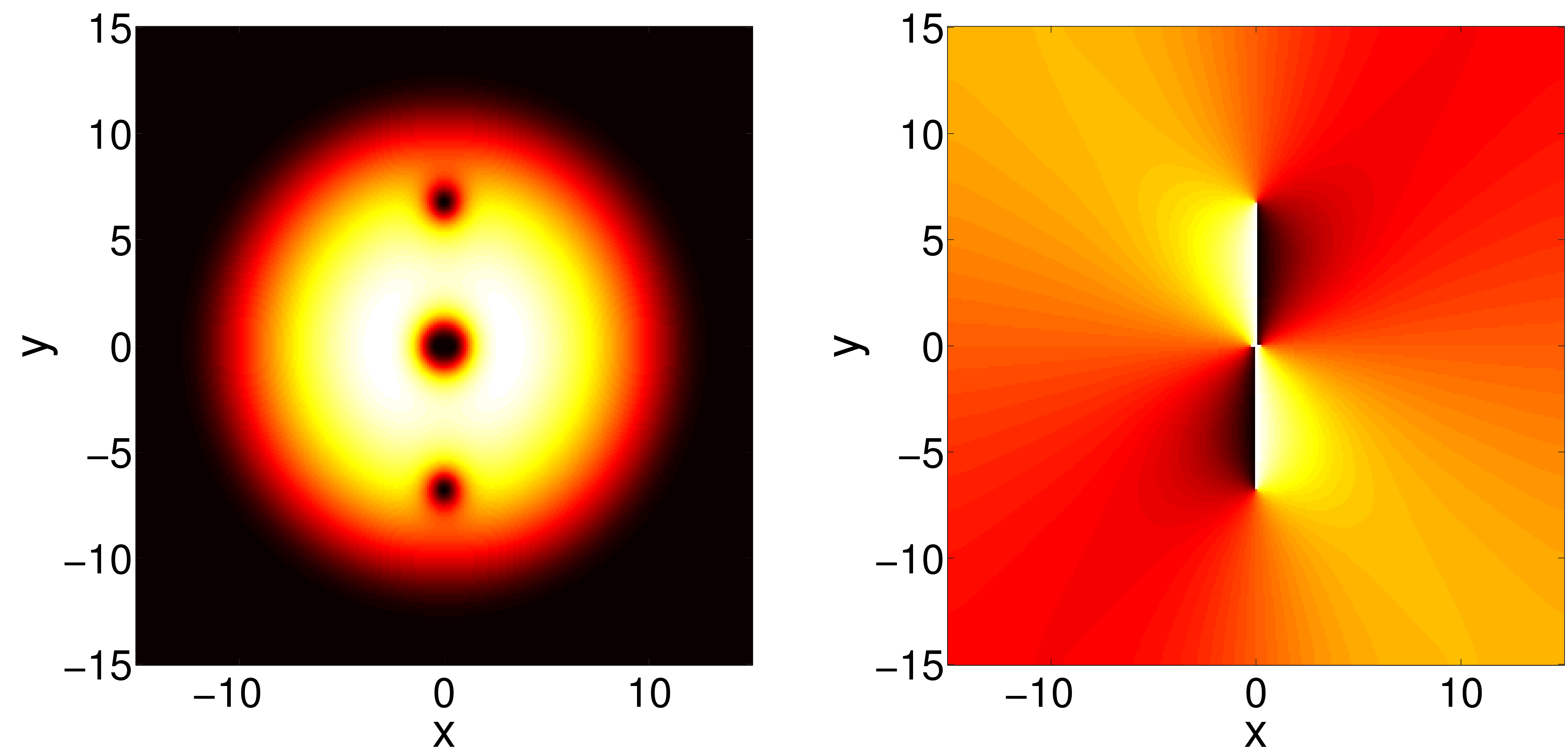}
\label{fig:20i11highmu}
}
\caption[Optional caption for list of figures]{\label{fig:states_20}Density and phase profiles of states emerging from the linear limit at $\mu=3\omega_r$. Chemical potential is chosen to be $\mu=1.2$  except where stated otherwise.}
\end{figure}

Before entering the discussion of bifurcation diagrams in anisotropic settings, in this section we will first take a detailed look at the isotropic case where $\omega_x = \omega_y \equiv \omega_r$.
Fig. \ref{fig:states_20} collects density and phase profiles of the different branches of states which emerge from the linear limit at a chemical potential of $\mu = 3 \omega_r$, i.e. which reduce to superpositions of harmonic oscillator states $\gamma_{20}$, $\gamma_{11}$, $\gamma_{02}$ as $N \rightarrow 0$.
$N(\mu)$ bifurcation diagrams including a number of these plus some additional branches (which emerge further away from the linear limit and are therefore not relevant for our present discussion) can be found in \cite{Middelkamp2010b}.

For $\alpha=1$ there are three different (apart from trivial rotations) solitonic branches emerging from the linear limit at $\mu=3\omega_r$:
The first of these is a state with two parallel soliton stripes, see Fig. \ref{fig:2sol}. It can be traced back to the harmonic oscillator eigenstate $\gamma_{20}$ as $N \rightarrow 0$, where again we fix the y-axis parallel to the density minima of the soliton stripes. 
The other two are the dark ring soliton, Fig. \ref{fig:ring}, and the two diagonally crossed dark solitons, Fig. \ref{fig:cross}, which can asymptotically (for small $N$) be decomposed as $\gamma_{20} + \gamma_{02}$ and $\gamma_{20} - \gamma_{02}$, respectively (where again overall normalization constants are omitted).
 Of the three solitonic states, two (namely, the ring and the cross) are unstable right from the linear limit, while the two soliton stripe state is stable for small $\mu$.
As $\mu$ is increased, vortex states bifurcate from all the solitonic branches
leading to (further) destabilization.

Additionally, there is a number of vortex states that also emerge from the 
linear limit at $\mu=3\omega_r$:
The most prominent cluster state of these is probably the (nonaligned) vortex quadrupole, consisting of four singly charged vortices of alternating vorticity which in the isotropic trap are located at the vertices of a square, and opposite vortices have the same charge.
This state has been identified and discussed in \cite{PhysRevA.74.023603,PhysRevA.71.033626,Kapitula2007112, PhysRevA.82.013646}.
Having in mind the extension to anisotropic settings, we already distinguish two different orientations of this vortex quadrupole, even if at $\alpha = 1$ they can be transformed into each other by a trivial rotation.
We speak of orientation A if the four vortices sit at the trap's $x$- and $y$-axes as in Fig. \ref{fig:quadA}.
In this case, the state's linear decomposition as $N \rightarrow 0$ is identified to be $\gamma_{20} + \gamma_{02} \pm \ii \sqrt{2} \gamma_{11}$, where we take the linear modes to be normalized to unity and omit the overall normalization constant.
On the other hand, if the vortices are located along the diagonals of the coordinate system as in Fig. \ref{fig:quadB}, we call this orientation B.
The corresponding linear limit reads $\gamma_{20} \pm \ii \gamma_{02}$.
For $\alpha = 1$, the quadrupole state (in any orientation) is stable for arbitrary chemical potentials (apart from an oscillatory instability window, see also \cite{PhysRevA.82.013646}), and no other stationary
states are found that bifurcate from it.

\begin{figure}[ht]
\centering
\includegraphics[width=0.45\textwidth]{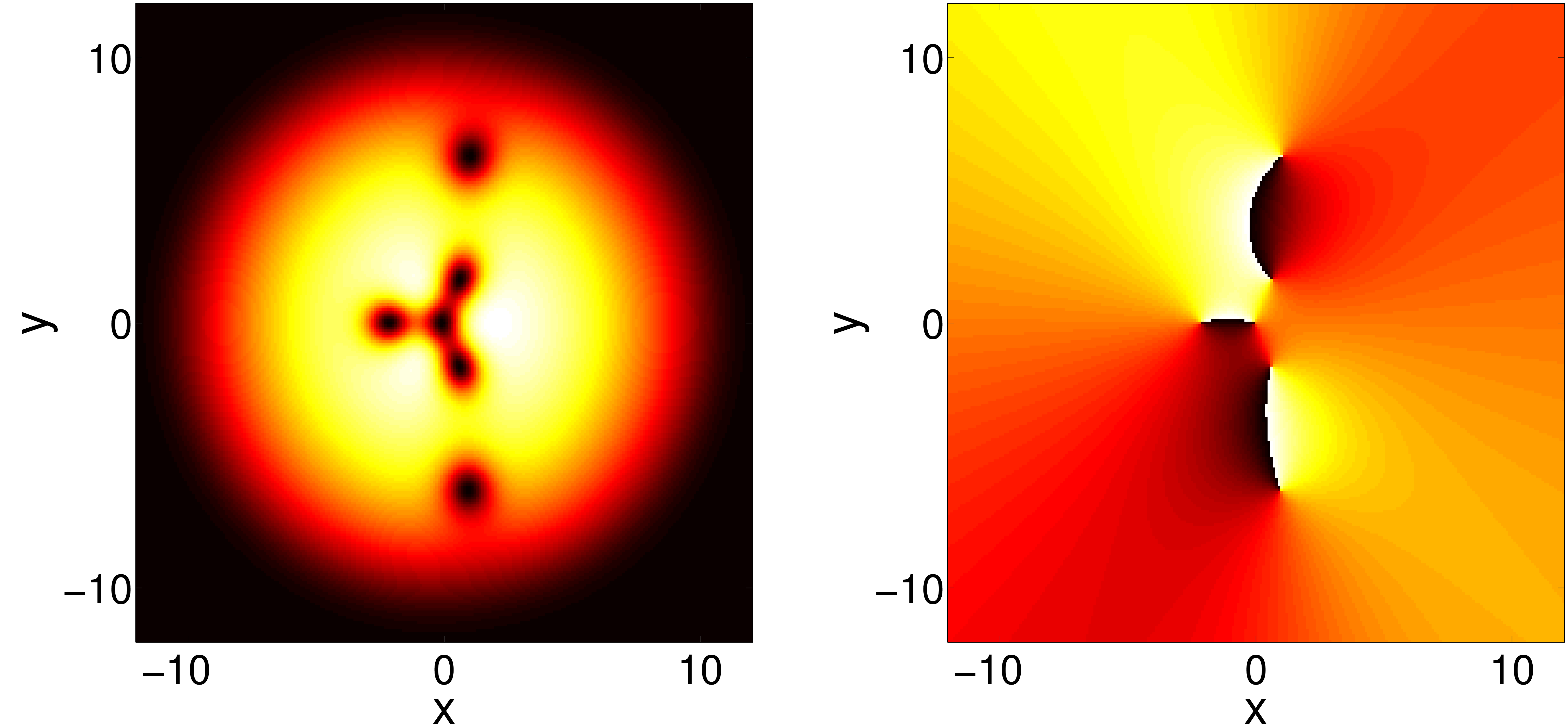}
\caption[Optional caption for list of figures]{\label{fig:biffrom20i11}Density and phase profiles at $\alpha = 1, \mu = 2.5$ of one of the symmetry-broken six vortex clusters that bifurcate from the tripole with the doubly charged center, destabilizing it.}
\end{figure}

Next, we turn to a branch of states that to our knowledge has not been described before, see Figs. \ref{fig:20i11} and \ref{fig:20i11highmu}.
Our findings indicate that it starts as $\gamma_{20} \pm \ii\gamma_{11}$ in the linear limit, and for small $N$ it shows some similarity to the vortex quadrupole in orientation A:
There are four singly charged vortices situated at the trap's axes.
In contrast to the quadrupole configuration, these four vortices do not form a square, however.
Instead, two of them (which have the same charge) are located closer to each other near the center of the trap, while the other two are further away from the trap center, in regions of low density.
Increasing the chemical potential, we find that the two central vortices finally merge to form one vortex of charge 2.
The whole configuration then has a tripole-like structure, where the charges of the three vortices in the tripole are given by $\mp1, \pm 2, \mp 1$.
Fig. \ref{fig:20i11highmu} illustrates this tripole profile at larger chemical potentials. 
Concerning stability, we find that in an isotropic setting this branch only suffers from relatively weak oscillatory instabilities when emerging from the linear limit, before at $\mu \approx 1.4$ one small purely imaginary mode arises in its BdG spectrum.
This destabilization is again due to a pair of symmetry-broken states bifurcating from the tripole with the doubly charged center, but in contrast to the prototypical bifurcations of vortex clusters from dark solitonic branches encountered so far, the symmetry that is broken is of a different kind: while the density of the tripole with the doubly charged center is symmetric with respect to reflections about both the $x$- and $y$-axes, the clusters of six vortices that bifurcate from it break the axial symmetry with respect to the $y$-direction, see Fig. \ref{fig:biffrom20i11}. To be more precise, the tripole state itself (not just its density) is invariant with respect to a combined reflection about $y$ and time-reversal transformation, and this symmetry is not shared by the vortex branches emerging in the bifurcation.
As can be seen in Fig. \ref{fig:biffrom20i11}, the newly found symmetry-broken clusters are made up of six singly charged vortices, three of each sign. One vortex is located in the trap center, surrounded by a triangular configuration of three oppositely charged vortices. The remaining two vortices are of the same charge as the central one and they are located close to the edge of the cloud near the $y$-axis equilibrating the whole structure. As $\mu$ is decreased towards the bifurcation point, the cluster straightens along the $y$-axis and a vortex-antivortex pair in the center vanishes in the zero density core of a third vortex, until at the critical value the precursor of the tripole with the doubly charged center as shown in Fig.~\ref{fig:20i11} is recovered.
As one would expect in this type of supercritical pitchfork bifurcation, the symmetry-broken six vortex clusters are stable in the sense that their BdG spectra do not exhibit purely imaginary modes as we have checked. Let us remark, however, that their spectra exhibit weak oscillatory instabilities to which our 
theoretical bifurcation analysis cannot provide access, 
but which can only be tracked numerically, 
see also the remark at the end of the preceding section.

The next (and final) branch emerging from the linear limit at $\mu=3\omega_r$ is again well-known: It is the charge 2-vortex branch, see Fig. \ref{fig:ch2vorpic}.
In the linear limit, this vortex state consists of $\gamma_{20} - \gamma_{02} \pm \ii \sqrt{2} \gamma_{11}$, which in polar coordinates $(r,\theta)$ leads to the expected azimuthal variable dependence $\propto \exp(\pm \ii 2 \theta)$ 
characteristic of a doubly charged vortex.
Apart from small oscillatory instability ``bubbles'' that arise due to subsequent collisions of positive and negative Krein signature modes, and then disappear and reappear again as $\mu$ is increased, we find the doubly charged vortex to be stable. This agrees with the results obtained in \cite{PhysRevA.59.1533}.
It should be noted here that the instability of such a higher charged 
vortex is towards splitting into lower charge vortices, a feature which
by now has been observed experimentally, see e.g.~\cite{PhysRevLett.93.160406}.

Having commented on all branches that emerge from the linear limit, let us now turn to the first bifurcations from them.
Increasing the chemical potential, we find bifurcations leading to vortex states for all three solitonic states.
 
For the two soliton stripe state, these bifurcations lead to pairs of aligned vortex configurations, analogous to those discussed in the one soliton case. 
Their emergence can be explained by subsequent admixtures of $\pm \ii \gamma_{0n}$, where $n \geq 2$.
In a degenerate sense, the vortex quadrupole in orientation B is the first 
example in this line of double aligned vortex states.
The lowest-lying bifurcation happening at finite $N$ leads to a double tripole (2x3).
The numerically found critical chemical potential is $\mu_\text{cr} \approx 0.87$, while the Galerkin approach predicts $\mu_\text{cr} = \frac{283}{67}\omega_r \approx 0.84$.
At this bifurcation, the two soliton stripe state is rendered unstable, while the 2x3 vortex state inherits its stability (that is, its BdG spectrum shows no purely imaginary mode).
All the higher double aligned vortex states then bifurcate from the destabilized two soliton branch and are thus unstable. None of them will be touched upon in this work.

The same goes for the vortex states bifurcating from the ring soliton branch. 
These are known as ``vortex necklace'' states \cite{Theocharis2003a}, and they are characterized by an even number (4, 6, ...) of alternately charged 
vortices -- hence resulting into no net topological charge --, 
located at the vertices of a regular polygon.
One can also consider the vortex quadrupole as the first, again degenerate,
example of such a necklace state, followed by a vortex hexagon, 
octagon, and so on.
Theoretically, the bifurcations of the necklace states have been studied in great detail in \cite{Kapitula2007112}.
In this work, we will omit a detailed analysis of the unstable necklace 
states. 

\begin{figure}[ht]
\centering
\subfigure[2x3 crossed vortex cluster]{
\includegraphics[width=0.45\textwidth]{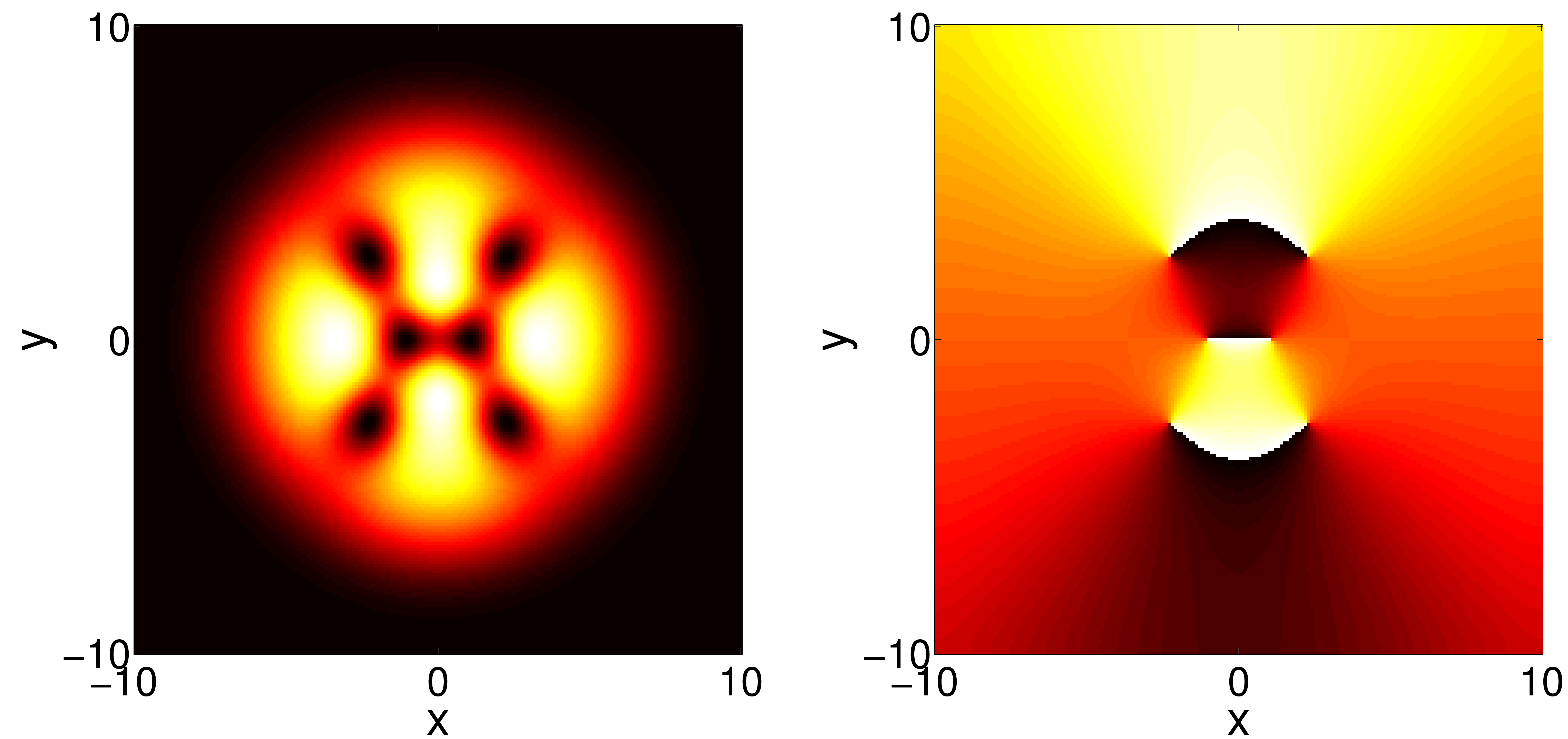}
\label{fig:2x3cross}
}
\subfigure[Hybrid one soliton/two vortex state]{
\includegraphics[width=0.45\textwidth]{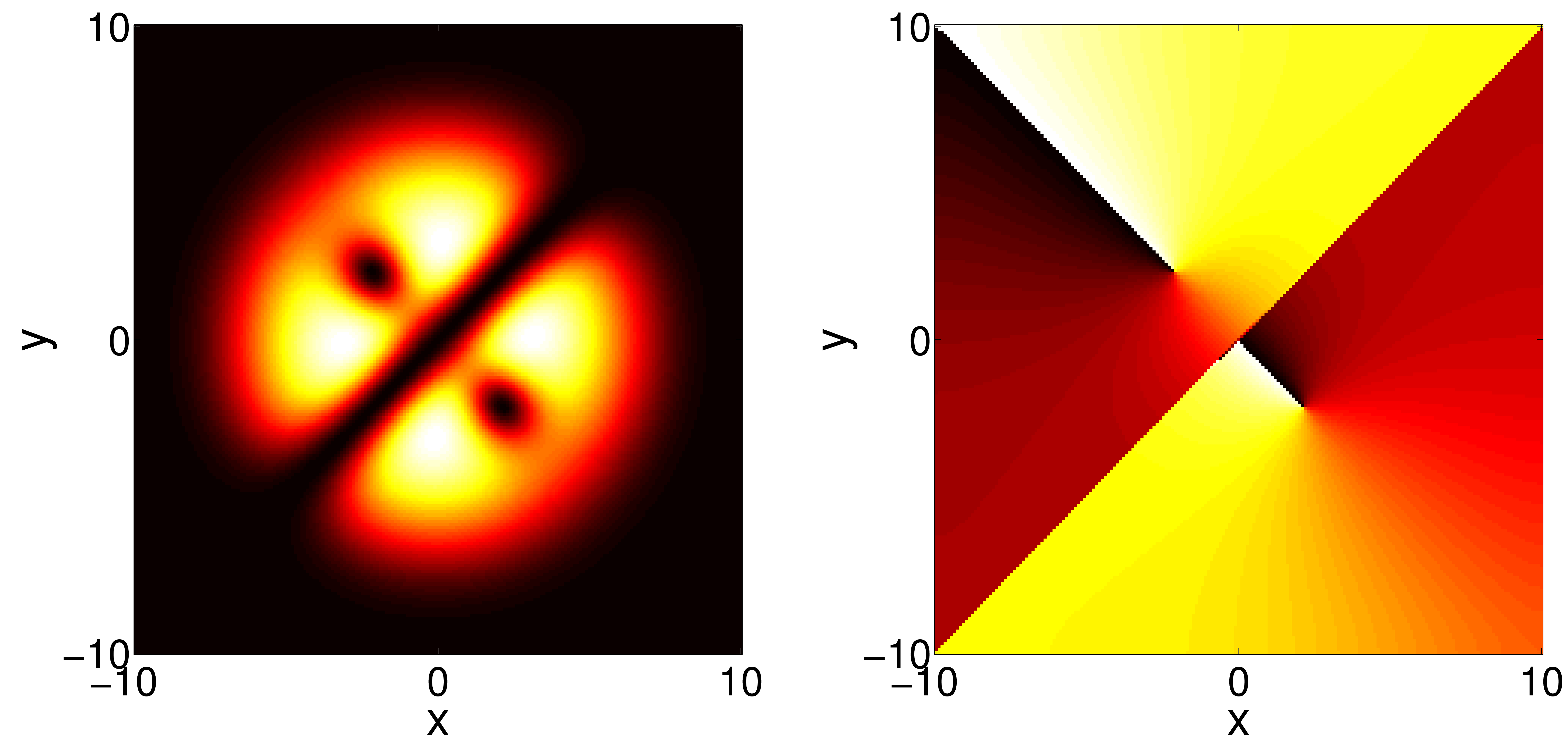}
\label{fig:2vor1sol}
}
\caption[Optional caption for list of figures]{\label{fig:bifoffcross}Density and phase profiles of states bifurcating from the soliton cross branch, $\mu = 1.3$.}
\end{figure}
Concerning the first bifurcations from the soliton cross, there are two branches of stationary states emerging from it at essentially the same critical value of $\mu$.
On the one hand, we find a six vortex state, similar to the double tripole bifurcating from the parallel soliton stripes, but with the vortices located along the former soliton cross, see Fig.~\ref{fig:2x3cross}.
Numerically, we find that this bifurcation happens at $\mu_\text{cr} \approx 0.92$.
In the linear picture, it is convenient to analyze this bifurcation in the rotated coordinate frame, where the soliton cross branch reduces to $\gamma_{11}$ in the linear limit.
Numerical decomposition of the numerically obtained branch of solutions into its harmonic oscillator components indicates that the emergence of the 2x3 crossed vortex state is then explained by an admixture of $\pm \ii (\gamma_{30}\pm\gamma_{03})$, where the relative sign between $\gamma_{03}$ and $\gamma_{30}$ decides along which direction the two central vortices in the cross configuration are located.
With these linear modes, the Galerkin equations yield $\mu_\text{cr} = \frac{111}{25} \omega_r \approx 0.89$.
Finally, a second branch (which also has not been identified before, 
to our knowledge) emerges from the soliton cross at the same critical chemical potential as the crossed 2x3 vortex branch.
This new branch is characterized by one of the soliton stripes formerly forming the cross staying intact while the other is replaced by two vortices, i.e.,
this is a ``hybrid'' state containing both solitonic stripes and vortex
waveforms as shown in Fig.~\ref{fig:2vor1sol}.
As in the symmetry-breaking bifurcation either of the soliton stripes can stay intact, there are actually again two different branches here, which can be transformed into each other by a rotation.
The numerically found critical chemical potential is $\mu_\text{cr} \approx 0.92$.
The bifurcation leading to this hybrid 
one soliton/two vortex state is also easier to understand if the axes are 
rotated by $\pi/4$, in which case the cross state coincides with $\gamma_{11}$ in the linear limit.
Our numerical data indicates that the linear admixture causing the bifurcation is then given by $\pm \ii \gamma_{03}$ or $\pm \ii \gamma_{30}$, depending on which of the soliton stripes is preserved and which is replaced by vortices.
In this rotated frame, evaluating the Galerkin equations is straightforward and again leads to $\mu_\text{cr} = \frac{111}{25}\omega_r \approx 0.89$, the same result as for the crossed 2x3 vortex branch.
Thus, the near-linear picture confirms that both branches should emerge at the same critical chemical potential, in agreement with our numerical findings.
As the soliton cross is unstable from the linear limit on, the bifurcating crossed 2x3 vortex branch and the one soliton/two vortex branch inherit this instability, and consequently their BdG spectra exhibit imaginary modes, a feature
that has been numerically checked.
It is worth noting that as $\mu$ is increased further, the 2x3 vortex states bifurcating from the parallel soliton stripes and the soliton cross, respectively, become more and more similar, and finally identical.
At a critical value of $\mu \approx 1.35$ a saddle-node bifurcation occurs and the two stationary states annihilate, as has also been observed in \cite{Middelkamp2010b}.
\section{Non-aligned vortex quadrupoles in the presence of anisotropy}
We now leave the isotropic limit and turn to cases where $\alpha \neq 1$.
Our primary (although not sole) focus will be on the vortex quadrupole state in its different orientations, as this state has received the most theoretical attention \cite{PhysRevA.71.033626,PhysRevA.74.023603} and with its comparably small number of vortices also seems experimentally more accessible than other, more complex non-aligned clusters.
In the first part of this section, we will therefore focus on the branches of states which are relevant for the emergence and stability of the quadrupole in its different orientations.

The first observation we make is that as soon as the rotational invariance is broken, the vortex quadrupole can no longer exist in arbitrary orientations.
In the presence of anisotropy, the four vortices can either be located at the trap's main axes, forming a rhombus centered at $x=y=0$
(this is the anisotropic generalization of orientation A), or
alternatively, the four vortices can form a rectangle centered at $x=y=0$ whose edges are parallel to the trap's main axes (this is the orientation B).

Let us now try to get insight into the bifurcation diagram including the two quadrupole branches.
First, it is useful to note that, due to their symmetry, studying the quadrupoles in the $\alpha > 1$ regime is in principle sufficient: 
the way we scan $\alpha$ (by keeping $\omega_x$ fixed and varying $\omega_y$), going from $\alpha$ to $1/\alpha$ for the quadrupoles merely corresponds to a rotation of the coordinate frame by $\pi/2$, followed by complex conjugation of the order parameter (flipping the signs of the vortex charges) and a rescaling of the overall trapping frequency (which, in turn, sets the scale for the chemical potential $\mu$).
Thus, all properties of the quadrupole solutions in the $\alpha < 1$ regime can be inferred from the results of the $\alpha > 1$ regime by accurately rescaling $\mu$ and $N$.
The BdG spectra of the quadrupoles shown below illustrate this symmetry property.

Having this in mind, we can restrict the discussion of the relevant bifurcations to $\alpha > 1$.
In the opposite regime of anisotropy, the topology of the bifurcation diagram (with all states rotated by $\pi/2$) is identical, even if the critical values where bifurcations occur are rescaled due to the different overall trapping.

In contrast to the isotropic case, for $\alpha \neq 1$ the harmonic oscillator states $\gamma_{02}$, $\gamma_{20}$ and $\gamma_{11}$ are no longer degenerate which leads to numerous modifications of the bifurcation diagram.
Specifically, for $\alpha >1$, $\gamma_{20}$ is energetically most 
favourable and thus exists for the smallest value of $\mu$.
Continuing this linear eigenstate to finite particle numbers, we find that 
the ensuing nonlinear mode progressively increasingly resembles a dark soliton ring. 
The additional two branches which become the two parallel dark soliton stripes and the diagonal (``cross'') configuration in the isotropic limit emerge in a saddle-node bifurcation detached from the above dark soliton ring branch. The critical particle number at which this saddle-node bifurcation occurs increases very rapidly as a function of $\alpha$, and both states emerging in it tend to be highly unstable for $\alpha > 1$.

Let us therefore in the following concentrate on the $\gamma_{20}$ branch, that near the linear limit is reminiscent of two dark solitons but as $\mu$ is increased progressively acquires a ring-shaped profile.
The first vortex cluster bifurcating from this initially stable solitonic branch is the quadrupole in configuration A. In linear terms, this bifurcation can be approximately understood by an admixture of $\pm \ii \gamma_{11}$, as for any $\alpha > 1$ this mode is lower in energy than $\gamma_{02}$ and can be expected to dominate the quadrupole's bifurcation, while the quadrupole's characteristic $\gamma_{02}$ component smoothly arises only well above the critical point. 
We remind the reader that the pure $\pm \ii \gamma_{11}$ admixture previously (i.e., in the isotropic case)
led to the modified tripole with the doubly charged vortex
in the center of Fig.~\ref{fig:states_20}(g), while the anisotropy
changes this scenario and our results indicate that only one pitchfork bifurcation from the $\gamma_{20}$ soliton branch involving the $\pm \ii \gamma_{11}$ admixture remains at $\alpha > 1$, namely the one leading to the quadrupole A.
The role of the tripole with the doubly charged center in the $\alpha > 1$ regime will be addressed below.

Again, the first supercritical pitchfork bifurcation renders the soliton branch unstable, while the quadrupole A inherits its stability.
The second bifurcation, then, leads to the quadrupole B, through a dominant admixture of $\pm \ii \gamma_{02}$, and this quadrupole configuration inherits the solitonic instability. 
From this analysis, we expect the quadrupole in configuration B to be unstable for any $\alpha > 1$, while configuration A should remain stable when leaving the isotropic limit. 

\begin{figure}[ht]
\centering
\includegraphics[width=0.8\textwidth]{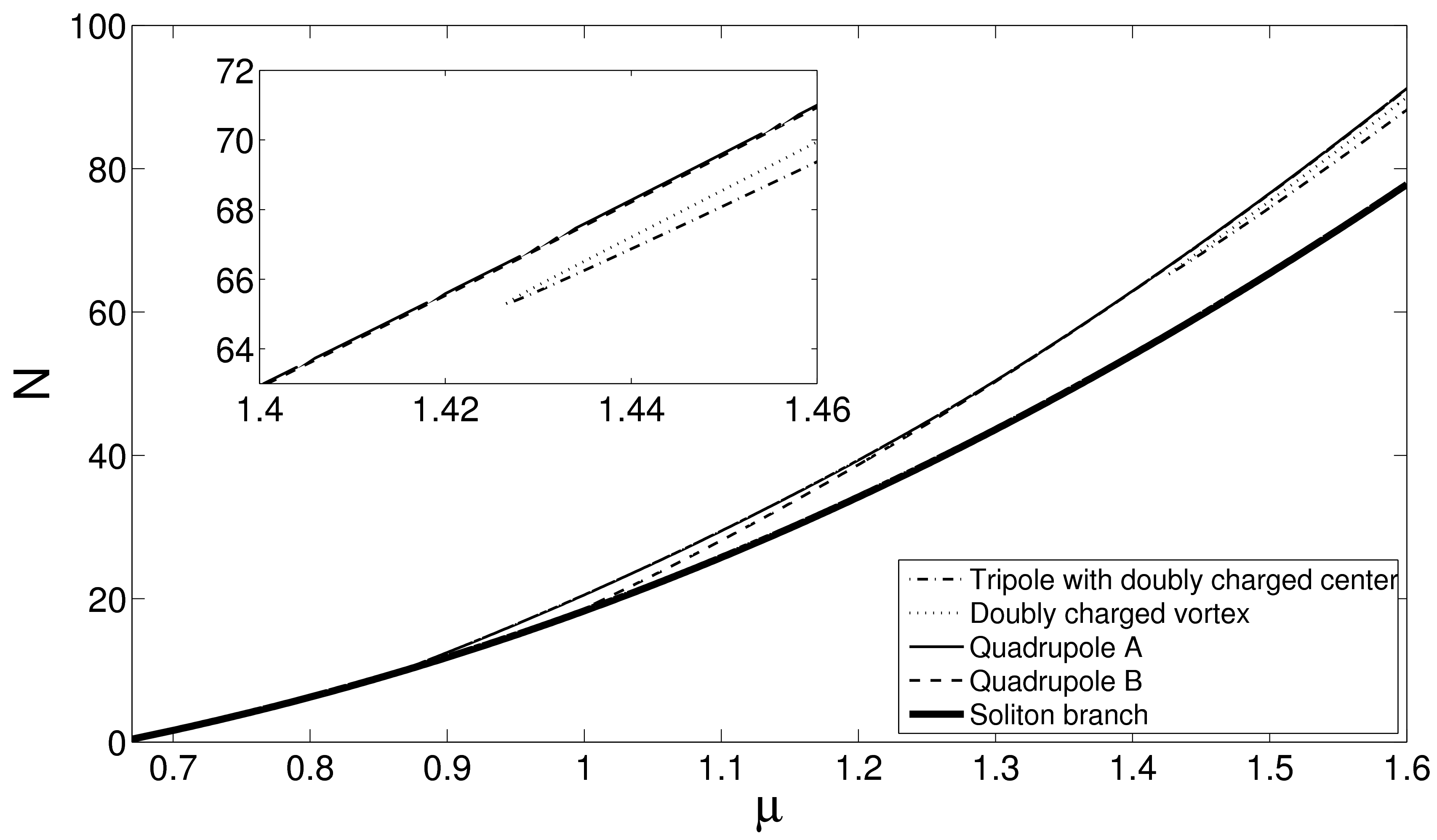}
\caption[Optional caption for list of figures]{\label{fig:bifinclsn}$N(\mu)$ bifurcation diagram at $\alpha=1.6$ including the two non-aligned quadrupoles (which for large $\mu$ can hardly be distinguished by their $N(\mu)$ dependence) and the additional two branches emerging from a nearby saddle-node bifurcation as seen in the blown up inset.}
\end{figure}

Let us at this point also address the other two vortex clusters shown in Fig. \ref{fig:states_20}, namely the doubly charged vortex and the tripole with the doubly charged center. Our findings indicate that these two branches also detach from the others as soon as $\alpha > 1$, and when lowering the particle number at some point they collide and vanish in a saddle-node bifurcation (with the doubly charged vortex playing the role of the more stable branch). 
Due to the different net topological charge of the doubly charged vortex and the tripole with the doubly charged center, it is interesting to study how these branches can become identical and annihilate. In fact, we observe that close to the bifurcation point both branches exhibit four vortex-type phase singularities, two of each sign, reminiscent of the profile shown in Fig.~\ref{fig:20i11}. In the tripole branch as $\mu$ is increased two of these merge to form the doubly charged center while the other two remain located at the $y$-axis.
In the doubly charged vortex branch, on the other hand, the central singularities merge to form the $s=2$ vortex, while the other two phase singularities are pushed out to regions of zero density and vanish there as $\mu$ is increased away from the bifurcation point.\\
In the immediate vicinity of the isotropic limit this bifurcation scenario is hard to clearly confirm by numerical simulations (as the expected saddle-node bifurcation still happens very close to the quadrupole branch), but for $\alpha=1.6$ the detaching of the two branches is clearly observable, see Fig. \ref{fig:bifinclsn}.

A major change in the bifurcation diagram occurs when $\alpha$ reaches the value $2$.
As discussed in the section on aligned clusters, above this critical value of the anisotropy the vortex dipole along the $x$-axis no longer bifurcates from the single soliton branch starting at $\gamma_{01}$, but instead from the $\gamma_{20}$ two soliton branch considered here.
This dipole branch is now the first to bifurcate from this solitonic branch, rendering it unstable and inheriting its stability.
From this observation one would expect the quadrupole A branch to be unstable for $\alpha > 2$, as in this regime it bifurcates from the destabilized 
solitonic branch.
By the same reasoning, the quadrupole B should have two imaginary 
eigenfrequencies in the $\alpha > 2$ regime (i.e., one more than in 
the $1<\alpha<2$ regime), as it is 
the third state to bifurcate from the $\gamma_{20}$ solitonic branch.
We will discuss these predictions in more detail below.

\begin{figure}[ht]
\centering
\includegraphics[width=0.6\textwidth]{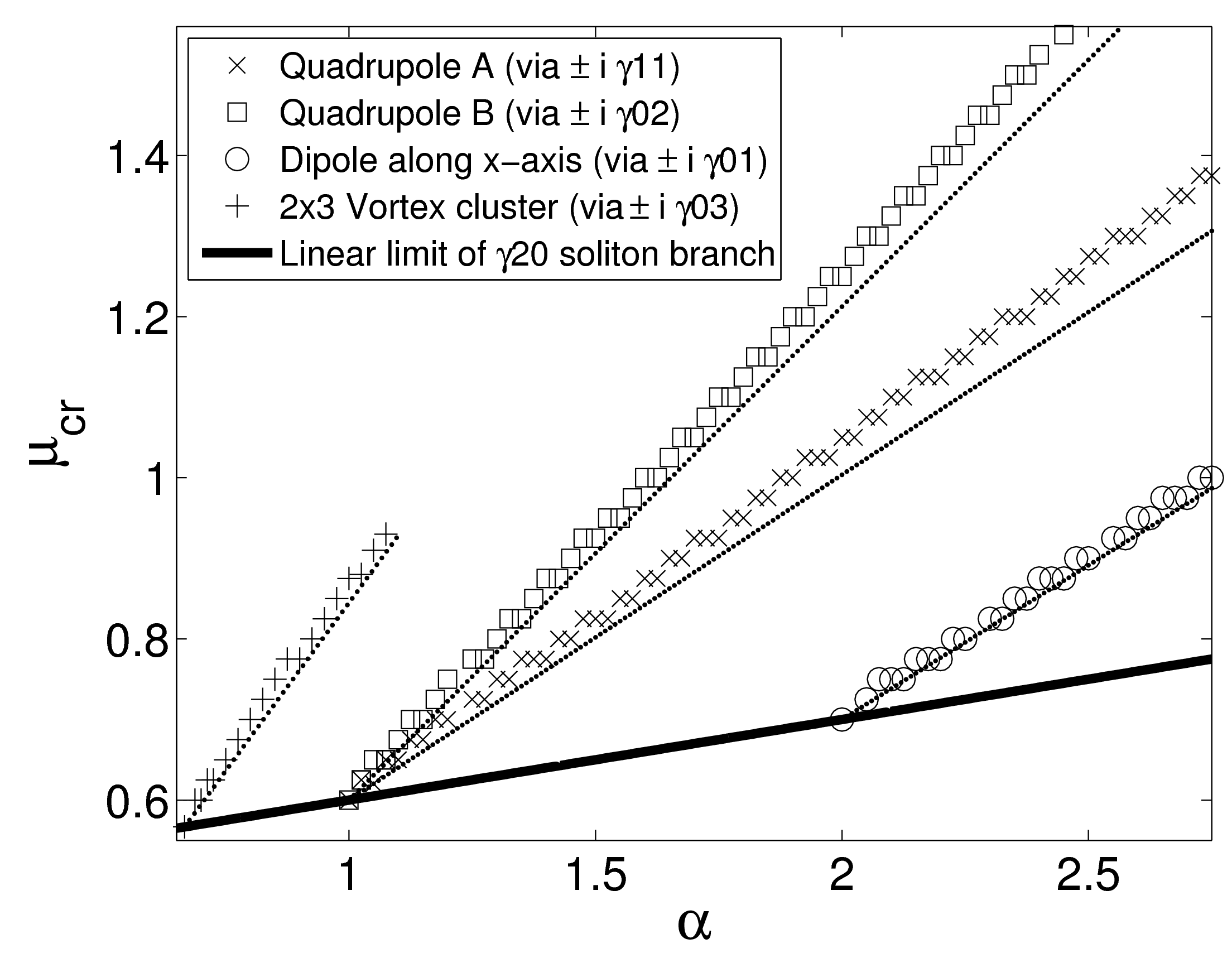}
\caption[Optional caption for list of figures]{\label{fig:bif_2sol_alpha}Critical values of the chemical potential $\mu$ at which bifurcations from the $\gamma_{20}$ soliton state occur, as a function of the anisotropy parameter $\alpha$: predictions of the Galerkin approach (thin dotted lines) and numerically found values (symbols as shown in legend). The fundamental change of stability properties of the quadrupole
states due to the (prior) bifurcation of the vortex dipole along the x-axis for
$\alpha>2$ is evident.}
\end{figure}

Numerically continuing the relevant branches of states in the parameters $\mu$ and $\alpha$, we can track the dependence on the anisotropy parameter $\alpha$ of the critical chemical potentials at which bifurcations from the $\gamma_{20}$ branch occur.
Fig. \ref{fig:bif_2sol_alpha} shows the numerically found critical values, together with the theoretical predictions from the Galerkin approach.
In addition to the vortex branches discussed above, it also depicts the bifurcation points of the 2x3 double tripole branch, which for $\alpha <1$ is the first to bifurcate from the $\gamma_{20}$ solitonic branch.

The numerical data shown in Fig. \ref{fig:bif_2sol_alpha} is obtained by stepwise continuations of the vortex quadrupole states, the 2x3 vortex state and the vortex dipole, respectively, from high values of $\mu$ to lower ones, until eventually the parental solitonic branch is reached. These scans over the parameter $\mu$ were repeated at different values of the anisotropy parameter $\alpha$.
Next, we employed two different methods for numerically identifying the critical chemical potentials.
On the one hand, we compared the particle number of the branch under consideration with the particle number of the $\gamma_{20}$ soliton branch and identified the bifurcation point as the parameter value of $\mu$ at which the difference in $N$ becomes nonzero.
On the other hand, we also made use of an indirect method, tracking down the emergence of the expected linear admixtures in the linear decomposition of the investigated branch.
Both approaches lead to consistent results.
Altogether, the data shown in Fig. \ref{fig:bif_2sol_alpha} supports our interpretation of the linear combinations underlying the bifurcating vortex states.
In general, the numerically found bifurcation points agree well with the theoretical predictions, especially when the bifurcations happen close to the linear limit.
For increasing critical chemical potential (and thus increasing critical particle number) the Galerkin approach is no longer exact and the deviations get larger.

As an interesting aside, let us comment on the two different 2x3-type vortex clusters that we discussed in the isotropic case, the one bifurcating from the $\gamma_{20}$ branch included in Fig. \ref{fig:bif_2sol_alpha}, and the one shown in Fig. \ref{fig:2x3cross} which bifurcates from the cross-like solitonic branch. As stated above, in the isotropic case an increase in $\mu$ eventually leads to a saddle-node bifurcation in which these two states collide and vanish. Going to $\alpha > 1$, we find that the critical values of $\mu$ where these two vortex clusters bifurcate from their respective solitonic parents increases. At the same time, the value of $\mu$ where they collide and annihilate in the saddle-node bifurcation decreases. Thus, the range of chemical potentials for which these two vortex branches exist becomes smaller as $\alpha$ is increased away from $1$, and eventually for $\alpha \gtrsim 1.1$ they do not exist for any value of $\mu$.

Finally, we discuss in a bit more detail whether the Galerkin approach can be expected to be applicable to the bifurcations encountered here.
To this end, remember that in deriving the expressions for $N_{\text{cr}}$ and $\mu_{\text{cr}}$, it is assumed that the integrals $A_{0001} = \int \varphi_0^3 \varphi_1 \text{d}x\text{d}y$ and $A_{0111} =\int \varphi_0 \varphi_1^3 \text{d}x\text{d}y$ vanish.
As the harmonic oscillator states are parity eigenstates and one typically integrates over a product of an odd and an even function, this is valid in most of the cases we discuss here.
For example, for the $\gamma_{10} \pm \ii \gamma_{0n}$ combinations discussed in the context of aligned vortex states, the integrals are zero due to the different parity of $\gamma_1(x)$ and $\gamma_0(x)$.
Now we encounter the first case where this assumption is not true, namely the quadrupole B composed of $\gamma_{20} \pm \ii \gamma_{02}$.
Explicitly, with $\varphi_0 = \gamma_{20}$, $\varphi_1 = \gamma_{02}$, the overlap integrals are calculated to be $A_{0000}=A_{1111}=\frac{41\sqrt{\alpha}}{128\pi}\omega_x$, $A_{0011}=\frac{9\sqrt{\alpha}}{128\pi}\omega_x$, $A_{0001}=A_{0111}=-\frac{\sqrt{\alpha}}{128\pi}\omega_x$. Note that the ``mixed'' 
integrals $A_{0001}$ and $A_{0111}$ do not vanish, yet are considerably smaller than the decisive integrals $A_{0000}$ and $A_{0011}$, which one may take as a hint that the results from the Galerkin equations can still be of use. In fact, as Fig. \ref{fig:bif_2sol_alpha} shows, this is justified, as the predictions from the Galerkin approach for the quadrupole B are in good agreement with the numerically found bifurcation points.

\subsection{Non-aligned quadrupoles in the particle picture}
Let us now employ the particle picture ODEs to obtain the equilibrium positions and linearization frequencies of the non-aligned quadrupoles.
Making an ansatz with rhombic symmetry, we find the vortices' equilibrium positions in orientation A to be given by 
\begin{align*}
  x_{1,3}=y_{2,4}&=0, \\
 x_2=-x_4&=\sqrt{\frac{B}{\omega_x^2 Q}} \cdot \frac{1}{\sqrt{3 \alpha^2-1 + \sqrt{9\alpha^4-14\alpha^2+9}}},\\ 
y_1=-y_3&=\sqrt{\frac{B}{\omega_x^2 Q}} \cdot \frac{1}{\sqrt{3 - \alpha^2 + \sqrt{9\alpha^4-14\alpha^2+9}}},
 \end{align*}
where the charges were taken to be $s_1 = s_3 = -s_2 = -s_4$.
We were not able to give analytical solutions for the linearization frequencies as a function of $\alpha$ for this state (although they can be
straightforwardly computed numerically).

On the other hand, the rectangular configuration of orientation B is captured by a fixed point of the particle picture system with
\begin{align*}
 y_{1,2}=-y_{3,4}=\sqrt{\frac{B}{4 \omega_x^2 Q}} \cdot \frac{1}{\sqrt{\alpha+\alpha^2}}, \quad
 x_{2,3}=-x_{1,4}=\sqrt{\frac{B}{4 \omega_x^2 Q}} \cdot \frac{1}{\sqrt{1+\alpha}}.
\end{align*}
Here, we can give analytical expressions for the linearization frequencies around this equilibrium position, namely
$\omega_{1,2} = \pm 2\sqrt{2}  \omega_{\text{pr}}$, 
$\omega_{3,4} = \pm \omega_{\text{pr}} \sqrt{2 - \alpha^{-2}}$, 
$\omega_{5,6} = \pm \omega_{\text{pr}} \sqrt{2 - \alpha^2}$, 
$\omega_{7,8} = \pm \ii \omega_{\text{pr}} \sqrt{(1 - 1/\alpha)^2 (1+4\alpha + \alpha^2)}$.
While the first mode $\omega_{1,2}$ stays real over the whole range of $\alpha$, $\omega_{7,8}$ (forming the zero mode due to rotational invariance in isotropic settings) becomes imaginary as soon as $\alpha \neq 1$.
The remaining modes $\omega_{3,4}$ and $\omega_{5,6}$ become purely imaginary for $\alpha<1/\sqrt{2}$ and $\alpha>\sqrt{2}$, respectively.

\begin{figure}
\centering
\subfigure[Quadrupole A]{\label{fig:eq_vqa}
 \includegraphics[height=0.3\textheight]{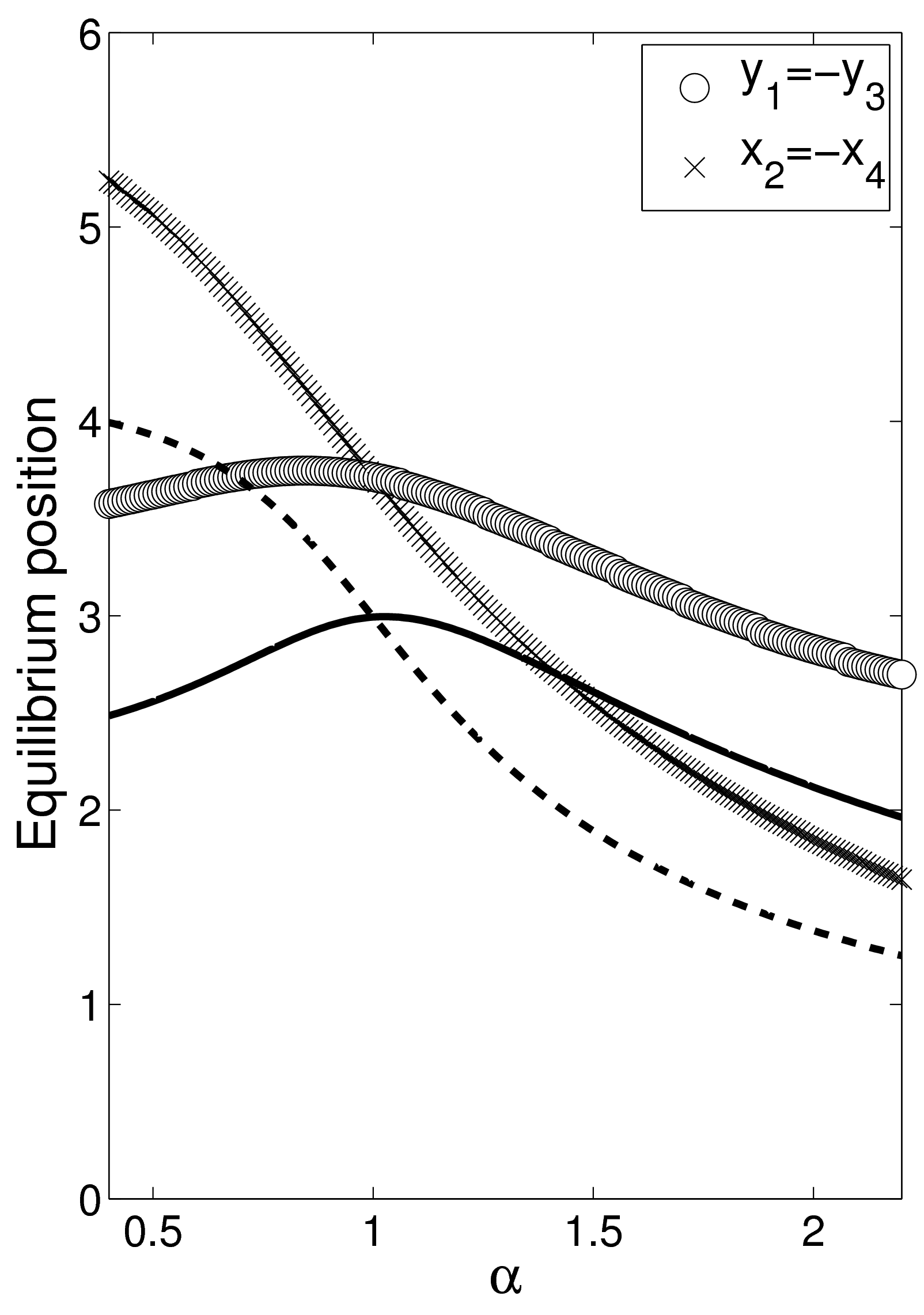}}
\subfigure[Quadrupole B]{\label{fig:eq_vqb}
 \includegraphics[height=0.3\textheight]{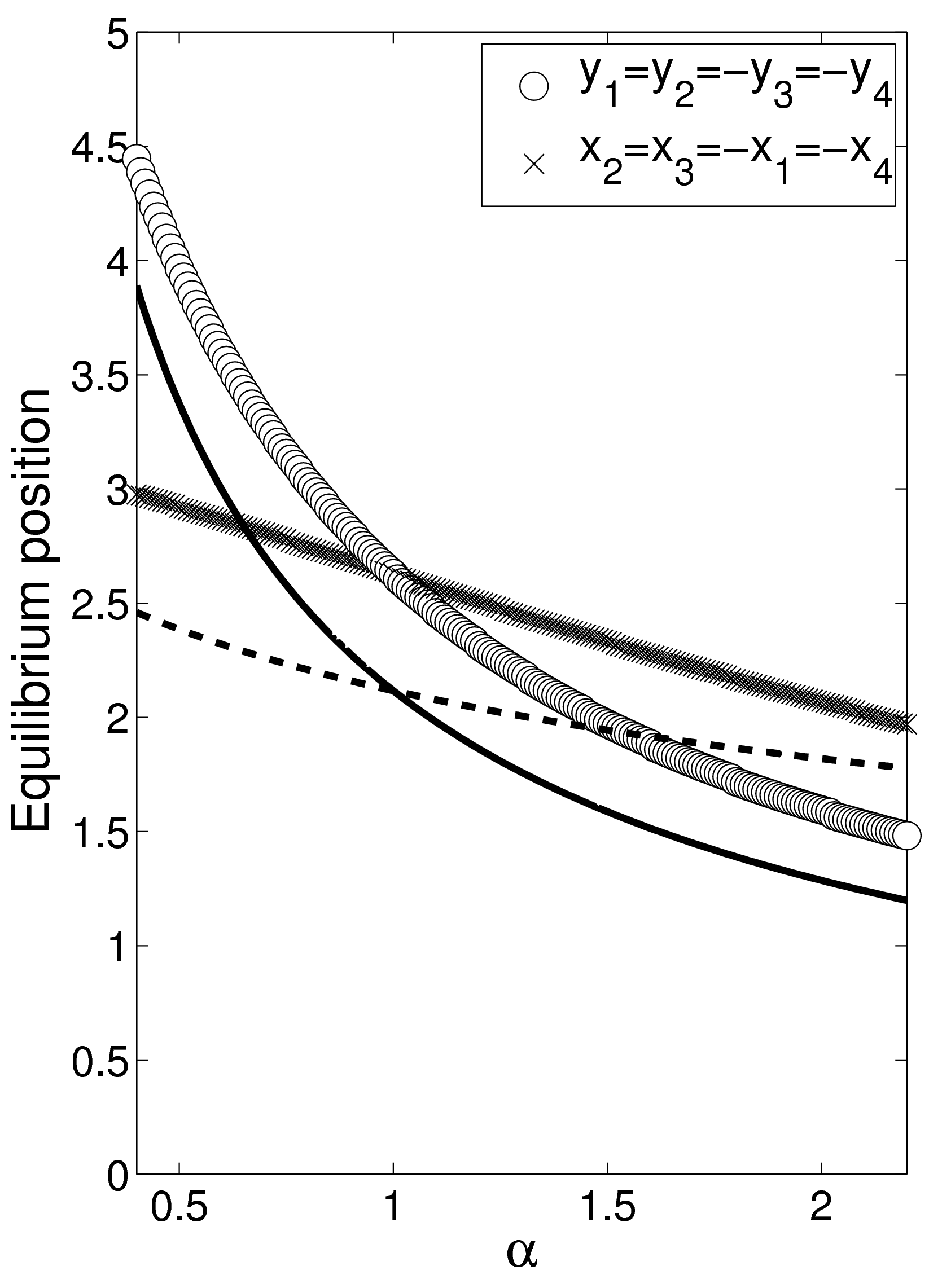}}
\caption{Equilibrium position of vortices in the non-aligned quadrupole configurations as a function of the 2D aspect ratio: numerical data (crosses/open circles) and ODE predictions (solid lines for the $y$-coordinates, dashed lines for $x$). The chemical potential is chosen as $\mu = 2.5$.}
\end{figure}

We now compare these predictions to the numerical results, obtained by solving the full GPE.
Figs. \ref{fig:eq_vqa} and \ref{fig:eq_vqb} show the numerically found equilibria positions of the vortices in the quadrupole states for different values of the anisotropy parameter $\alpha$, together with the fixed points of the ODE system calculated above. 
Generally, both quadrupole configurations are qualitatively, and to a large extent also quantitatively, accurately described by the particle picture.
As an aside, we remark that the quantitative agreement becomes better if the non-modified value of $B=1.95$, valid for vortex interaction on a homogeneous condensate background, is used for the particle picture interaction constant (data not shown), again indicating that the vortex-vortex interaction on the 
inhomogeneous background is not fully accurately modeled by our ODEs.

\begin{figure}[ht]
\centering
\includegraphics[width=0.55\textwidth]{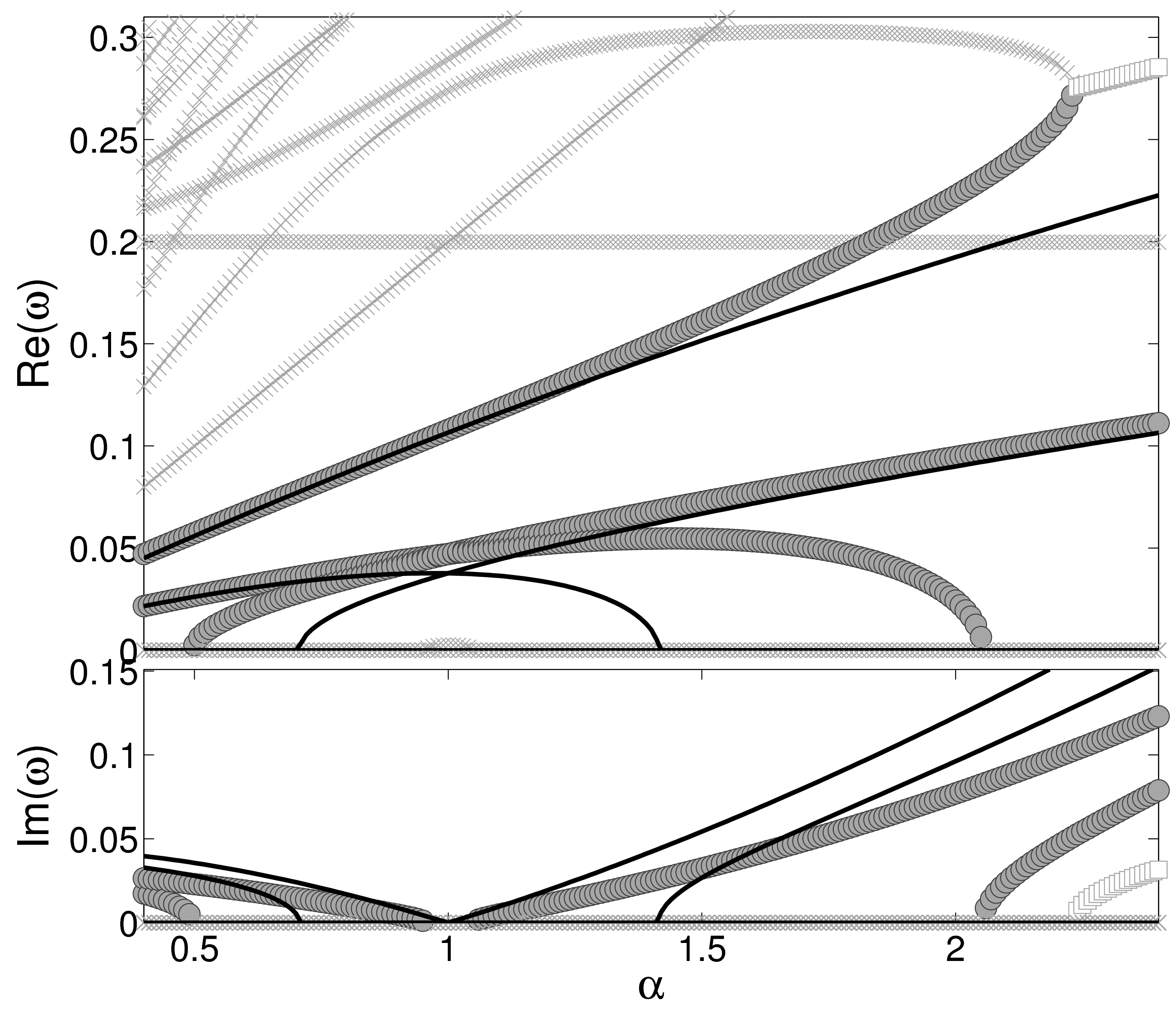}
\caption[Optional caption for list of figures]{\label{fig:bdg_vqb}BdG spectra of the vortex quadrupole B as a function of $\alpha$. 
The predictions for the anomalous and purely imaginary modes given by the particle picture are shown in black.}
\end{figure}

Let us now turn to the BdG spectra. Fig. \ref{fig:bdg_vqb} shows the numerically obtained spectrum of the quadrupole in orientation B as a function of $\alpha$.
The chemical potential $\mu$ is fixed to $\mu = 2.5$.
The particle picture again captures the main features: 
For any $\alpha \neq 1$, the existence of a purely imaginary mode is correctly predicted. 
Furthermore, there are additional imaginary BdG modes arising in regimes of larger anisotropy. While the particle picture predicts these to occur at $\alpha>\sqrt{2}$ and $\alpha<1/\sqrt{2}$, respectively, numerically they are found even further away from the linear limit, at $\alpha \gtrsim 2$ and $\alpha \lesssim 1/2$.
Finally, there is one purely real anomalous mode eigenfrequency 
whose general functional dependence on $\alpha$ is correctly 
captured by the ODE description.
In the spectrum shown here, this anomalous mode collides with a background mode at $\alpha \approx 2.2$, resulting in the emergence of a complex quartet (represented by open square symbols).
This process is not contained in the particle picture predictions, and it cannot be expected to be:
in fact, this oscillatory instability in the quadrupole's spectrum at large $\alpha$ is only present for comparably small values of the chemical potential, and the fact that it shows up at large $\alpha$ here merely indicates that for the particle picture to hold in the $\alpha \gtrsim 2$ regime, one has to study the spectrum at a higher value of $\mu$ (remember that the particle picture ODEs are derived under the assumption of the large $\mu$ Thomas-Fermi limit, where
the highly localized vortex structures can be identified as individual
particles).

We decided to still show the spectrum containing the oscillatory mode here because it illustrates our introductory remarks on the quadrupole's symmetry with respect to $\alpha \rightarrow 1/\alpha$.
In general, this transformation is an exact symmetry if it is accompanied by an appropriate rescaling of the other variables, in particular of the particle number and the chemical potential.
In the special case encountered here, the spectrum's symmetry around $\alpha=1$ (up to an overall scaling of the modes' numerical values) is clearly visible.
However, as we keep $\mu$ fixed while scanning, the underlying symmetry is not obvious in our numerical data: 
Increasing $\alpha$ effectively increases the trapping frequency, and consequently the effective chemical potential (measured in units of the trapping frequency) becomes smaller. Eventually, the oscillatory instabilities appear, which are not captured by the particle picture as they vanish when further approaching the appropriate Thomas-Fermi limit.

Before turning to the other quadrupole configuration, let us recall the bifurcation picture discussed above. 
In the $\alpha > 1$ regime we have seen there that for $1 < \alpha < 2$ the quadrupole B is the second branch to bifurcate from the $\gamma_{20}$ soliton branch, thus inheriting one unstable mode. For $\alpha \geq 2$ there is also the 
vortex dipole along the $x$-axis which bifurcates from the  $\gamma_{20}$ soliton branch, thus inducing a second instability which is passed on to the quadrupole B. By symmetry of the quadrupole, the same effects must occur in the $\alpha < 1$ regime: For $1/2 < \alpha < 1$, the quadrupole A branch is the first to bifurcate from the $\gamma_{02}$ soliton branch, before the quadrupole B, and for $\alpha \leq 1/2$ additionally the vortex dipole along the $y$-axis becomes relevant.
Thus, the discussion of the bifurcation diagram (and its approximate analytical description in terms of the Galerkin approach) correctly describes the imaginary modes arising in the spectrum of the quadrupole B in different regimes of anisotropy.

\begin{figure}[ht]
\centering
\subfigure[$\mu=2.5$]{
\includegraphics[width=0.48\textwidth]{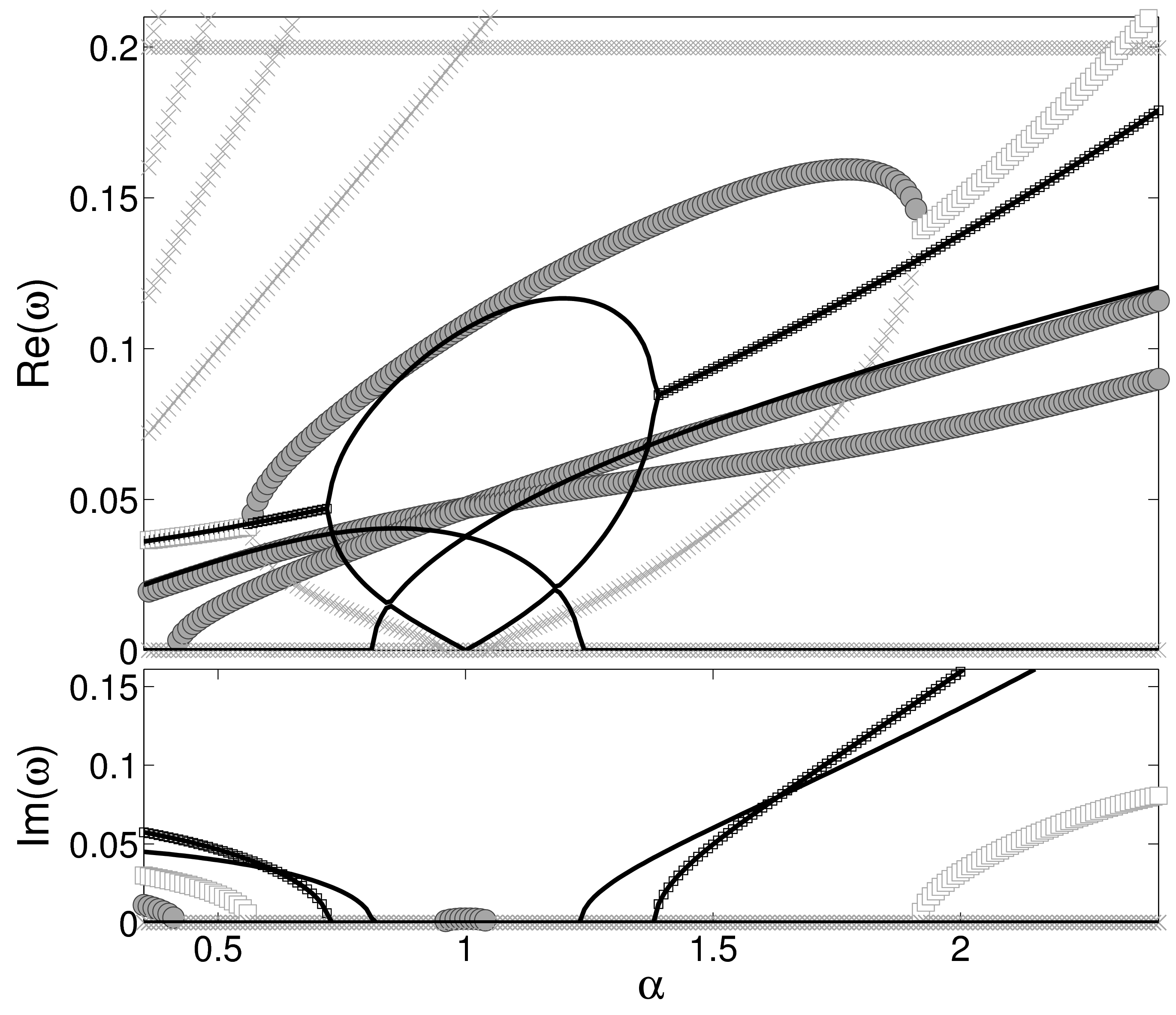}
\label{fig:bdg_vqa_mu2-5}
}
\subfigure[$\mu=6.5$]{
\includegraphics[width=0.48\textwidth]{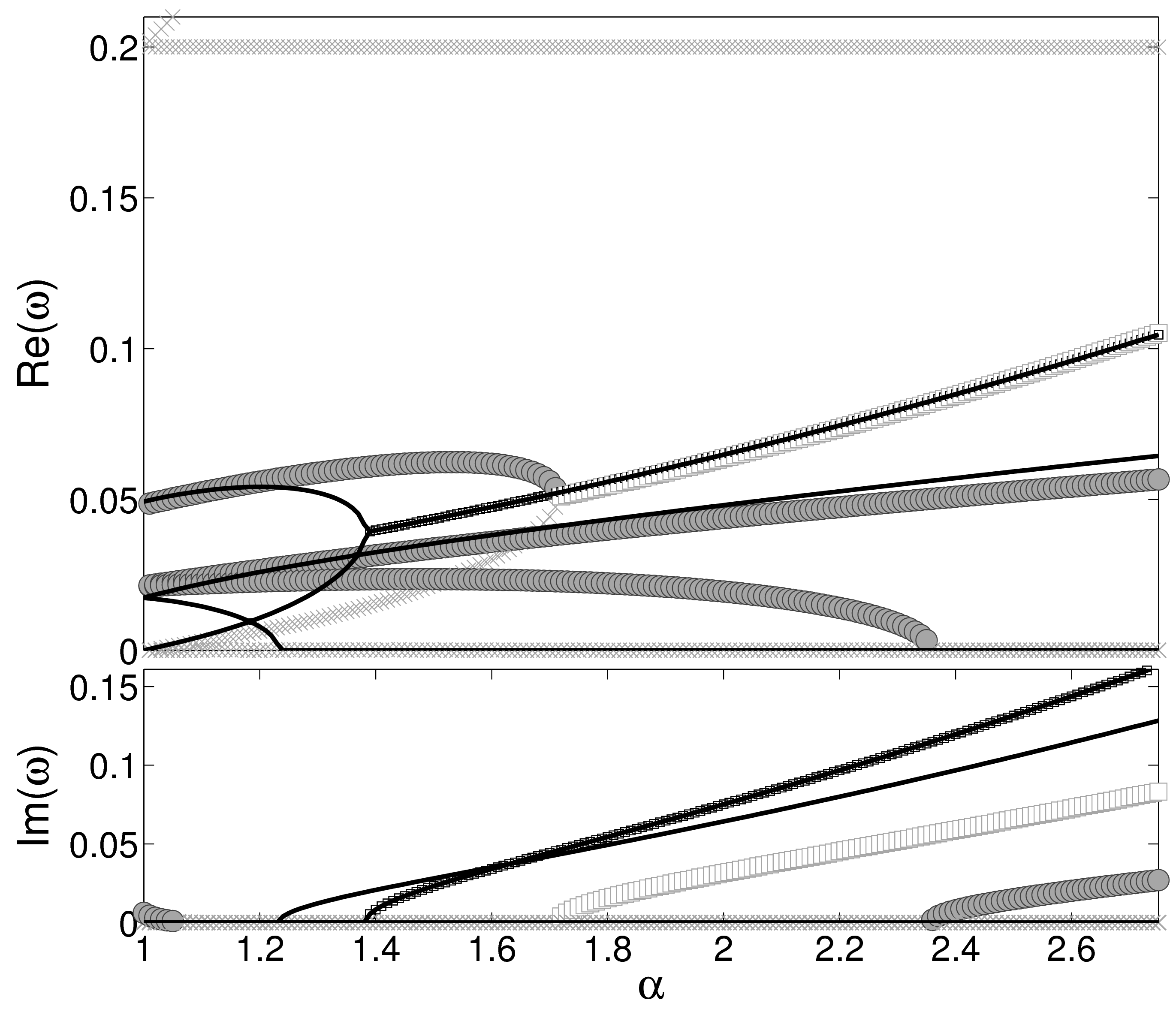}
\label{fig:bdg_vqa_mu6-5}}
\caption[Optional caption for list of figures]{\label{fig:bdg_vqa}BdG spectra of the vortex quadrupole A as a function of $\alpha$, predictions from the particle picture in black.}
\end{figure}

Next, let us discuss the spectrum of the quadrupole A branch of 
Fig.~\ref{fig:bdg_vqa}.
Interestingly, for this configuration (where we had to find the roots of the characteristic polynomial numerically) the particle picture predicts not only the onset of purely imaginary modes, but also the existence of oscillatory instabilities in anisotropic regimes, namely for $\alpha \gtrsim 1.4$, $\alpha \lesssim 0.7$ in our case. In Fig.~\ref{fig:bdg_vqa} the modes predicted by the ODEs are shown as black solid lines, with the oscillatorily unstable modes distinguished by small additional square markers.
In the numerically found BdG spectrum such oscillatory instabilies emerge at $\alpha \approx 1.9$ and $\alpha \approx 0.58$, respectively, where a positive energy mode (which forms the Goldstone mode associated with the rotational invariance 
at $\alpha = 1$) and an anomalous mode collide as qualitatively correctly predicted by the particle picture. 
As a final feature, the ODEs predict purely imaginary modes for $\alpha \gtrsim 1.22$, $\alpha \lesssim 0.8$.  Indeed we observe the presence of such an imaginary mode in the full BdG spectrum below $\alpha \approx 0.4$, in qualitative agreement with, but quantitatively quite far away from the particle picture prediction.
In the $\alpha > 1$ part of the spectrum, a corresponding imaginary mode at $\mu = 2.5$ cannot be found, which can again be attributed to the effective rescaling of the chemical potential due to our way of scanning $\alpha$: Scanning the $\alpha > 1$ regime at $\mu=6.5$, the imaginary mode is present, and furthermore the agreement between the numerical data and the other predicted modes is better than at $\mu=2.5$, which is expected when approaching the Thomas-Fermi limit, see Fig. \ref{fig:bdg_vqa_mu6-5}.
Finally, let us remark that the very small purely imaginary mode that seems to be present in the above spectra near $\alpha = 1$ is only due to the finite resolution of our numerical grid, we have observed that it diminishes the smaller the grid spacing is chosen.

Generally, we can conclude from the above that the particle picture still gives qualitatively and, to some extent, also quantitatively accurate results when applied to the non-aligned quadrupole configurations.
In comparison to the aligned vortex states, the predictions are less exact, which can be attributed to a number of possible sources of error.
On the one hand, in the quadrupole A and B all four vortices are located off-center which leads to an effective change in their precession frequency that we do not take into account.
Furthermore, in these regions of relatively low and inhomogeneous densities, the assumption of an undisturbed velocity field around the individual vortices that implicitly underlies our modeling of the vortex-vortex interaction is no longer justified.
Both effects must be expected to lead to a modification of the effective vortex dynamics, which our simple ODE system does not appropriately correct for.
Amending these aspects of the model would be a natural direction for
future work that could improve the quantitative agreement with the
full model of the Gross-Pitaevskii equation.

\begin{figure}[ht]
  \centering
  \subfigure[\label{fig:bdgvqamu}Vortex quadrupole A, $\alpha=2.5$]{
  \includegraphics[width=0.4\textwidth]{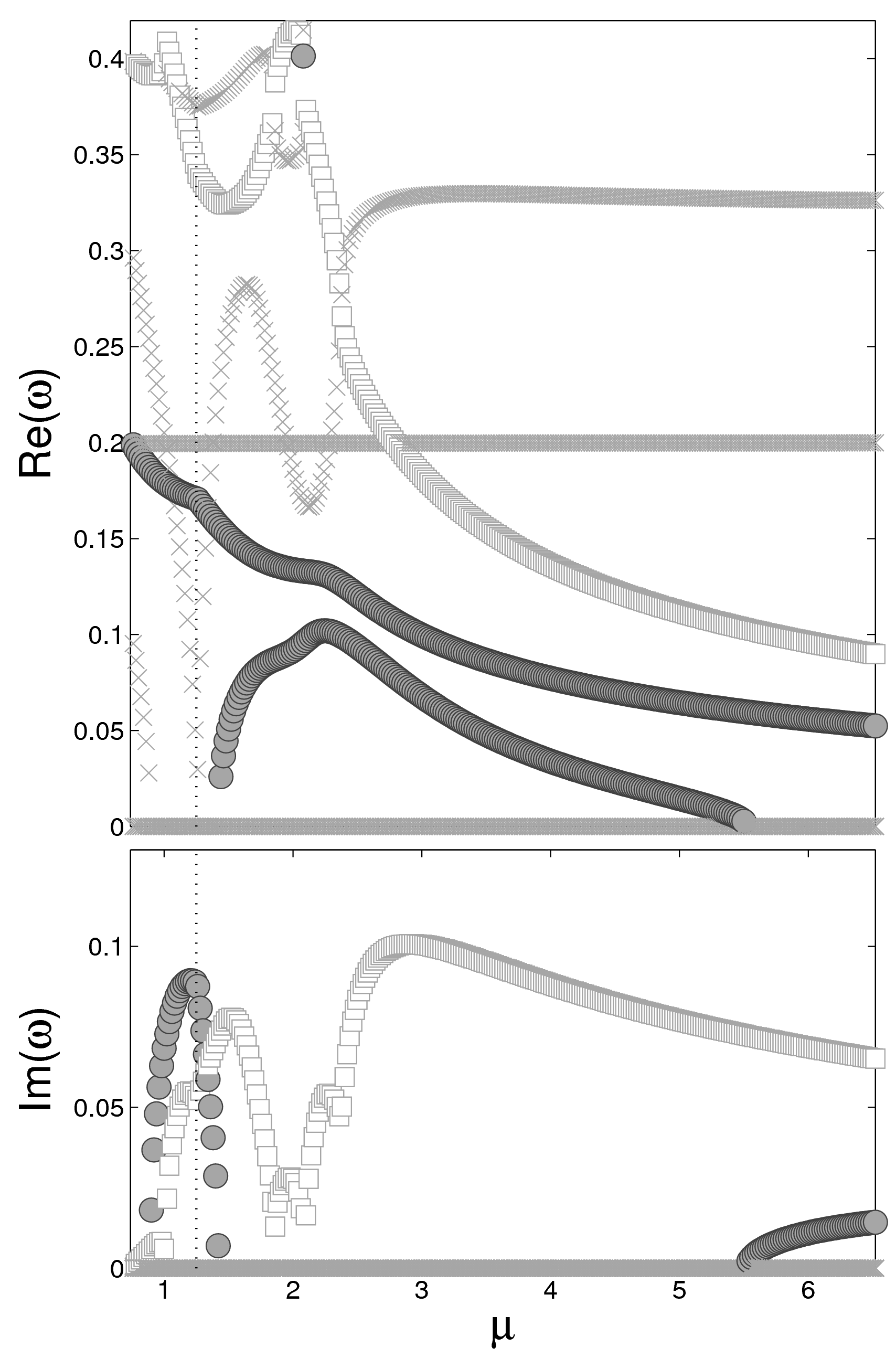}
  }
  \subfigure[\label{fig:bdgvqbmu}Vortex quadrupole B, $\alpha=1.5$]{
  \includegraphics[width=0.4\textwidth]{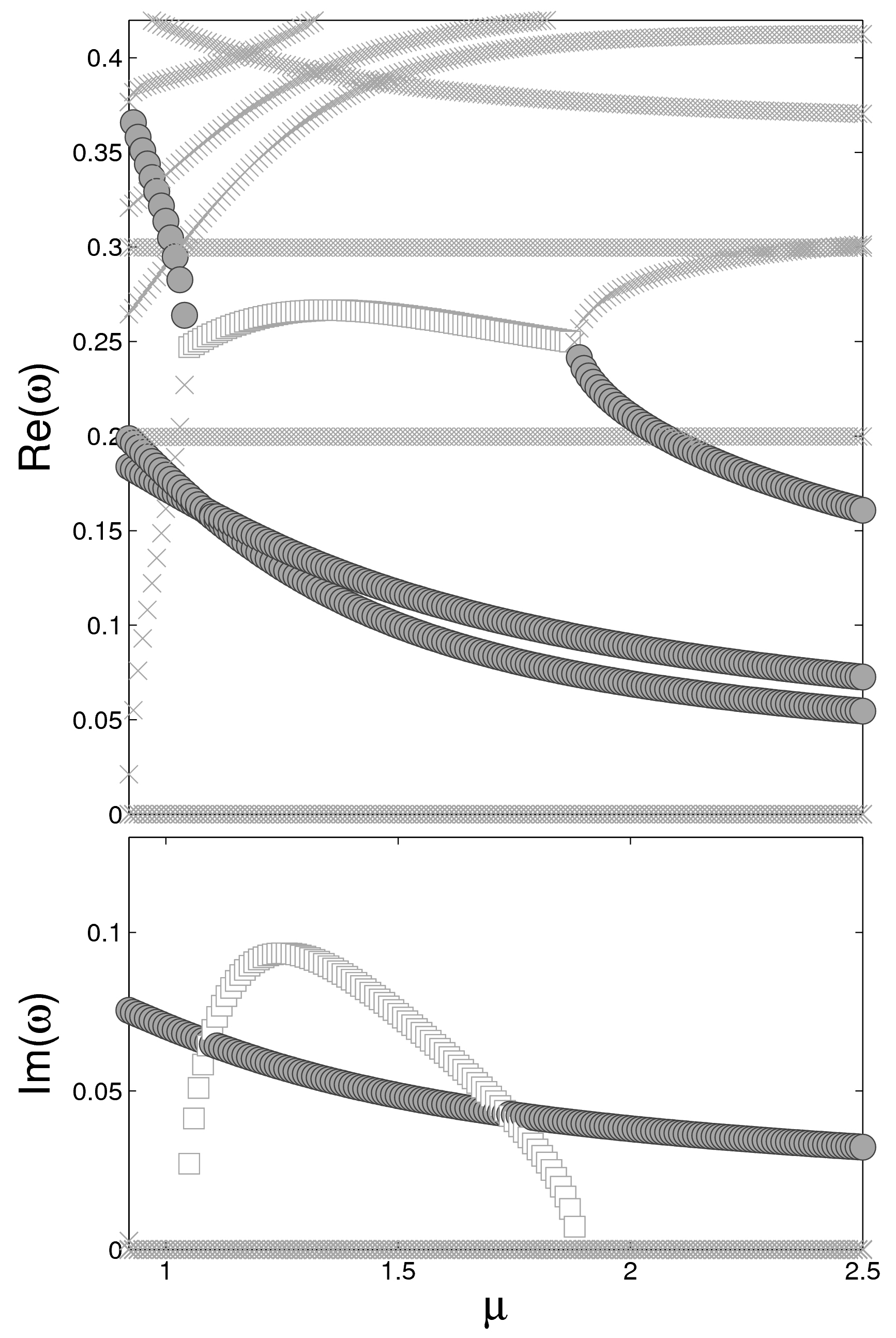}
  }
  \caption[Optional caption for list of figures]{\label{fig:bdgvqmu}BdG spectra of the vortex quadrupoles as a function of $\mu$ at fixed $\alpha$. The quadrupole A branch is continued down to its solitonic origin at small $\mu$: The dipole's bifurcation from the branch at $\mu_\text{cr} \approx 0.9$ induces an imaginary mode in the spectrum. At $\mu_\text{cr} \approx 1.25$ (indicated by the dotted vertical lines) the quadrupole A itself arises, inheriting this unstable mode which however becomes real again very quickly as $\mu$ is further increased. Only for large chemical potentials it turns imaginary again as is correctly predicted by the particle picture ODEs. In contrast, for quadrupole B (where the solitonic lower part of the branch is not shown) the imaginary mode induced by the first bifurcation from the two soliton branch is present for all values of $\mu$ that we scanned.}
  \end{figure}

We have already made a remark concerning the imaginary mode in the $\alpha > 1$ regime of quadrupole A which indicates that our previous strategy of explaining the emergence of instabilities at different values of $\alpha$ from the bifurcation diagram will not be straightforward to apply here.
The bifurcation analysis allows statements about the number of unstable modes in a state's spectrum right after its bifurcation from the parental branch.
No predictions can be made on what happens to these modes as $\mu$ is increased, away from the critical value.
In the case of quadrupole A, the fact that at $\mu=6.5$ an imaginary mode is present at $\alpha = 2.5$ which cannot be observed at $\mu = 2.5$ implies that there is a substantial dependence of the number of imaginary modes on the chemical potential in this branch.
Indeed, we found direct numerical proof for this shifting of the imaginary modes as a function of $\mu$ at fixed anisotropy:
For the quadrupole A at $\alpha=2.5$, Fig. \ref{fig:bdgvqamu}, we expect one imaginary mode at low $\mu$, caused by the dipole's bifurcation from the $\gamma_{20}$ branch and then passed on to the quadrupole. Indeed, such a mode is present, but as $\mu$ is increased, it quickly crosses to the real axis again. Only at large $\mu$ an imaginary mode reappears, which is the one that we have seen in the spectrum as a function of $\alpha$, Fig. \ref{fig:bdg_vqa_mu6-5}, and which is contained in the particle picture's predictions.
In contrast, for the quadrupole B at $\alpha=1.5$, Fig. \ref{fig:bdgvqbmu}, the imaginary mode that is present for small $\mu$ and explained by the previous bifurcations stays present over the whole range of chemical potentials.

\begin{figure}[ht]
\centering
\fbox{
\subfigure[Orientation A]{
\includegraphics[width=0.7\textwidth]{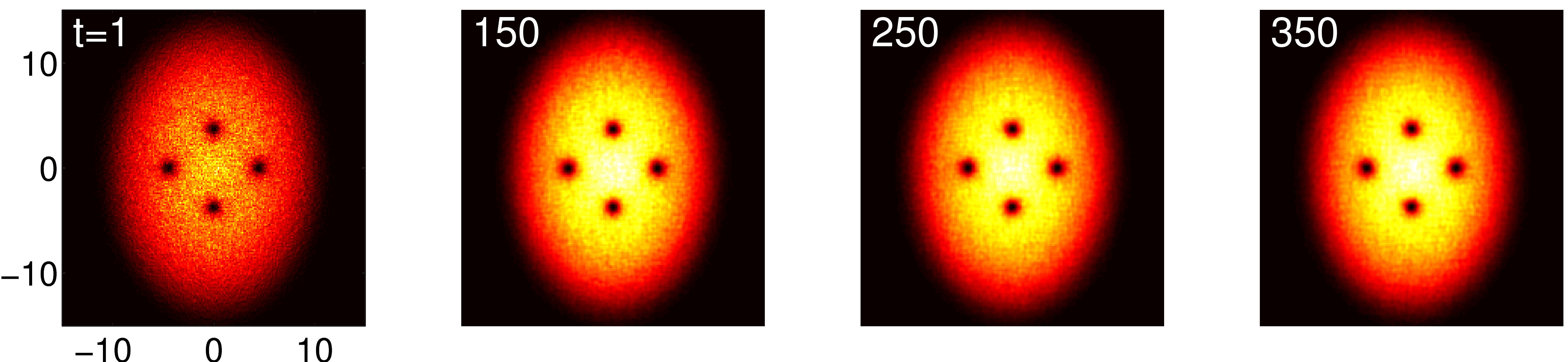}
}}
\fbox{
\subfigure[Orientation B]{
\includegraphics[width=0.7\textwidth]{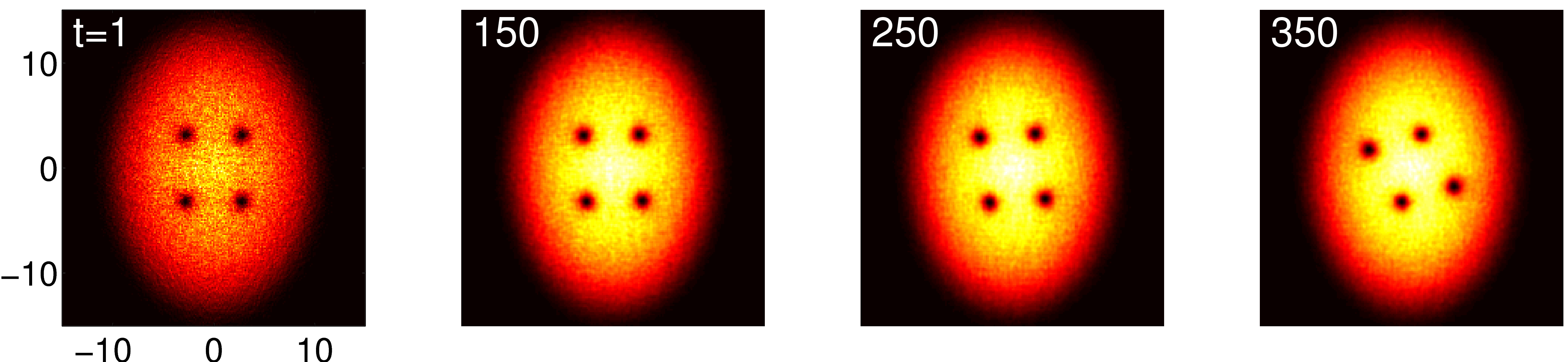}
}}
\caption[Optional caption for list of figures]{\label{fig:prop_vq_0-75}Time propagation of the vortex quadrupoles, seeded with white noise at $\alpha=0.75$, initial $\mu=2.5$. While configuration A is stable, configuration B is dynamically unstable and starts to rotate.}
\end{figure}

\begin{figure}[ht]
\centering
\fbox{\includegraphics[width=0.72\textwidth]{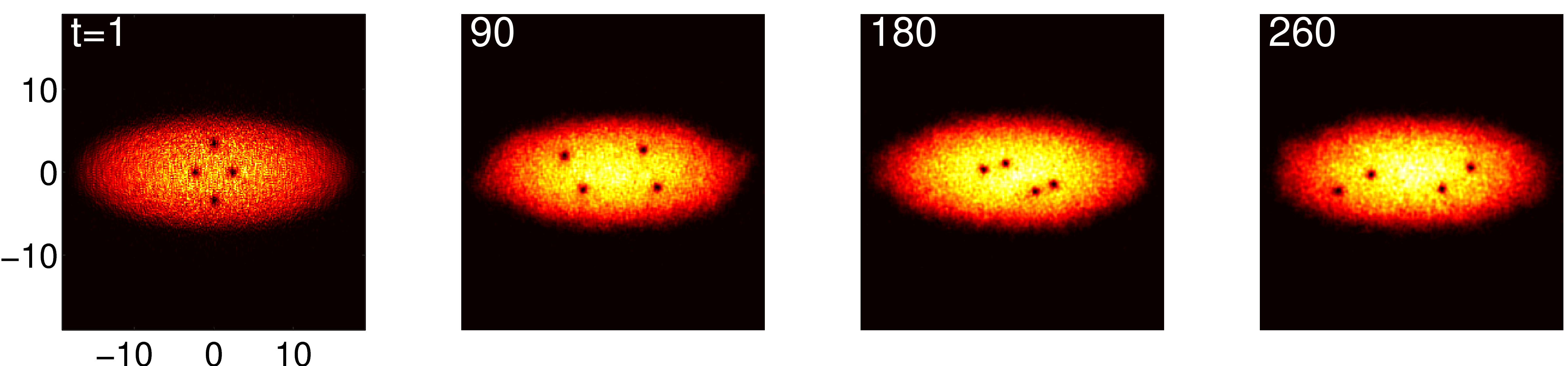}}
\caption[Optional caption for list of figures]{\label{fig:prop_vqa_2-5}Time propagation of vortex quadrupole A, seeded with white noise at $\alpha=2.5$, initial $\mu=6.5$. The strong anisotropy of the trap renders orientation A unstable as well.}
\end{figure}

Finally, we have probed the stability of vortex quadrupoles in different regimes of anisotropy by simulating their time-evolution according to the full Gross-Pitaevskii equation.
The results are as expected: For $\alpha = 0.75$, the quadrupole in orientation A is fully stable, while the quadrupole in orientation B is weakly unstable and starts to rotate if initially disturbed, see Fig. \ref{fig:prop_vq_0-75}.
For much larger values of $\alpha$, the quadrupole A is unstable as well, and adding a noise signal leads to the onset of vortex dynamics.
As can be seen in Fig. \ref{fig:prop_vqa_2-5}, when moving through the condensate the four vortices have a tendency to form dipole-like 
vortex-antivortex pairs. 

\clearpage
\section{Conclusions and Outlook}

In this chapter, we illustrated how dramatically the dynamics of 
vortices may be affected by the presence of anisotropy in Bose-Einstein
condensates. Although the single vortex remains dynamically robust
under the effect of anisotropic trapping, all other encountered
configurations are subject to fundamental changes in their stability
and nonlinear dynamics through the critical handle of distinct trapping
strengths in the different axial directions. More specifically, it
was found that compressions parallel to the axis of an aligned vortex
cluster always destabilize it by breaking its symmetry of rotational
invariance, while compressions perpendicular to the axis of
the cluster may eventually stabilize even highly unstable vortex
clusters. It was demonstrated how these aligned vortex states arise 
from nonlinear variants of linear states of the system through
symmetry breaking bifurcations, and that an analysis of these bifurcations can shed light on their observed (de)stabilization due to anisotropy. 
The phenomenology is even more
complex for the multitude of non-aligned multi-vortex solutions
identified (many of which are generically unstable). However,
the critical role of the anisotropy in improving or completely
eliminating the stability of such states was confirmed in this
case as well.

This investigation indicates a high level of experimental
control that can be achieved in the coherent multi-vortex
states of BECs. This control could be used for transport, manipulation,
dynamical localization and numerous other similar scopes both
in this more pristine context but also in other related fields, e.g. in nonlinear optics. A natural direction for extending
these investigations is the context of multi-component condensates.
There, novel states including vortex-bright solitary waves
can be seen to arise~\cite{Law2010} and understanding the
dynamics and interactions of multiple ones of these structures
is far more complex (due to the ``dual'' character -- soliton and
vortical -- of the interactions). On the other hand, another
natural extension consists of the three-dimensional generalizations
of the present states, namely of vortex rings which have 
already been experimentally 
observed~\cite{PhysRevLett.86.2926,PhysRevLett.94.040403}. Yet, it
would be of particular interest to devise a description of both their
near linear as well as especially of their highly nonlinear phenomenology
analogous to the one presented herein, both in isotropic and in
anisotropic settings. Such themes are presently under investigation
and will be reported in future publications.
\bibliographystyle{phjcp}

\bibliography{references}

\begin{thebibliography}{10}

\bibitem{Donnelly2005}
{\sc R.~J. Donnelly},
\newblock {\em Quantized Vortices in Helium II},
\newblock Cambridge University Press, Cambridge, 2005.

\bibitem{Blatter1994}
{\sc G.~Blatter}, {\sc M.~V. Feigel'man}, {\sc V.~B. Geshkenbein}, {\sc A.~I.
  Larkin}, and {\sc V.~M. Vinokur},
\newblock {\em Rev. Mod. Phys.} {\bf 66}, 1125 (1994).

\bibitem{Kivshar1998}
{\sc Y.~S. Kivshar} and {\sc B.~Luther-Davies},
\newblock {\em Phys. Rep.} {\bf 298}, 81 (1998).

\bibitem{desya}
{\sc A.~Desyatnikov}, {\sc Y.~Kivshar}, and {\sc L.~Torner},
\newblock {\em Prog. Opt.} {\bf 47}, 291 (2005).

\bibitem{9780198507192}
{\sc L.~Pitaevskii} and {\sc S.~Stringari},
\newblock {\em Bose-Einstein Condensation},
\newblock Clarendon Press, Oxford, 2003.

\bibitem{pethick}
{\sc C.~J. Pethick} and {\sc H.~Smith},
\newblock {\em Bose-Einstein Condensation in Dilute Gases},
\newblock Cambridge University Press, Cambridge, 2008.

\bibitem{book_kevrekidis}
{\sc P.~G. Kevrekidis}, {\sc D.~J. Frantzeskakis}, and {\sc
  R.~Carretero-Gonz\'alez},
\newblock {\em Emergent Nonlinear Phenomena in Bose-Einstein Condensates},
\newblock Springer, Berlin, 2008.

\bibitem{book_pismen}
{\sc L.~M. Pismen},
\newblock {\em Vortices in nonlinear fields},
\newblock Clarendon Press, Oxford, 1999.

\bibitem{PhysRevE.66.036612}
{\sc L.-C. Crasovan}, {\sc G.~Molina-Terriza}, {\sc J.~P. Torres}, {\sc
  L.~Torner}, {\sc V.~M. P\'erez-Garc\'\i{}a}, and {\sc D.~Mihalache},
\newblock {\em Phys. Rev. E} {\bf 66}, 036612 (2002).

\bibitem{PhysRevA.68.063609}
{\sc L.-C. Crasovan}, {\sc V.~Vekslerchik}, {\sc V.~M. P\'erez-Garc\'\i{}a},
  {\sc J.~P. Torres}, {\sc D.~Mihalache}, and {\sc L.~Torner},
\newblock {\em Phys. Rev. A} {\bf 68}, 063609 (2003).

\bibitem{PhysRevA.65.043612}
{\sc J.~Brand} and {\sc W.~P. Reinhardt},
\newblock {\em Phys. Rev. A} {\bf 65}, 043612 (2002).

\bibitem{PhysRevA.71.033626}
{\sc M.~M\"ott\"onen}, {\sc S.~M.~M. Virtanen}, {\sc T.~Isoshima}, and {\sc
  M.~M. Salomaa},
\newblock {\em Phys. Rev. A} {\bf 71}, 033626 (2005).

\bibitem{PhysRevA.74.023603}
{\sc V.~Pietil\"a}, {\sc M.~M\"ott\"onen}, {\sc T.~Isoshima}, {\sc J.~A.~M.
  Huhtam\"aki}, and {\sc S.~M.~M. Virtanen},
\newblock {\em Phys. Rev. A} {\bf 74}, 023603 (2006).

\bibitem{PhysRevA.83.011603}
{\sc P.~Kuopanportti}, {\sc J.~A.~M. Huhtam\"aki}, and {\sc M.~M\"ott\"onen},
\newblock {\em Phys. Rev. A} {\bf 83}, 011603 (2011).

\bibitem{PhysRevA.77.053610}
{\sc W.~Li}, {\sc M.~Haque}, and {\sc S.~Komineas},
\newblock {\em Phys. Rev. A} {\bf 77}, 053610 (2008).

\bibitem{PhysRevA.82.013646}
{\sc S.~Middelkamp}, {\sc P.~G. Kevrekidis}, {\sc D.~J. Frantzeskakis}, {\sc
  R.~Carretero-Gonz\'alez}, and {\sc P.~Schmelcher},
\newblock {\em Phys. Rev. A} {\bf 82}, 013646 (2010).

\bibitem{Torres20113044}
{\sc P.~Torres}, {\sc P.~Kevrekidis}, {\sc D.~Frantzeskakis}, {\sc
  R.~Carretero-González}, {\sc P.~Schmelcher}, and {\sc D.~Hall},
\newblock {\em Phys. Lett. A} {\bf 375}, 3044  (2011).

\bibitem{PhysRevLett.104.160401}
{\sc T.~W. Neely}, {\sc E.~C. Samson}, {\sc A.~S. Bradley}, {\sc M.~J. Davis},
  and {\sc B.~P. Anderson},
\newblock {\em Phys. Rev. Lett.} {\bf 104}, 160401 (2010).

\bibitem{D.V.Freilich09032010}
{\sc D.~V. Freilich}, {\sc D.~M. Bianchi}, {\sc A.~M. Kaufman}, {\sc T.~K.
  Langin}, and {\sc D.~S. Hall},
\newblock {\em Science} {\bf 329}, 1182 (2010).

\bibitem{PhysRevA.84.011605}
{\sc S.~Middelkamp}, {\sc P.~J. Torres}, {\sc P.~G. Kevrekidis}, {\sc D.~J.
  Frantzeskakis}, {\sc R.~Carretero-Gonz\'alez}, {\sc P.~Schmelcher}, {\sc
  D.~V. Freilich}, and {\sc D.~S. Hall},
\newblock {\em Phys. Rev. A} {\bf 84}, 011605 (2011).

\bibitem{Middelkamp2010b}
{\sc S.~Middelkamp}, {\sc P.~Kevrekidis}, {\sc D.~Frantzeskakis}, {\sc
  R.~Carretero-González}, and {\sc P.~Schmelcher},
\newblock {\em Physica D} {\bf 240}, 1449  (2011).

\bibitem{Seman2009}
{\sc J.~A. Seman}, {\sc E.~A.~L. Henn}, {\sc M.~Haque}, {\sc R.~F. Shiozaki},
  {\sc E.~R.~F. Ramos}, {\sc M.~Caracanhas}, {\sc P.~Castilho}, {\sc
  C.~Castelo~Branco}, {\sc P.~E.~S. Tavares}, {\sc F.~J. Poveda-Cuevas}, {\sc
  G.~Roati}, {\sc K.~M.~F. Magalh\~aes}, and {\sc V.~S. Bagnato},
\newblock {\em Phys. Rev. A} {\bf 82}, 033616 (2010).

\bibitem{Mayteevarunyoo20091439}
{\sc T.~Mayteevarunyoo}, {\sc B.~A. Malomed}, {\sc B.~B. Baizakov}, and {\sc
  M.~Salerno},
\newblock {\em Physica D} {\bf 238}, 1439  (2009).

\bibitem{sakaguchi2005}
{\sc H.~Sakaguchi} and {\sc B.~Malomed},
\newblock {\em Europhys. Lett.} {\bf 72}, 698 (2005).

\bibitem{PhysRevLett.95.203904}
{\sc A.~S. Desyatnikov}, {\sc A.~A. Sukhorukov}, and {\sc Y.~S. Kivshar},
\newblock {\em Phys. Rev. Lett.} {\bf 95}, 203904 (2005).

\bibitem{PhysRevA.77.025602}
{\sc V.~M. Lashkin},
\newblock {\em Phys. Rev. A} {\bf 77}, 025602 (2008).

\bibitem{PhysRevA.85.013620}
{\sc V.~M. Lashkin}, {\sc A.~S. Desyatnikov}, {\sc E.~A. Ostrovskaya}, and {\sc
  Y.~S. Kivshar},
\newblock {\em Phys. Rev. A} {\bf 85}, 013620 (2012).

\bibitem{aniso}
{\sc J.~Stockhofe}, {\sc S.~Middelkamp}, {\sc P.~G. Kevrekidis}, and {\sc
  P.~Schmelcher},
\newblock {\em Europhys. Lett.} {\bf 93}, 20008 (2011).

\bibitem{PhysRevA.79.053616}
{\sc S.~McEndoo} and {\sc T.~Busch},
\newblock {\em Phys. Rev. A} {\bf 79}, 053616 (2009).

\bibitem{PhysRevA.82.013628}
{\sc S.~McEndoo} and {\sc T.~Busch},
\newblock {\em Phys. Rev. A} {\bf 82}, 013628 (2010).

\bibitem{PhysRevA.83.053612}
{\sc N.~Lo~Gullo}, {\sc T.~Busch}, and {\sc M.~Paternostro},
\newblock {\em Phys. Rev. A} {\bf 83}, 053612 (2011).

\bibitem{Kapitula2004263}
{\sc T.~Kapitula}, {\sc P.~G. Kevrekidis}, and {\sc B.~Sandstede},
\newblock {\em Physica D} {\bf 195}, 263  (2004).

\bibitem{PhysRevA.63.013602}
{\sc D.~V. Skryabin},
\newblock {\em Phys. Rev. A} {\bf 63}, 013602 (2000).

\bibitem{MacKay}
{\sc R.~S. MacKay} and {\sc J.~Meiss},
\newblock {\em Hamiltonian Dynamical Systems},
\newblock Hilger, Bristol, 1987.

\bibitem{Fetter2001}
{\sc A.~L. Fetter} and {\sc A.~A. Svidzinsky},
\newblock {\em J. Phys.: Cond. Matt.} {\bf 13}, R135 (2001).

\bibitem{Svidzinsky2000}
{\sc A.~A. Svidzinsky} and {\sc A.~L. Fetter},
\newblock {\em Phys. Rev. Lett.} {\bf 84}, 5919 (2000).

\bibitem{0953-4075-43-15-155303}
{\sc S.~Middelkamp}, {\sc P.~G. Kevrekidis}, {\sc D.~J. Frantzeskakis}, {\sc
  R.~Carretero-Gonz\'{a}lez}, and {\sc P.~Schmelcher},
\newblock {\em J. Phys. B} {\bf 43}, 155303 (2010).

\bibitem{Kevrekidis2004c}
{\sc P.~G. Kevrekidis}, {\sc R.~Carretero-Gonz\'alez}, {\sc D.~J.
  Frantzeskakis}, and {\sc I.~G. Kevrekidis},
\newblock {\em Mod. Phys. B} {\bf 18}, 1481 (2004).

\bibitem{Newton2009}
{\sc P.~K. Newton} and {\sc G.~Chamoun},
\newblock {\em Siam Review} {\bf 51}, 501 (2009).

\bibitem{0295-5075-55-1-045}
{\sc M.~S. Jean}, {\sc C.~Even}, and {\sc C.~Guthmann},
\newblock {\em Europhys. Lett.} {\bf 55}, 45 (2001).

\bibitem{PhysRevE.72.046122}
{\sc S.~W.~S. Apolinario}, {\sc B.~Partoens}, and {\sc F.~M. Peeters},
\newblock {\em Phys. Rev. E} {\bf 72}, 046122 (2005).

\bibitem{kelley}
{\sc C.~T. Kelley},
\newblock {\em Solving Nonlinear Equations with Newton's Method},
\newblock Society for Industrial and Applied Mathematics, Philadelphia, 2003.

\bibitem{PhysRevLett.86.564}
{\sc D.~L. Feder}, {\sc A.~A. Svidzinsky}, {\sc A.~L. Fetter}, and {\sc C.~W.
  Clark},
\newblock {\em Phys. Rev. Lett.} {\bf 86}, 564 (2001).

\bibitem{PhysRevA.85.043627}
{\sc B.~Nowak}, {\sc J.~Schole}, {\sc D.~Sexty}, and {\sc T.~Gasenzer},
\newblock {\em Phys. Rev. A} {\bf 85}, 043627 (2012).

\bibitem{Frantzeskakis2010}
{\sc D.~J. Frantzeskakis},
\newblock {\em J. Phys. A} {\bf 43}, 213001 (2010).

\bibitem{PhysRevE.74.056608}
{\sc G.~Theocharis}, {\sc P.~G. Kevrekidis}, {\sc D.~J. Frantzeskakis}, and
  {\sc P.~Schmelcher},
\newblock {\em Phys. Rev. E} {\bf 74}, 056608 (2006).

\bibitem{PhysRevA.73.043615}
{\sc D.~Mihalache}, {\sc D.~Mazilu}, {\sc B.~A. Malomed}, and {\sc F.~Lederer},
\newblock {\em Phys. Rev. A} {\bf 73}, 043615 (2006).

\bibitem{Kapitula2007112}
{\sc T.~Kapitula}, {\sc P.~Kevrekidis}, and {\sc R.~Carretero-Gonz\'alez},
\newblock {\em Physica D} {\bf 233}, 112  (2007).

\bibitem{PhysRevA.59.1533}
{\sc H.~Pu}, {\sc C.~K. Law}, {\sc J.~H. Eberly}, and {\sc N.~P. Bigelow},
\newblock {\em Phys. Rev. A} {\bf 59}, 1533 (1999).

\bibitem{PhysRevLett.93.160406}
{\sc Y.~Shin}, {\sc M.~Saba}, {\sc M.~Vengalattore}, {\sc T.~A. Pasquini}, {\sc
  C.~Sanner}, {\sc A.~E. Leanhardt}, {\sc M.~Prentiss}, {\sc D.~E. Pritchard},
  and {\sc W.~Ketterle},
\newblock {\em Phys. Rev. Lett.} {\bf 93}, 160406 (2004).

\bibitem{Theocharis2003a}
{\sc G.~Theocharis}, {\sc D.~J. Frantzeskakis}, {\sc P.~G. Kevrekidis}, {\sc
  B.~A. Malomed}, and {\sc Y.~S. Kivshar},
\newblock {\em Phys. Rev. Lett.} {\bf 90}, 120403 (2003).

\bibitem{Law2010}
{\sc K.~J.~H. Law}, {\sc P.~G. Kevrekidis}, and {\sc L.~S. Tuckerman},
\newblock {\em Phys. Rev. Lett.} {\bf 105}, 160405 (2010).

\bibitem{PhysRevLett.86.2926}
{\sc B.~P. Anderson}, {\sc P.~C. Haljan}, {\sc C.~A. Regal}, {\sc D.~L. Feder},
  {\sc L.~A. Collins}, {\sc C.~W. Clark}, and {\sc E.~A. Cornell},
\newblock {\em Phys. Rev. Lett.} {\bf 86}, 2926 (2001).

\bibitem{PhysRevLett.94.040403}
{\sc N.~S. Ginsberg}, {\sc J.~Brand}, and {\sc L.~V. Hau},
\newblock {\em Phys. Rev. Lett.} {\bf 94}, 040403 (2005).

\end{thebibliography}

\end{document}